\documentclass[]{aa}
\usepackage[varg]{txfonts}
\usepackage{graphicx}
\usepackage{booktabs}
\usepackage{amsmath}
\usepackage{colortbl}
\usepackage{subfig}
\usepackage{rotating}
\usepackage{amssymb}
\usepackage{latexsym}
\usepackage{ifthen}
\usepackage{enumitem}
\usepackage[colorlinks=true,allcolors=blue]{hyperref}
\usepackage{hyperref}
\hypersetup{colorlinks = true, citecolor = {blue}}

\usepackage{psfrag,color,ulem,gensymb}
\definecolor{deepblue}{rgb}{0,0,0.5}

\begin{document}

\title{MUSE view of Arp220: Kpc-scale multi-phase outflow and evidence for positive feedback} 
\titlerunning{MUSE IFS of multi-phase outflow in Arp220} 

\author{M. Perna
                \inst{\ref{i1},\ref{i2}}\thanks{E-mail: mperna@cab.inta-csic.es}
                \and 
        S. Arribas
                \inst{\ref{i1}} 
                \and
        C. Catal\'an-Torrecilla
                \inst{\ref{i1}} 
                \and        	
        L. Colina
        		\inst{\ref{i1}}
		\and
	E. Bellocchi
		\inst{\ref{i3}}
		\and
	A. Fluetsch
		\inst{\ref{i4}, \ref{i5}}
		\and
	R. Maiolino
		\inst{\ref{i4}, \ref{i5}}
		\and
	S. Cazzoli
        		\inst{\ref{i6}}
		\and
	A. Hern\'an Caballero
		\inst{\ref{i7}}
		\and
	M. Pereira Santaella
	        \inst{\ref{i1}} 
                \and
        J. Piqueras L\'opez
                \inst{\ref{i1}} 
                \and
         B. Rodr\'iguez del Pino 
                \inst{\ref{i1}} 
}


\institute{Centro de Astrobiolog\'ia (CAB, CSIC--INTA), Departamento de Astrof\'\i sica, Cra. de Ajalvir Km.~4, 28850 -- Torrej\'on de Ardoz, Madrid, Spain\label{i1}
       \and
       INAF - Osservatorio Astrofisico di Arcetri, Largo Enrico Fermi 5, I-50125 Firenze, Italy\label{i2}
       \and
       Centro de Astrobiolog\'ia (CSIC-INTA), ESAC Campus, 28692 Villanueva de la Ca\~nada, Madrid, Spain\label{i3}
       \and
       University of Cambridge, Cavendish Laboratory, Cambridge CB3 0HE, UK\label{i4}
       \and
       University of Cambridge, Kavli Institute for Cosmology, Cambridge CB3 0HE, UK\label{i5}
       \and
       IAA - Instituto de Astrofísica de Andalucía (CSIC), Apdo. 3004, 18008, Granada, Spain\label{i6}
       \and
       Centro de Estudios de F\'isica del Cosmos de Arag\'on (CEFCA), Plaza de San Juan, 1, 44001 Teruel, Spain\label{i7}
}

\date{Received 2 November 1992 / Accepted 7 January 1993}

\abstract 
{ 
Arp220 is the nearest and prototypical ULIRG, and shows evidence of pc-scale molecular outflows in its nuclear regions and strongly perturbed ionised gas kinematics on kpc scales. It is therefore the ideal system for investigating outflow mechanisms and feedback phenomena in details. 
}  
{ 
We investigate the feedback effects on the Arp220 interstellar medium (ISM), deriving a detailed picture of the atomic gas in terms of  physical and kinematic properties, with a spatial resolution never obtained before ($0.56''$, i.e. $\sim 210$ pc).
} 
{ 
We use optical integral-field spectroscopic (IFS) observations from VLT/MUSE-AO to obtain spatially resolved stellar and gas kinematics, for both ionised ([N {\small II}]$\lambda6583$) and neutral (Na {\small ID}$\lambda\lambda5891,96$) components; we also derive dust attenuation, electron density, ionisation conditions and hydrogen column density maps to characterise the ISM properties. 
} 
{
Arp220 kinematics reveal the presence of a disturbed, kpc-scale disk in the innermost nuclear regions, and highly perturbed, multi-phase (neutral and ionised) gas along the minor-axis of the disk, which we interpret as a galactic-scale outflow emerging from the Arp220 eastern nucleus. This outflow involves velocities up to $\sim 1000$ km/s at galactocentric distances of $\approx 5$ kpc, and has a mass rate of $\sim 50$ M$_\odot$/yr, and kinetic and momentum power of $\sim 10^{43}$ erg/s and $\sim 10^{35}$ dyne, respectively. The inferred energetics do not allow us to distinguish the origin of the outflows, i.e. whether they are AGN-driven or starburst-driven. We also present evidence for enhanced star formation at the edges of - and within - the outflow, with a star formation rate SFR $\sim 5$ M$_\odot$/yr (i.e. $\sim 2\%$ of the total SFR). 
}
{
Our findings suggest the presence of powerful winds in Arp220: they might be capable of removing or heating large amounts of gas from the host (``negative feedback''), but could be also responsible for  triggering star formation (``positive feedback’').
}

\keywords{galaxies:active - galaxies: starburst - galaxies: individual: Arp 220 - galaxies:ISM}
\maketitle

\section[Introduction]{Introduction}

The formation and evolution of galaxies are profoundly coupled with the growth of supermassive black holes (BH) sitting in their centre (e.g. \citealt{Kormendy2013}).  To explain this coupling, it has been proposed that both the stellar population and the central BH of a galaxy 
grow and evolve by the merging of smaller gas-rich systems (\citealt{Sanders1988,DiMatteo2005,Hopkins2008}). In this scenario,  ultra-luminous infrared galaxies (ULIRGs, $L_{8-1000\mu m} > 10^{12}$ L$_\odot$) appear during the final coalescence of the galaxies, when massive inflows of cool material trigger intense starbursts (SBs) in the nuclear regions, while the BH may be buried by  dust and gas which feed the BH at high rates, causing the birth of an obscured active galactic nucleus (AGN).

Powerful outflows are routinely invoked to explain the tight BH-galaxy scaling relations: these phenomena are expected to affect the physical and dynamical conditions of the interstellar medium (ISM) and thus regulate with a feedback mechanism the formation of new stars and the accretion onto the BH (e.g. \citealt{Somerville2015}). 
Outflows are thought to be originated  either from BH accretion disc winds (AGN-driven outflows; e.g. \citealt{King2015}), during vigorous growth phases of the BH (i.e. at  high AGN luminosities and Eddington ratios; e.g. \citealt{Fiore2017,Perna2017a,Villar2020}), or from stellar winds and
supernovae (SNe) explosions (SB-driven outflows; e.g. \citealt{Hopkins2012}; but see also e.g. \citealt{Naab2017} and \citealt{Vogelsberger2020}). Consequently, outflows might play a crucial role in the evolution of galaxies, especially at high redshifts ($z >1$), where episodes of intense star-formation (SF) and strong AGN activity were very common (e.g. \citealt{Brusa2015,Talia2017,Schreiber2018,Kakkad2020}).

Local ULIRGs, which are powered by strong SB (\citealt{Genzel1998}) and/or AGN (\citealt{Nardini2010}), present some properties similar to those of high-$z$ luminous and dusty star-forming galaxies (e.g. \citealt{Arribas2012}, \citealt{Hung2014}; but see also e.g. \citealt{Tacconi2018}), opening therefore the opportunity of investigating the properties of outflows and their feedback effects at the relevant scales (i.e. $<$ kpc), and much better than achievable at high-z. In the past, several studies have demonstrated the presence of multi-phase (ionised, neutral, molecular) gas outflows in ULIRGs (e.g., \citealt{Westmoquette2012, Bellocchi2013,Veilleux2013, Arribas2014, Cicone2014, Cazzoli2016,Fluetsch2019,Fluetsch2020}). Most of these works, however, have focused on the study of a specific gas phase and, more critically, have resolutions ($>$kpc) unable to trace in detail the outflow structure. High resolution and multi-wavelength observations are therefore required to study outflows in detail (e.g. \citealt{Emonts2017,PereiraSantaella2018,Cicone2020}).

We have recently started a project aimed at studying at sub-kpc scales the 2D, multi-phase outflow structure in a representative volume-limited ($z < 0.165$) sample of 24 local ULIRGs, combining the capabilities offered by ALMA, required to trace the molecular gas, and VLT/MUSE-AO, needed to trace the atomic neutral and ionised gas. The selection criteria, the main properties of the sample, and the path chosen to analyse and investigate multi-phase outflows will be discussed in a forthcoming paper. 
In this paper, we present MUSE observations of the archetypal target Arp220, whose nuclear molecular outflow has already been studied with ALMA at $\sim 0.1''$ resolution (\citealt{BarcosMunoz2018, Wheeler2020}). 
This manuscript is organised as follows. In Sect. \ref{Sanchillary} we present the ancillary target data. Sect. \ref{reduction} presents the MUSE observations and data reduction. Sect. \ref{Sanalysis} presents the spectral fitting analysis. 
Sect. \ref{Sgaskinematics} reports the spatially resolved kinematics of the atomic gas. In Sect. \ref{ionisationconditions} we study the ionisation mechanisms for the emitting gas; Sects. \ref{Sne}, \ref{Sextinction} and \ref{Snh} present the main ISM properties derived from standard optical line diagnostics. Finally, in Sects. \ref{Soutflow} and \ref{Ssf} we characterise the kpc-scale outflow identified in the MUSE field and its effects on the ISM. Sect. \ref{Sconclusions} summarises our conclusions. 
Throughout this paper, we adopt the cosmological parameters $H_0 =$ 70 km/s/Mpc, $\Omega_m$ = 0.3 and $\Omega_\Lambda =$ 0.7, and  a Salpeter initial mass function.

\section{Arp220 properties}\label{Sanchillary}

\begin{figure}[t]
\centering
\includegraphics[width=9.cm,trim= 0 20 3 20,clip]{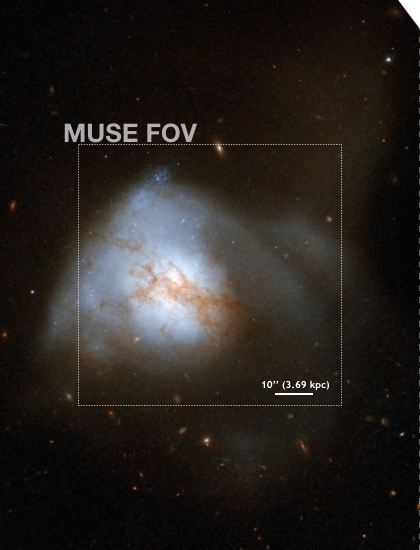}
\caption{\small   Three-colour optical image of Arp220 obtained combining HST observations performed through three different filters (B, B+I, I). The white box indicates the region analysed in this work, corresponding to the MUSE FOV. North is up.  Credit: NASA, ESA, the Hubble Heritage Team (STScI/AURA)-ESA/Hubble Collaboration and A. Evans (University of Virginia, Charlottesville/NRAO/Stony Brook University).
}
\label{HST}
\end{figure}

Arp220 is the nearest ULIRG ($D_L \approx 77$ Mpc) and is thus an ideal laboratory to study the processes taking place in SB galaxies. It is a late-stage merger with a peculiar morphology and a heavily obscured central region (Fig. \ref{HST}). Near-IR and radio observations revealed two bright sources about $1''$ (370 pc) apart, thought to be the nuclei of the merging galaxies (e.g. \citealt{Scoville1997}). Molecular gas emission revealed the presence of two discs around the nuclei, with radii of $\lesssim 100$pc (e.g. \citealt{Scoville1997,BarcosMunoz2015}), and an outer disc likely formed from the gas of the progenitor galaxies, with a radius of $\sim 1$ kpc (e.g. \citealt{Scoville1997,Sakamoto1999}). The spin axis of the eastern (E) nucleus is aligned with that of the kpc-scale disk, while the western (W) nucleus rotates in the opposite direction. These observations suggest that Arp220 is the product of a prograde-retrograde merger of two gas-rich spiral galaxies (\citealt{Scoville1997}; see also \citealt{Hibbard2000}).  

\begin{table*}
\footnotesize
\begin{minipage}[!h]{1\linewidth}
\setlength{\tabcolsep}{9pt}
\centering
\caption{Sources in the MUSE FoV towards Arp220}
\begin{tabular}{lccccc}
source ID    &            RA              & DEC &  $z$        & wavelength range  & resolution\\
             & ($^h:\ ^m:\ ^s$) & ($^{\circ}:\ ':\ '')$ & &                  ($\AA$)                   & ($''$ / kpc)\\
\scriptsize{(1)} & \scriptsize{(2)}   &\scriptsize{(3)}   & \scriptsize{(4)}   &\scriptsize{(5)}   & \scriptsize{(6)}   \\
\hline
Arp220  & 15:34:57.28 &+23:30:11.64& $0.0181 \pm 0.0001$ & 4640-9100    & 0.56 / 0.21\\
id2070   & 15:34:56.47 &+23:30:43.60& $0.0901\pm 0.0022$ & 4310-8575     & 0.56 / 0.94\\
id1644   & 15:34:57.15 &+23:29:44.60& $0.4987\pm 0.0001$ & 3170-6237     & 0.56 / 3.45\\ 
Gal. I     & 15:34:56.99 &+23:29:50.51& $0.4992 \pm 0.0001$ & 3170-6237    & 0.56 / 3.45\\ 
Gal. II    & 15:34:56.45 &+23:30:01.06& $0.5636 \pm 0.0001$ & 3500-5980    & 0.56/ 3.67\\ 
id1641   & 15:34:54.60 &+23:30:27.21& $0.7279\pm 0.0001$ & 2270-5410      & 0.56 / 4.11\\ 
Gal. III   & 15:34:54.98 &+23:30:21.89& $0.7277\pm 0.0001$ & 2270-5410      & 0.56 / 4.11\\ 
Gal. IV   & 15:34:55.33 &+23:30:18.69& $1.0128\pm 0.0004$ & 2335-4644      & 0.56 / 4.56\\ 
id1679   & 15:34:54.45 &+23:29:51.41& $1.2979\pm 0.0001$ & 2046-4070      & 0.56 / 4.76\\ 

\hline
\end{tabular}
\label{tab1}
\vspace{0.2cm}
\end{minipage}
{\small
{\it Note.} Column (1): target name (id from DECaLS DR7 catalog for known sources; 'Gal. \#', with \# from I to IV for new discovered targets; see Appendix \ref{Abkg}); (2) and (3): RA and DEC coordinates; (4): spectroscopic redshift; (5):  Rest-frame wavelength range covered by MUSE IFS; (6): angular resolution and physical scale at the redshift of the detected target. 
}
\end{table*}

Arp220 experiences a powerful SB at each of the nuclei, which results in the IR prominence ($L_{IR} \sim 5.6\times 10^{45}$ erg/s; \citealt{Nardini2010}) and multiple radio SNe and SN remnants at each of the nuclei (\citealt{Lonsdale2006,Varenius2019}). There is still no convincing direct evidence, from radio to hard
X-ray wavelengths, for AGN in Arp 220 (e.g. \citealt{Sakamoto2008,Scoville2015,Aalto2015,Teng2015,Paggi2017}). However, indirect arguments suggest that an active nucleus produces a significant fraction of the radiated power in the W nucleus (Wilson et al. 2014; Rangwala et al. 2015). Additional evidence for the presence of an AGN in the latter nucleus has been recently provided by {\it Fermi} detection of high-energy $\gamma$-rays (\citealt{YoastHull2017}). If an AGN is present in Arp220, it is highly Compton-thick (CT; see e.g. \citealt{Teng2015}). 
X-ray and IR data have been used to evaluate the AGN luminosity in Arp220: from the 2-10 keV emitted luminosity, \citet{Paggi2017} estimated (lower limit) AGN bolometric luminosities $\sim 8.3\times 10^{43}$ erg/s (E) and $\sim 2.5\times 10^{43}$ erg/s (W nucleus), representing only $\sim 1\%$ of the bolometric luminosity ($L_{bol} \sim 6\times 10^{45}$ erg/s, \citealt{Sanders1988}); from  $5-8$ $\mu$m spectral analysis of {\it Spitzer} data, \citet{Nardini2010} estimated an AGN luminosity of the order of $\sim 17\%$ of $L_{bol}$. \citet{Veilleux2009} quantified the AGN contribution to $L_{bol}$ using six independent methods based on $5-35\mu$m  {\it Spitzer} data, obtaining a range of values from 0\% to $<37\%$, and an average contribution of  $\sim 18.5\%$ of $L_{bol}$. These results confirm that overall the emission of Arp220 is likely dominated by the SB component. 

Whatever the energy source, Arp220 is of considerable astrophysical significance. It is a recent example of a short-lived burst of assembly and nuclear activity that are common at $z> 1$, and can be used to shed light on the main physical processes governing the baryon cycle, from the fuelling of nuclear SBs and AGN with dust and gas  to the feedback mechanisms that return part of this material into the circum-galactic environment.    
Evidence of molecular outflows in Arp220 have been inferred from the analysis of the integrated spectra of both nuclei 
(e.g. \citealt{BarcosMunoz2018} and references therein). In particular, \citet{BarcosMunoz2018} reported the first spatially ($\sim 0.1''$) and spectrally resolved image of the molecular outflow in the W nucleus, using ALMA observations of the HCN (1-0) transition. This outflow is compact and collimated, with a bipolar morphology and an extension $\lesssim 120$ pc (along the north-south direction). 
Recently, \citet{Wheeler2020} presented new observational evidence of collimated outflows in both nuclei, using ALMA observations of carbon monoxide transition, revealing for the first time an outflow in the E nucleus oriented along the kinematic minor axis of the circumnuclear molecular disk.

The ionised gas emission in Arp220 is spatially extended. XMM-Newton (\citealt{Iwasawa2005}) and Chandra (\citealt{McDowell2003,Paggi2017}) data show gas emission from several structures along the direction of the minor axis of the kpc-scale nuclear disk, on scales from $1''$ (in hard X-ray) to several arcminutes (in soft X-ray; see e.g. Fig. 1 in \citealt{McDowell2003}). Very similar structures are observed in narrow-band H$\alpha$ images (e.g. \citealt{Heckman1987,Taniguchi2012}).  
 This ionised gas is more extended than the region from which IR luminosity originates (\citealt{Scoville1998}), and the possible connection between the nuclear and kpc-scale emitting ISM has not been investigated in detail yet. Early integral field spectroscopy (IFS) with INTEGRAL (\citealt{Arribas2001,Colina2004}) has shown complex kinematics in H$\alpha$ and [N {\small II}]$\lambda 6583$ ionised gas along the same direction of soft X-ray. 
 These results were interpreted as indicative of a biconical outflow,  related to a galactic wind driven by a central SB and/or AGN  (see also e.g. \citealt{Heckman1990,Lockhart2015}).  
 
However, these early IFS observations were designed to map the Arp220 large scale structures (with pixel scales $>1''$), making difficult to explain the kinematics observed in the different sub-structures cospatial with X-ray emission. Moreover, these IFS data suffer heavy obscuration by dust, preventing the detection of [O {\small III}]$\lambda5007$ and H$\beta$ in most of the regions and, hence, the use of standard diagnostics to characterise the physical conditions of the ISM. The MUSE IFS observations presented in this paper overcome these limitations. 

\begin{figure*}[h]
\centering
\includegraphics[width=17.3cm,trim=20 40 60 30,clip]{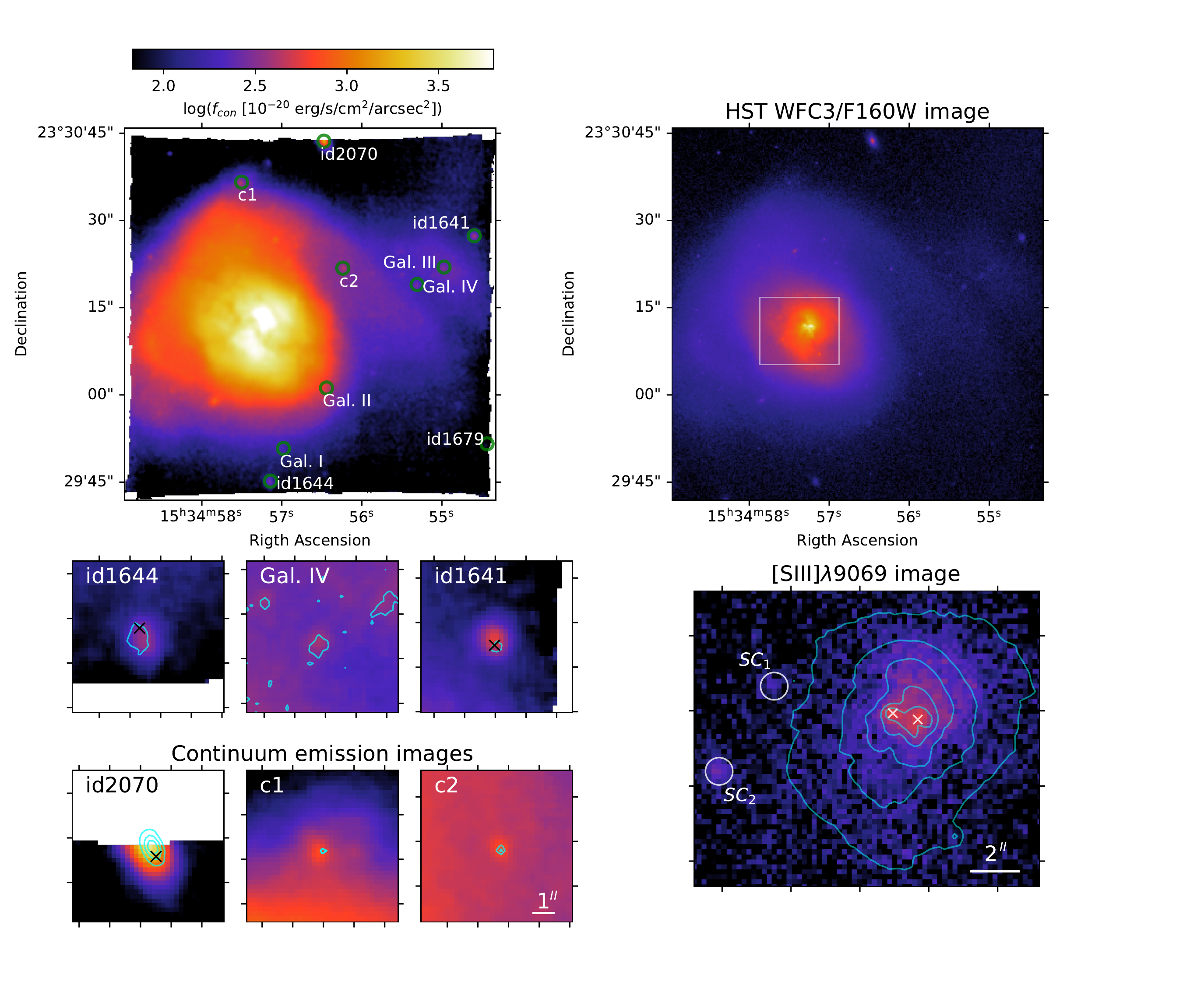}
\caption{\small  {\it Upper left:} Continuum emission from MUSE data cube, having collapsed the data cube in the range  $8025-8125\AA$ (rest-frame). A few bright Arp220 clumps (c1 and c2) and background galaxies  are labeled (see Appendix \ref{Abkg} and Table \ref{tab1}); these knots are used to perform a bona fide astrometric registration of the MUSE data (see Sect. \ref{Satrometry}). {\it Upper right:} HST/WFC3 F160W image from HST archive (total exposure time of 172 s and pixel scale of 0.13$''$; PI: Larson). The image shows the region analysed in this work; the white box indicates the portion displayed in the bottom right panel. {\it Bottom left:} Zoom-in insets of $7''\times 7''$  showing the continuum emission map around some of the sources selected in the upper left panel. The overlaid cyan contours represent the NIR emission from the HST image. For the sources in the DECaLS DR7 catalog (\citealt{Dey2018}), we mark the DECaLS positions with black crosses. {\it Bottom-right}: [S {\small III}]$\lambda9069$ image obtained integrating the flux in the wavelength range $9065-9083\AA$, after subtracting the continuum emission, with overlaid contours of the NIR emission (upper right panel). Crosses mark the [S {\small III}] nuclear peaks; white circles mark the position of two star-forming clumps identified in this work (see Sect. \ref{ionisationconditions}). 
North is up. 
}
\label{FIG1}
\end{figure*}

\section{MUSE observations and data reduction}\label{reduction}

Arp220 observations were conducted as part of our programme ``Sub-kpc multi-phase gas structure of massive outflows in Ultraluminous Infrared Galaxies'' (ESO projects 0103.B-0391(A) and  0104.B-0151(A), PI: Arribas). The observations presented in this paper were carried out with MUSE in ESO Period 103, with its Adaptive Optics Wide Field Mode (AO-WFM; \citealt{Bacon2010}). 
MUSE covers a $60''$$\times 60''$ field of view (FOV) with a sampling of $0.2''$$\times 0.2''$, resulting in a massive dataset of $\sim 90000$ individual spectra. We used the nominal instrument setup, with a spectral coverage from 4750 to 9350$\AA$ and a mean resolution of 2.65 $\AA$ (FWHM). Because of the use of AO with sodium laser guide system, the wavelength range from $\approx$ 5800 to 5970$\AA$ is blocked, to avoid contamination and saturation of the detector by sodium light.  

The requested observations were distributed in three 40-minutes Observing Blocks (OBs) with a total integration time of 2 hours on source. We split the exposures in each OB into four (dithered
and 90$^{\circ}$-rotated) frames of 585 s; because Arp220 fills most of the FOV, we also observed a black sky field for 29s. 
Observations were performed with a clear sky transparency and a seeing $\approx 1''$; the rms of the flux variation was $\approx 1\%$, as measured during the night by the VLT DIMM station. Unfortunately, only one OB was observed, for a total observing time of 40 min on source. 

Arp220 observations were reduced using the MUSE esoreflex pipeline recipes (muse - 2.6.2), which provides a fully calibrated and combined MUSE data cube. In the last step of our data reduction, we identified and subtracted the residual sky contamination in the final data cube using the Zurich Atmosphere Purge (ZAP) software package (Soto et al.
2016). The residual contamination, due to the time difference between the sky and target acquisitions, was derived from the outermost regions of the MUSE object cube, where the sky is dominant, after masking Arp220 emission as well as background sources in the FOV.

\subsection{Arp220 continuum emission and background galaxies}
Figure \ref{FIG1} shows an image of the Arp220 continuum emission, obtained by collapsing the MUSE data cube in the wavelength range $8025-8125\AA$ (rest-frame). This range is not affected by strong ISM emission lines and stellar absorption features, and is free of sky lines; moreover, it is less affected by dust absorption with respect to shorter wavelengths, and can provide a good tracer for the continuum emission in Arp220.  A few compact sources can be identified within this MUSE narrow-band image (insets in Fig. \ref{FIG1}, bottom left). Most of them are resolved, and are associated with strong optical emission (e.g. H$\alpha$, [N {\small II}]) and absorption (e.g. Na {\small ID}$\lambda\lambda5891,96$) lines at the same redshift of Arp220; others are associated with background galaxies. 
In Table \ref{tab1} we report the coordinates of these sources, and the spectroscopic redshifts derived in this work (see  Appendix \ref{Abkg}).

\begin{figure*}[h]
\centering
\includegraphics[width=17.cm,trim=0 30 0 40,clip]{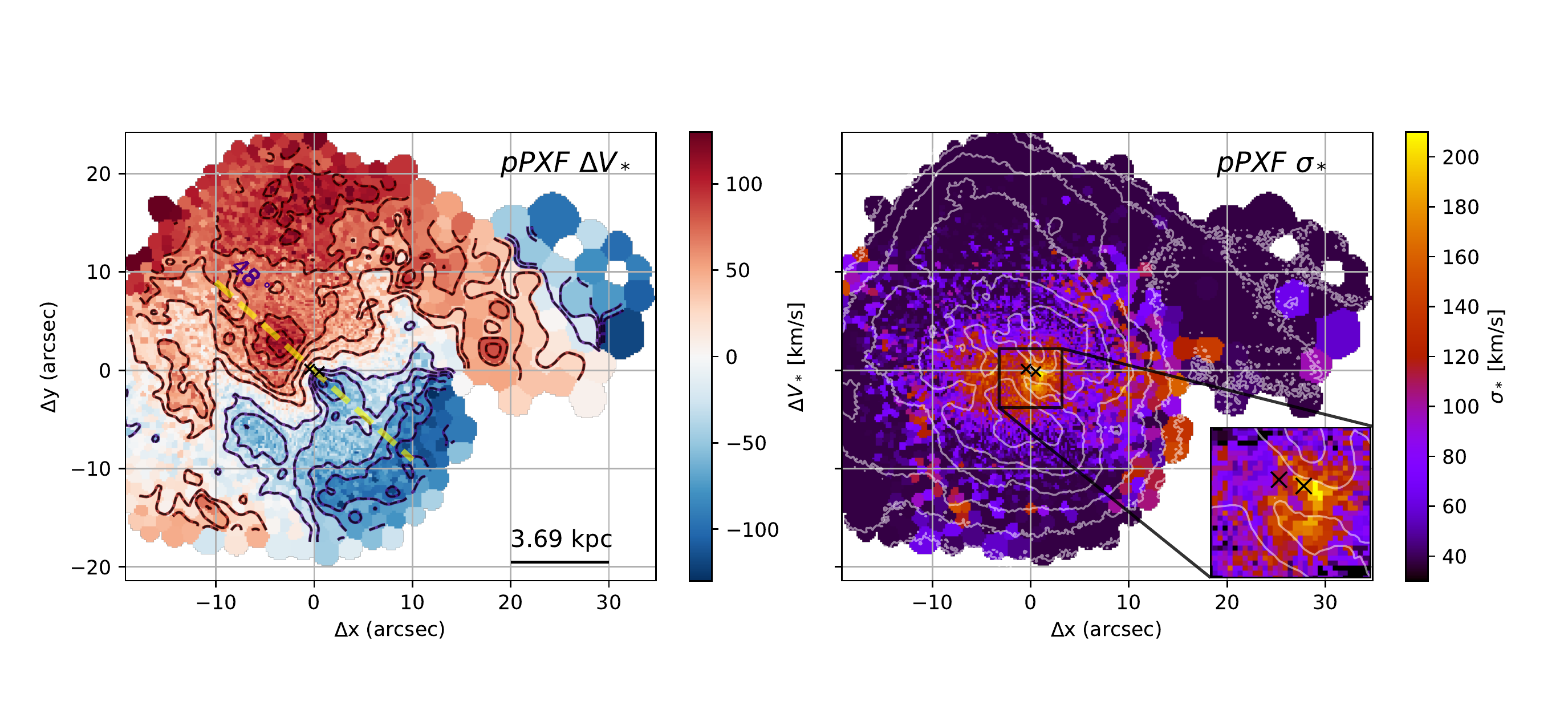}
\caption{\small  Arp220 stellar kinematic maps from our pPXF analysis. The contours in the left panel show the negative (dark-blue) and positive (dark-red) isovelocity curves, from $-100$ to $+100$ km/s, equally spaced in steps of 20 km/s; the contours in the right panel are derived from the continuum emission image shown in Fig. \ref{FIG1} (top left), and are equally spaced in steps of 0.25 dex, starting from  $1.3 \times 10^{-16}$ erg/s/cm$^2$/arcsec$^2$. The crosses mark the two nuclei;  yellow dashed line in the left panel shows PA $=48^\circ$, corresponding to the 'major axis' of Arp220. In the right panel, the inset shows a zoom-in in the innermost nuclear regions, highlighting the ring-like shape with high $\sigma_*$ values (using a slightly different colour-bar); the position of the two nuclei is shown with black crosses. Stellar velocity dispersions are not corrected for the instrumental broadening. North is up. 
}
\label{FIG2}
\end{figure*}

\subsection{Astrometric registration and angular resolution}\label{Satrometry}
Since no bright point sources are present in the FOV, we cannot estimate a proper spatial resolution nor obtain an astrometric registration with foreground stars. We took advantage of available near-infrared HST image (Fig \ref{FIG1}, right) to match the position of the individual sources in the FOV (Fig \ref{FIG1}, bottom-left insets). We also derived a MUSE equivalent narrow-band image for the [S {\small III}]$\lambda$9069 emission line, after subtracting the continuum emission using the adjacent regions at shorter and longer wavelengths with respect to the sulfur line systemic. The regions with bright [S {\small III}]$\lambda$9069 emission perfectly resemble the NIR HST flux distribution in the nuclear regions of Arp220 (Fig \ref{FIG1}, bottom-right inset). We therefore obtained a bona fide astrometric registration matching all the sources reported in the insets of Fig. 1 with those in the HST image\footnote{Starting from the astrometry derived from esoreflex pipeline, we had to apply a correction of $\Delta RA = -2.1''$ and $\Delta DEC = 0.8''$; no rotational term has been taken into account. }. We verified our astrometry by comparing the new coordinates of the background galaxies with those in the DECaLS Survey DR7 catalogue (\citealt{Dey2018}; black crosses in Fig \ref{FIG1}, bottom-left insets), obtaining a good match. 

The spatial resolution of MUSE data can be roughly estimated from the [S {\small III}]$\lambda$9069 emission peaks in Fig \ref{FIG1}, bottom-right panel, and the (apparently) point source galaxy at $z \sim 0.56$ (Fig. \ref{aFIG2}). The two brightest peaks in the [S {\small III}]$\lambda$9069 map (crosses in Fig. \ref{FIG1}, bottom right panel) can be reasonably associated with the position of the two nuclei of Arp220, since they perfectly match the NIR emission, which in turn match the location of strong X-ray emission (e.g. \citealt{Lockhart2015}; see also e.g. \citealt{Paggi2017}). We stress here that [S {\small III}]$\lambda$9069 emission provides the only diagnostic in MUSE data-cube to locate the two nuclei: in fact, all emission lines detected at shorter wavelengths and the continuum emission are strongly absorbed by dust. 
The two additional, compact clumps with strong [S {\small III}] emission in Fig. \ref{FIG1} present instead properties attributable to SF, according to their optical line ratio diagnostics and kinematics (Sects. \ref{ionisationconditions}, \ref{Ssf}); they are therefore labeled in this work as star-forming clumps (SC hereinafter). 

For each (apparently) point source in the MUSE FOV, we performed a 2D gaussian fit, and derived an estimate for the angular resolution from the FWHM of the Gaussian fit. We obtained a resolution of $\sim 0.56''$, corresponding to $\sim 0.21$ kpc at the distance of Arp220.


\section{Spectral fitting analysis}\label{Sanalysis}

\subsection{Stellar component modelling}

\begin{figure*}[h]
\centering
\includegraphics[width=17.cm,trim=0 0 0 0,clip]{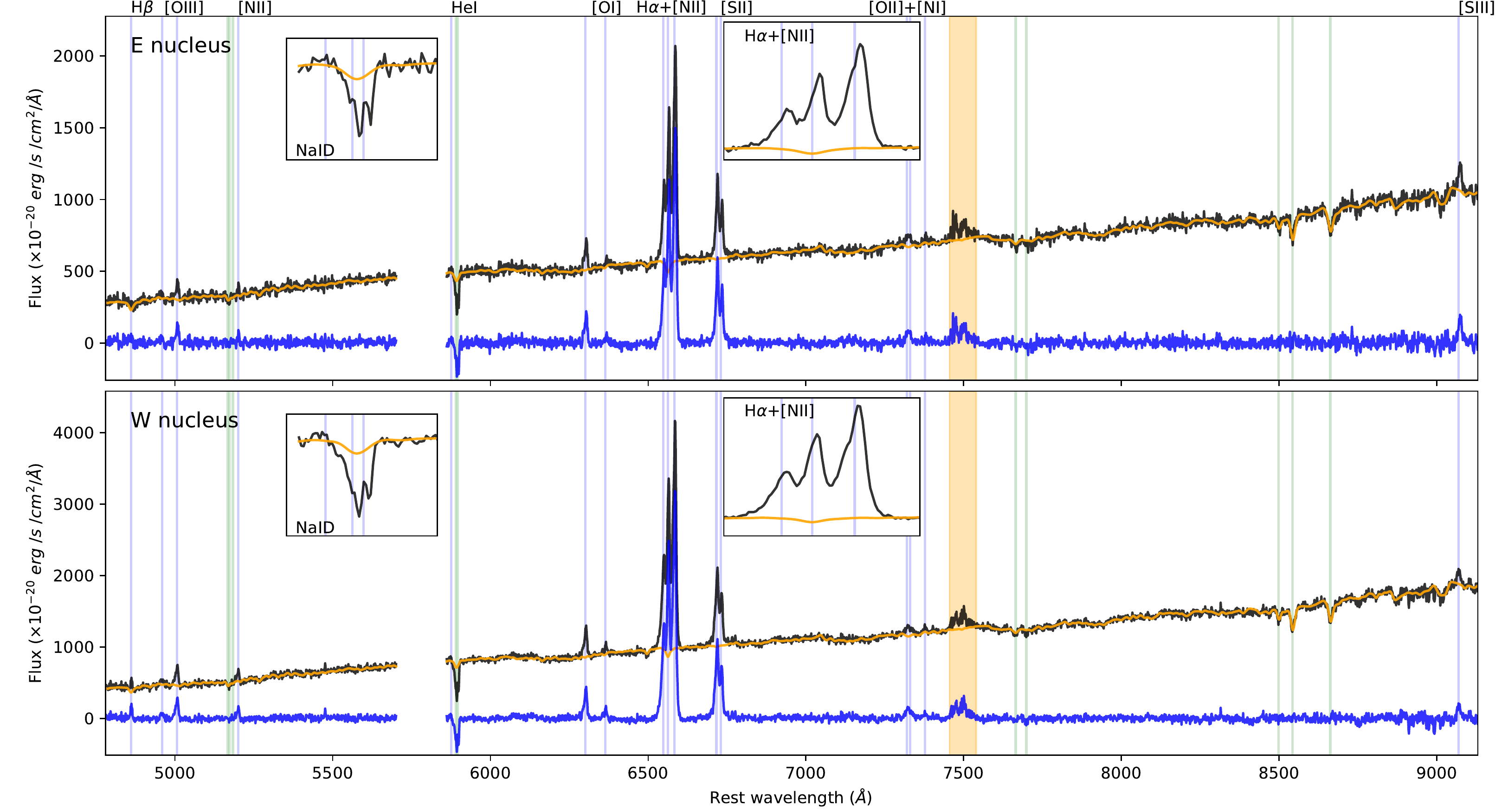}
\caption{\small  Eastern (top panel) and western (bottom) nucleus spectra (black curves), extracted from $2\times 2$ pixel regions. The corresponding pPXF best-fit model profiles are shown with orange curves. The pure emission/absorption ISM spectra (blue curves) are obtained subtracting the best-fit stellar contribution from the original spectra. The insets show the spectra and stellar models around Na {\small ID} and the H$\alpha$+[N {\small II}] complex. The blue vertical lines mark the wavelengths of the emission lines detected in the two spectra; the green lines mark the position of stellar absorption systems (i.e. from left to right: MgI triplet, Na {\small ID} and KI doublets, CaII triplet). The region excluded from the pPXF fits and corresponding to the most intense sky line residuals are highlighted as orange shaded areas; the portion of the spectra around $5700\AA$ is missing, because of a filter blocking the laser contamination. 
}
\label{FIG3}
\end{figure*}

Prior to modelling the Arp220 ISM features, we used the penalized pixel fitting routines (pPXF; \citealt{Cappellari2004,Cappellari2017}) to extract the stellar kinematics. 
The entire wavelength range covered with MUSE (i.e. 4640-9100$\AA$, rest-frame) was used in our analysis to model the continuum emission and recover the stellar kinematics from stellar absorption lines, after masking all optical emission lines detected in MUSE data cube (see e.g. Fig. \ref{FIG3}). In addition, we masked the resonant Na {\small ID} transitions, as both stellar and interstellar absorption can be at the origin of these lines. Finally, we excluded from this analysis a narrow wavelength region at $\approx  7630\AA$ (observer frame; see e.g. Fig. \ref{FIG3}), associated with strong sky-subtraction residuals, and the region 5800-5970$\AA$ which is blocked by a filter to avoid contamination from the Lasers (see Sect. \ref{reduction}). 

We made use of the Indo-U.S. Coud\'e Feed Spectral Library (\citealt{Valdes2004}) as stellar spectral templates to model the stellar continuum emission and absorption line systems. The models, with a spectral resolution of 1.35$\AA$, were broadened to the (wavelength dependent) spectral resolution of the MUSE data ($\sim 2.6 - 2.9\AA$) before the fitting process (see e.g. \citealt{Husser2016}).  pPXF fit was performed on binned spaxels using a Voronoi tessellation (\citealt{Cappellari2003}) to achieve a minimum signal-to-noise SNR $> 16$ per wavelength channel on the continuum in the [$5250,5450$]$\AA$ rest frame wavelength range.

During the fitting procedure, we used fourth-order multiplicative Legendre polynomials to match the overall spectral shape of the data. These polynomials are generally used instead of an extinction law, which was found to produce worse fits to the stellar continuum, and allow to correct for small inaccuracies in the flux calibration (e.g. \citealt{Belfiore2019}).

\subsection{Stellar line-of-sight velocity distributions}

Figure \ref{FIG2} shows the Arp220 stellar kinematic maps (for a better visual output, all the maps presented from now on are centred on RA: 15:34:57.28, DEC: +23:30:11.87, corresponding to the intermediate position between the two nuclei). The systemic redshift of Arp220, $z = 0.0181 \pm 0.0001$, has been chosen to obtain a symmetric stellar velocity gradient in the Arp220 central regions, and corresponds to the one at the position of the W nucleus (the error has been obtained translating the positional error into velocity uncertainty).

The stellar velocity map (left panel) allows us to define a `major axis' as the PA along which the velocity shear is the maximum (e.g. \citealt{Harrison2014}). The `minor axis' is defined as perpendicular to the former one. The Arp220 `major axis' corresponds to the  north-east -- south-west direction (PA $\sim 48^\circ$), with a distinct velocity gradient from $\sim -100$ km/s to $+100$ km/s in $\sim 3$ kpc; the stellar velocities along the `minor axis' (PA $\sim 138^\circ$)  are instead lower (in the range $\sim [-40,40] $ km/s) and do not show clear velocity gradients. 
The $V_*$ map also show different velocity-coherent structures in the outermost regions (r $\gtrsim 10''$), possibly associated with Arp220 tidal tails: in particular, the structure  extending from the north towards the south-west direction and associated with positive velocities, and the western structure with negative velocities, could be associated with the two tidal tails structures emerging from the merger simulations by \citet{Konig2012}.

The stellar velocity dispersion map (right panel) shows the presence of high $\sigma_*$ ($\gtrsim 150$ km/s) in the innermost nuclear regions, but with an irregular distribution. In this region, we can recognise a ring-like structure with enhanced $\sigma_*$ (see zoom-in inset). In order to define the reliability of this feature, we performed Monte-Carlo (MC) analysis using 50 simulations. In each simulation, the input spectrum was randomly perturbed within the 1$\sigma$ flux errors\footnote{The STAT data-cube produced by the esoreflex pipeline is used to define the flux errors.} and the pPXF fits were recalculated. To simplify the calculation and save computational time, only the innermost regions (in the zoom-in inset in Fig. \ref{FIG2}) were analysed with MC trials.  The uncertainty of $\sigma_*$ was taken as the standard error of the fitted moment from the MC trials; its median value is 7 km/s. Therefore, MC trials confirms the presence of enhanced $\sigma_*$ in the ring-like feature (see Appendix \ref{Asigmastar}).
In order to quantify systematic errors in the velocity maps across the entire FOV, we also performed an independent fit with pPXF, using identical tessellation and  wavelength coverage but different stellar templates. This analysis will be presented in Catal\'an Torrecilla et al. (in prep.), together with the derived stellar population parameters. Here we briefly mention that the best-fit stellar kinematics and continuum models obtained with the two approaches are consistent (within a few $ \%$); in particular, the independent analysis confirmed the presence of the ring-like region and the high velocity dispersions in several Voronoi bins out to $\sim 15''$ from the two nuclei. Finally, we note that our reconstructed stellar kinematics are consistent with those presented in \citet{FalconBarroso2017} and \citet{deAmorin2017}, obtained from CALIFA spectroscopic data (i.e. with poorer spatial information). 

As reported in \citet{Harrison2014}, the `major axis' corresponds to the kinematic major axis of a galaxy when the velocity map traces galactic rotation.  
The presence of kinematically disturbed regions, curved tidal tails and the ring-like feature suggest that Arp220 has not yet reached its dynamically relaxed configuration. The two PA associated with the `major' and `minor axis', therefore, cannot properly associated with the Arp220 kinematic axes. Nevertheless, they present well defined and distinct properties in terms of stellar and gas kinematics, as also shown in the next sections.

\subsection{Extraction of pure ISM emission and absorption contribution} 

\begin{figure*}[h]
\centering
\includegraphics[width=17.cm,trim=0 0 0 20,clip]{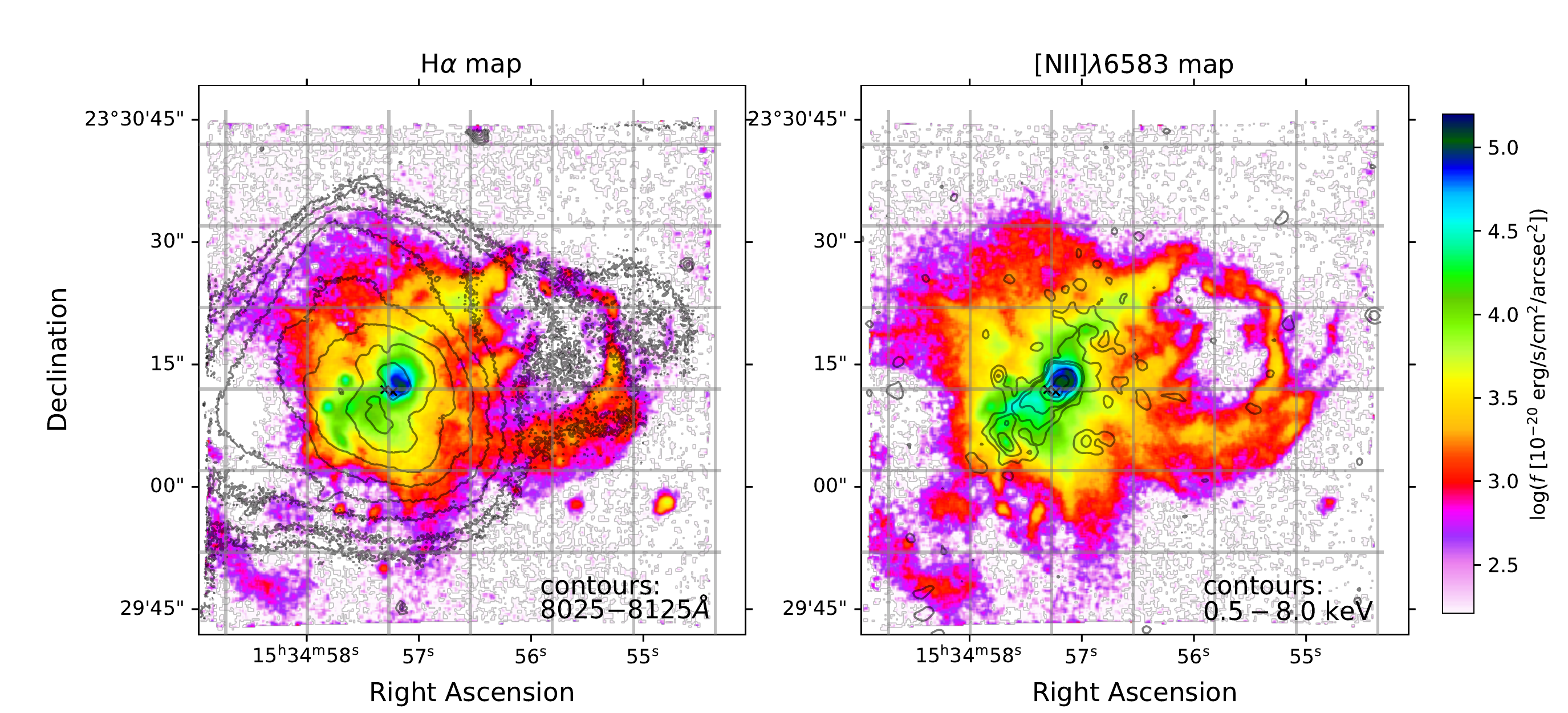}
\caption{\small  H$\alpha$ and [N {\small II}]$\lambda6583$ channel maps, obtained collapsing the ISM data-cube on the emission lines core (i.e. velocity channels within [$-150, +150$] km/s from the systemic).  Contours in the left panel are derived from the continuum emission image shown in Fig. \ref{FIG2}, top left; contours in the right panel show the $0.5-8$ keV emission from Chandra-ACIS observations (OBSID 16092; \citealt{Paggi2017}).}
\label{Hachannels}
\end{figure*}

We used the pPXF best-fit models to subtract the stellar contribution from the original data cube, and derive a pure emission and absorption ISM line data cube. More in detail, we subtracted the stellar emission from the
original unbinned data cube, by rescaling the fitted stellar contribution (constant within each Voronoi bin) to the original unbinned spectrum of each spaxel before subtracting it (e.g. \citealt{Venturi2018}).
The separation between stellar and ISM contributions is especially needed to properly recover the Balmer line fluxes (e.g. \citealt{Belfiore2019,Perna2017a}), and measure the kinematics and physical properties of neutral gas in the ISM (e.g. \citealt{Rupke2005b}).

In Fig. \ref{FIG3} we report the spectra extracted from 2$\times2$ pixel regions  associated with the W and E nuclei, with the corresponding pPXF best-fit model profiles (orange curves). We also report the pure emission/absorption ISM spectra (blue curves), obtained subtracting the best-fit stellar contribution. The two spectra are highly obscured (stellar colour excess $E(B-V)_* \approx 1.0$  from Catal\'an Torrecilla et al., in prep.), and are characterised by strong H$\alpha$, [N {\small II}] and [S {\small II}] lines with very asymmetric and complex profiles (see zoom-in insets). The [N {\small II}]/H$\alpha$ and [S {\small III}]/[S {\small II}] line ratios suggest a high ionisation, typical of AGN (e.g. \citealt{Kewley2001}), and/or shocks (e.g. \citealt{Diaz1985}) produced by fast outflows (e.g. \citealt{Mingozzi2019}). The spectra also show neutral Na {\small ID} absorbing gas, with broad and blueshifted components indicative of the presence of neutral outflows (see zoom-in insets).

Prior to modelling the ISM features and deriving spatially resolved kinematic and physical properties of Arp220 gas, we generated H$\alpha$ and [N {\small II}]$\lambda$6583 emission line images collapsing the ISM data-cube  on the core of the two emission lines ([-150,+150] km/s in velocity space, considering the H$\alpha$ and [N {\small II}]$\lambda$6583 systemics as zero-velocity). Assuming a similar kinematic behaviour of gas and stars, this emission should trace the less perturbed gas in the merging system, being the stellar velocities in the range $\approx$ [$-130, +130$] km/s. The  two maps are reported in Fig. \ref{Hachannels}; for easier comparison with stellar kinematic maps, in the left panel we show the stellar continuum emission (contours as in Fig. \ref{FIG2}). 
Large scale rings, arcs and filaments, as well as diffuse line emission can be observed across the MUSE field; several clumps are also observed, especially along PA $\sim 138^\circ$ (north-west to south-east direction). We note that MUSE observations do not cover the entire extension of the ionised gas in Arp220, with its well-known `8-shaped' structure due to bipolar lobes (e.g. \citealt{Heckman1987}), being the east arc  mostly outside the MUSE FOV (see also Fig. \ref{HST}). 

The H$\alpha$ and [N {\small II}]$\lambda$6583 maps display similar spatial distributions, although [N {\small II}] emission appears more extended and brighter, especially along  PA $\sim 138^\circ$ and in the west lobe. 
The match between optical line and X-ray emission along  PA $\sim 138^\circ$ is made explicit in the right panel of  Fig. \ref{Hachannels}, where we reported with black contours the broadband $0.5-8$ keV emission  from  Chandra-ACIS (\citealt{Paggi2017}).

\begin{figure*}[h]
\centering
\includegraphics[width=17.cm,trim=0 0 0 0,clip]{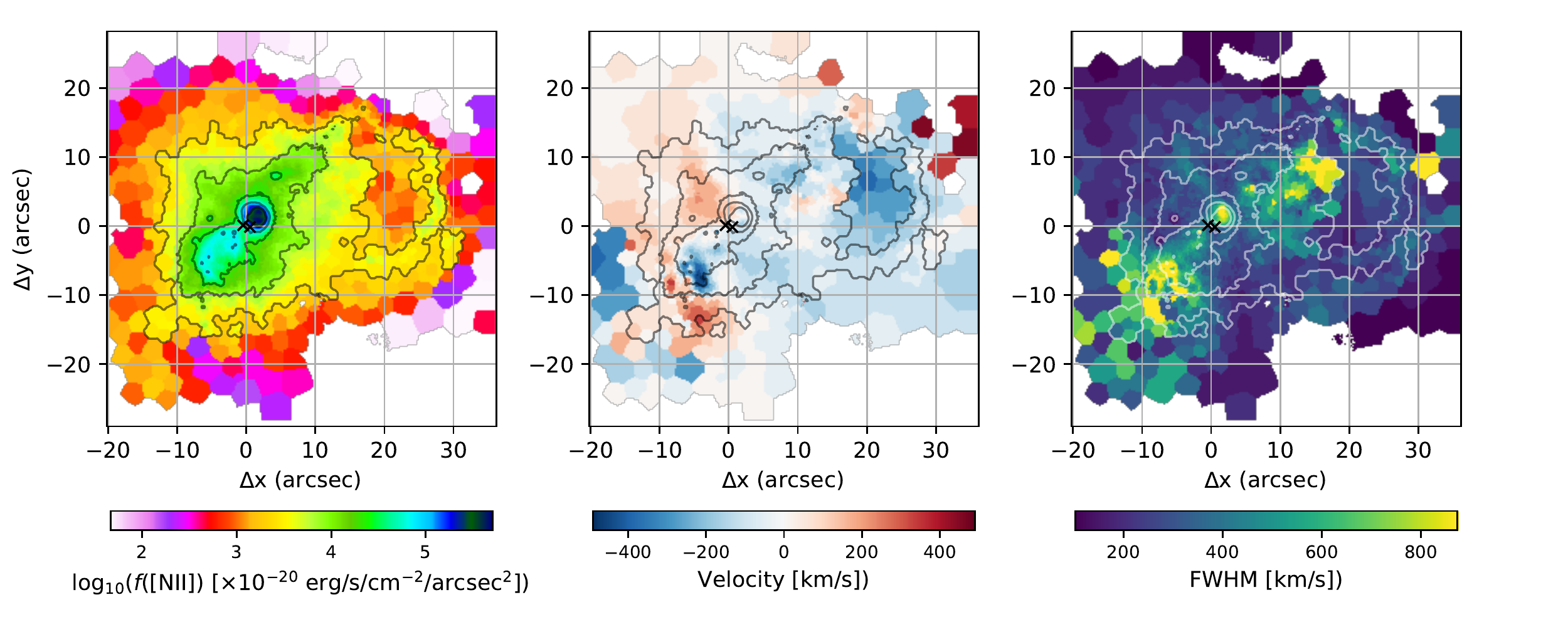}
\caption{\small Left: [N {\small II}] integrated flux obtained from single Gaussian fits. The first solid contour is $3\times 10^{-17}$ erg/s/cm$^2$/arcsec$^2$, and the jump is 0.5 dex. Centre and right: velocity and FWHM maps obtained from single component fit. Solid contours from the left panel. The crosses mark the two nuclei.
}
\label{Ngau1}
\end{figure*}

\subsection{Ionised gas features modelling}

The emission line profiles in Arp220 are very complex, probably because of the superposition of several kinematic components associated with tidal tails (e.g. \citealt{Konig2012}), compact star-forming clumps (e.g. \citealt{Wilson2006,Varenius2019}), diffuse emission and, probably, inflows and outflows (\citealt{Arribas2001,Colina2004}). The H$\alpha$ and [N {\small II}] lines, the brightest features in Arp220 optical spectra, generally show double peaks and prominent blue and/or red wings. 

As a first step, we fitted single Gaussians to the relevant emission line tracers of the gas. In particular, we modelled the H$\beta$ and H$\alpha$ lines, the [O {\small III}]$\lambda\lambda$4959,5007, [N {\small II}]$\lambda\lambda$6548,83 and [S {\small II}]$\lambda\lambda$6716,31 doublets, and the [O {\small I}] emission lines at 6300 and 6364$\AA$.  We constrained the wavelength separation between emission lines in accordance with atomic physics; moreover, we fixed the FWHM to be the same for all the emission lines. Finally, the relative flux of the two [N II] and [O III] components was fixed to 2.99, that of the two [O {\small I}] lines to 3.13, and the [S II] flux ratio was required to be within the range $0.44< f$($\lambda$6716)/$f$($\lambda$6731) $<1.42$ (\citealt{Osterbrock2006}).

Before proceeding with the fit, we derived a second Voronoi tessellation to achieve a minimum SNR = 7 of the [O {\small III}]$\lambda$5007 line for each bin. This feature is generally very faint across the FOV, because of the significant dust reddening. The H$\beta$ is even fainter and is undetected in several regions even after this tessellation. However, we chose to use the [O {\small III}] line as a reference for the tesselation, to limit the loss of important spatial information (both for kinematics  and emission lines structures). This, of course, will restrict our knowledge about the ISM physical conditions (e.g. dust attenuation, ionisation conditions), generally derived through emission line ratios involving the H$\beta$ flux. 

All emission lines were fitted simultaneously with our own suite of python scripts, and using the Levenberg–Markwardt least-squares fitting code CAP-MPFIT (\citealt{Cappellari2017}).
In this first step, we do not model the Na {\small ID} and [S {\small III}] lines. The former would require the use of multiple components to trace the emitting and absorbing contributions, and will be analysed in the next steps (Sect. \ref{naidfit}). The sulfur line is detected only in the innermost nuclear part (see Fig. \ref{FIG1}). This feature, generally used to constrain the ionisation conditions of warm material (e.g. \citealt{Kewley2002,Cresci2017,Mingozzi2020}), 
will be analysed in Sect. \ref{ionisationconditions}.

\subsubsection{Single Gaussian fit results}

Figure \ref{Ngau1} shows the best-fit results from a single component parametrisation. The left panel shows the [N {\small II}]$\lambda$6583 integrated flux, clearly resembling the main features observed in Fig. \ref{Hachannels}, although with some loss of spatial information. The central panel shows the ionised gas velocity (associated with the centroid of the Gaussian profile); a clear velocity gradient can be observed along PA $\sim 48^\circ$, as observed in the stellar velocity map (Fig. \ref{FIG2}) and in the CO molecular gas (e.g. Figs. 5, 6 in \citealt{Scoville1997}). More disturbed kinematics are found along the perpendicular direction (PA $\sim 138^\circ$). 
Gas and stars in the innermost nuclear regions are kinematically aligned (with a 'major axis' $\sim 48^\circ$), but the gas velocity amplitude is significantly higher when compared to that of the stars. 

The right panel of Fig. \ref{Ngau1} shows the $FWHM$ map. It further highlights the presence of highly perturbed gas along  PA $\sim 138^\circ$; moderate line widths can also  be observed along the bright arcs locate at about $25''$ west. 

There are at least four regions along  PA $\sim 138^\circ$ with $FWHM \gtrsim 800$ km/s. The first one, at about $2''$ northwest from the central position, has been associated with a bubble in H$\alpha$+[N {\small II}] of $\sim 1.6''$ (600 pc) in diameter by \citet{Lockhart2015}. They used HST/WFC3 narrowband filters to create high spatial resolution emission line maps, but without resolving the individual contributions from H$\alpha$ and [N {\small II}] (the lines are within the same WFC3 filter), nor inferring kinematic information. Additional narrowband filters centred on the H$\beta$ and [O {\small III}] lines were used to map the [O {\small III}]/H$\beta$ flux ratios; the bubble was associated with the highest flux ratios in the field (log [O {\small III}]/H$\beta \approx 0.2-0.3$)\footnote{[O {\small III}] and H$\beta$ emission along the dust lane and at $r \gtrsim 10''$ from the nuclei could be below the sensitivity of the HST images used to derive the [O {\small III}]/H$\beta$ map in \citet{Lockhart2015}.}, hence indicating high ionisation conditions. They also found a clear spatial correlation with X-ray soft emission (see e.g. Fig. \ref{Hachannels}) and suggested that an AGN jet or a (AGN- or SB-driven) outflow could be responsible of the bubble. This interpretation is also consistent with the presence of high-velocity H$_2 1-0$ S(1) line emission at the base of the bubble (with $FWHM\approx 600$ km/s), observed with VLT/SINFONI by \citet{Engel2011}. 
The extreme kinematics we observe in the optical lines (Fig. \ref{Ngau1}) are totally consistent with this scenario.  
Moving north-west along  PA $\sim 138^\circ$, we find another region with high-velocity dispersion, possibly associated with the over-position of the emission from the main system and the inner west arm.  The last two areas with extreme line widths are located at $\sim 10''$ south-east. They have a significant velocity offset with respect to the systemic of Arp220, of $\approx -400$ km/s (inner part) and $\approx +300$ km/s (outer part; see Fig. \ref{Ngau1}, centre), and are associated with intense X-ray soft emission (Fig. \ref{Hachannels}). 

To summarise, these preliminary analysis results support the hypothesis of a kpc-scale (AGN- or SB-driven) outflow along  PA $\sim 138^\circ$. A more detailed kinematical and physical analysis testing this scenario is presented in the next sections.

\subsubsection{Emission line decomposition}

\begin{figure*}[t]
\centering

\includegraphics[width=17.1 cm,height = 8.5 cm,trim= 7 0 7 0,clip]{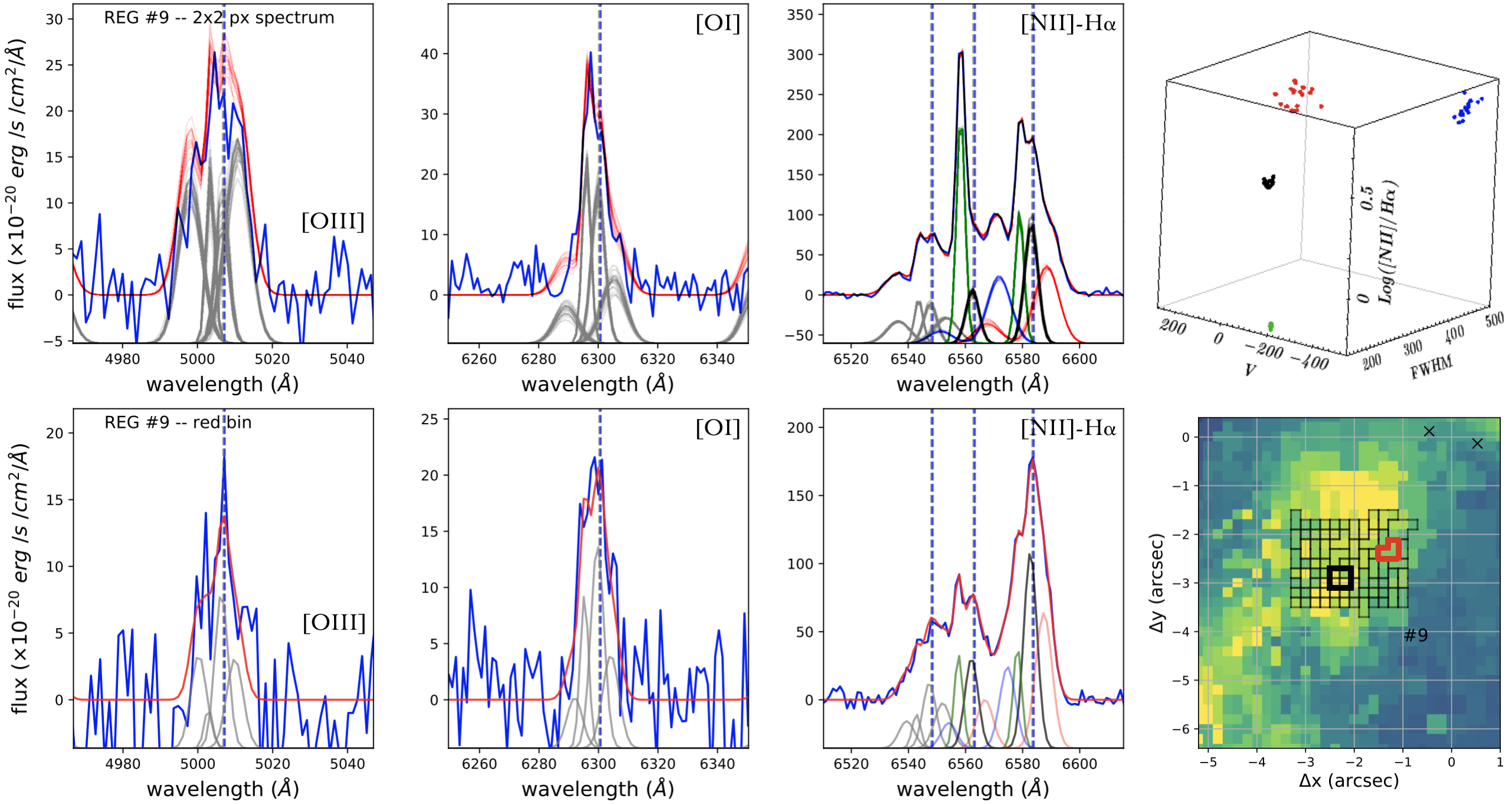}

\caption{\small  Illustration of the profile decomposition method. {\it Top}: the first three panels show the spectra in the vicinity of the [O {\small III}] (left), [O {\small I}]$\lambda$6300 (centre) and [N {\small II}]+H$\alpha$ lines (right). The red curves represent the best-fit models obtained from 500 MC trials. For all but [N {\small II}]$\lambda6583$ and H$\alpha$ lines, we show the Gaussian profiles used to reproduce the spectrum with grey curves. Different colours are instead used for  the H$\alpha$ and [N {\small II}]$\lambda6583$ lines, to distinguish the different kinematic components. Grey and blue dashed lines mark the Arp220 systemic and the local stellar velocity. In the right panel, we show for the same kinematic components the parameter space $V-FWHM-$ [N {\small II}]/H$\alpha$, highlighting the clear separation between the four Gaussian sets.  
{\it Bottom}: best fit results for one of the nearby Voronoi bins, obtained with the kinematic constraints from the 2x2 spectra shown in the top panels. The bottom right panel shows the velocity dispersion map with the Voronoi bins (black outlines) associated with the above defined kinematic constraints; the black (red) contours mark the spaxels from which the top (bottom) spectra have been extracted. }
\label{decomposition}
\end{figure*}

\begin{figure*}[t]
\centering
\includegraphics[width=17.1 cm,height = 8.5 cm, trim= 7 0 9 0,clip]{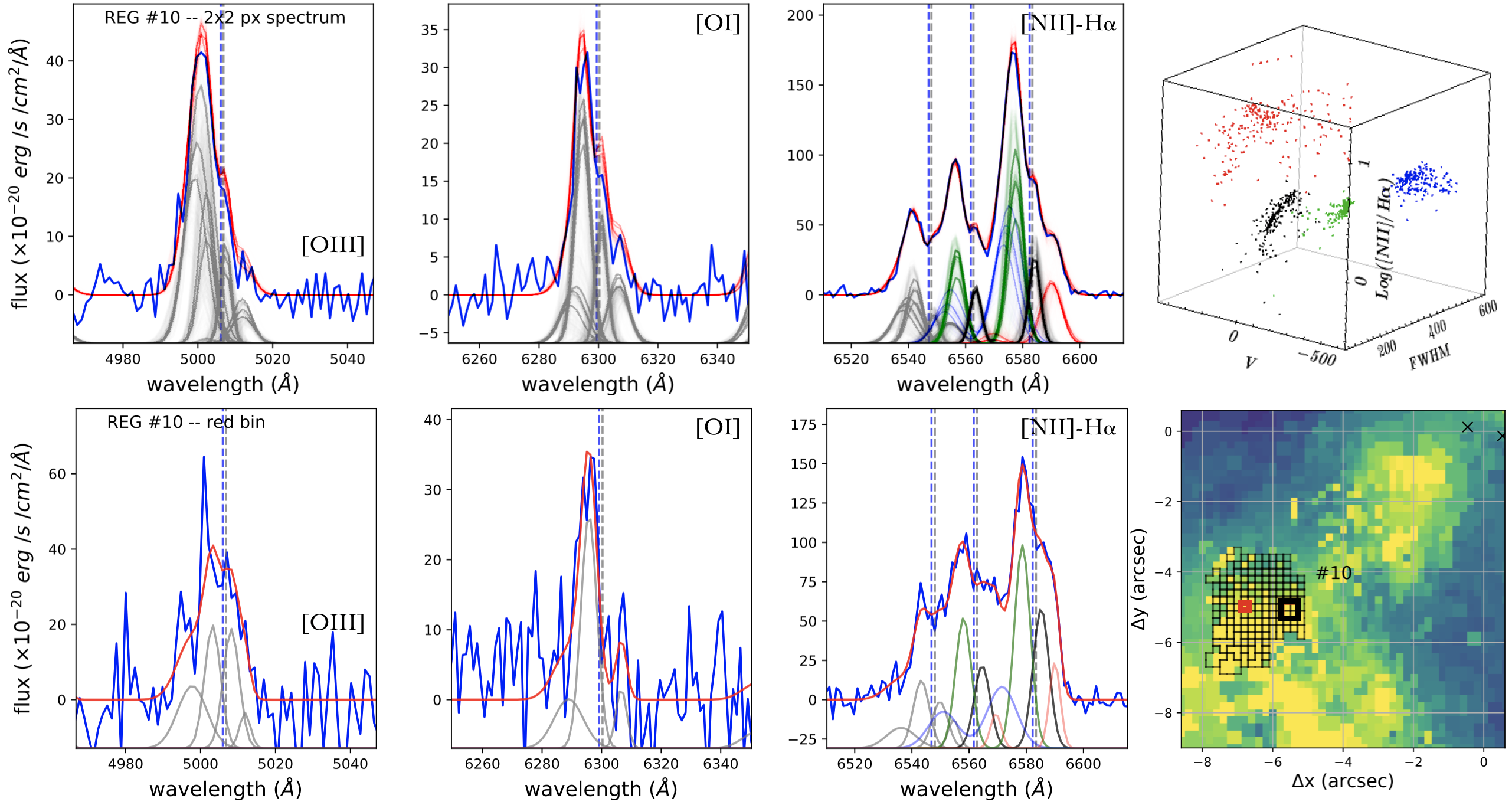}
\caption{\small  Illustration of the profile decomposition method. See Fig. \ref{decomposition} for details. In this region, the velocities of the blue and green Gaussian sets overlap; a clear separation between the two is however assured due to their distinct FWHMs. }
\label{decomposition2}
\end{figure*}

Because of the extremely complex structure of Arp220, the spaxel-by-spaxel fit with a standard multi-component approach (e.g. \citealt{Perna2019,Venturi2018}) leads, in some regions, to unphysical discontinuities in the derived kinematic components and flux distributions of individual components. 
We therefore decided to apply a different method, which allows a more appropriate regularisation of the best-fit parameters (see also e.g. \citealt{Shimizu2019} for a similar approach). 

A visual inspection of the spectra associated with the more disturbed kinematics in Fig \ref{Ngau1} revealed the presence of multiple components, which contribute to the increase of the measured velocity dispersion in the single Gaussian fits. 
These kinematic components could be related to either strong SF or AGN emission, tidal streams, gas inflows and outflow, and a combination of them.

We started by visually selecting the 2$\times$$2$ pixel integrated spectra showing H$\alpha$+[N {\small II}] complex with a clear evidence of additional peaks, and/or inflection points that, at first sight, could be related to the presence of several distinct kinematic components. A couple of spectra are shown in Figs. \ref{decomposition} and \ref{decomposition2} (top panels). 
Then, for each spectrum, we performed a second fit using four (at maximum) Gaussian profiles per emission lines; in the following, all lines profiles  associated with given kinematic parameters (i.e. velocity and FWHM) will be referred to as Gaussian set. For each set of Gaussian functions, we considered the same constraints mentioned in the previous section. The number of Gaussian sets (i.e. distinct kinematic components) used to model the spectra was derived on the basis of the Bayesian information criterion (BIC; Schwarz 1978), which uses differences in $\chi^2$ that penalise models with more free parameters (see e.g. \citealt{Harrison2016,Concas2019}). 

We then used a statistical approach to study the possible degeneracy between the different Gaussian sets. We used 500 Monte Carlo trials of mock spectra obtained from the best fit models, adding Gaussian random noise\footnote{The STAT data-cube produced by the esoreflex pipeline is used to set the standard deviations of the Gaussian distributions from with a random noise is extracted.} (e.g. \citealt{Perna2019}), to obtain a distribution for the kinematic parameters associated with each Gaussian set (namely, the velocity $V$ and the FWHM). The distribution in the 2D  parameter space $V$ vs. FWHM is then used to check that there is no overlap between the high-probability regions of the distinct Gaussian sets, considering the  95\% occurrence intervals for each kinematic parameter. 
In Figs. \ref{decomposition} and \ref{decomposition2} (top right) we show the $V-FWHM-$[N {\small II}]/H$\alpha$ parameters space obtained from the MC trials, showing that the decomposition, sometimes, is also able to distinguish between distinct emission line ratios.

After identifying the individual  kinematic components in a given 2$\times$2 pixel spatial region, we fitted the adjacent Voronoi bins performing a parameter tuning using the best-fit kinematic components from the original 2x2 spectrum, but leaving the velocity parameters free to vary within the ranges defined with our MC trials in the previous step (considering 95\% occurrence intervals). 
In a few cases - i.e. when the MC trials give back tight high-probability regions - the lower and upper bounds associated with $V$ and $FWHM$ have been opportunely refined during this step, but still avoiding an overlap between different components in the kinematic space, according to the above mentioned criterium. 
A chi-square goodness of the fit test determines if the new Voronoi bin can be associated with these kinematic components.

In Figs. \ref{decomposition} and \ref{decomposition2} (lower panels) we  show the best fit in one of the nearby bins obtained with the kinematic constraints from the 2x2 spectra. These figures show that the spectra, regardless the apparent diversity in the (H$\alpha$+[N {\small II}]) line profiles, can be reproduced with similar kinematics. The lower-right panels also indicate the regions which kinematic properties can be obtained using the constraints from the initial 2x2 spectrum.

We note that the converged solutions (Figs. \ref{decomposition}, \ref{decomposition2}, top right) are not claimed to represent the actual physical components responsible of the observed line profiles. In fact, for instance, our results are still based on the assumption that emission lines can be represented by a combination of Gaussian profiles (but see e.g. \citealt{Liu2013, Cresci2015}), and that the same kinematic component can be associated with both low- and high-ionisation lines (i.e. that there is no gas stratification with respect to the ionisation source, see e.g. \citealt{DeRobertis1986}). Nevertheless, this approach allows us to avoid  unphysical discontinuities in those regions with more complex kinematic structures.

With this approach, we selected 11 spatial regions with well-determined kinematic properties (and, in a few cases, [N {\small II}]/H$\alpha$ flux ratios; see Sect. \ref{Ssf}). In Appendix \ref{Adecomposition} we report all the 2x2 pixel spectra as well as the location and the extent of the selected regions.

\subsection{Neutral gas features modelling}\label{naidfit}

\begin{figure}[t]
\centering
\includegraphics[width=8.cm,trim=0 40 0 0,clip]{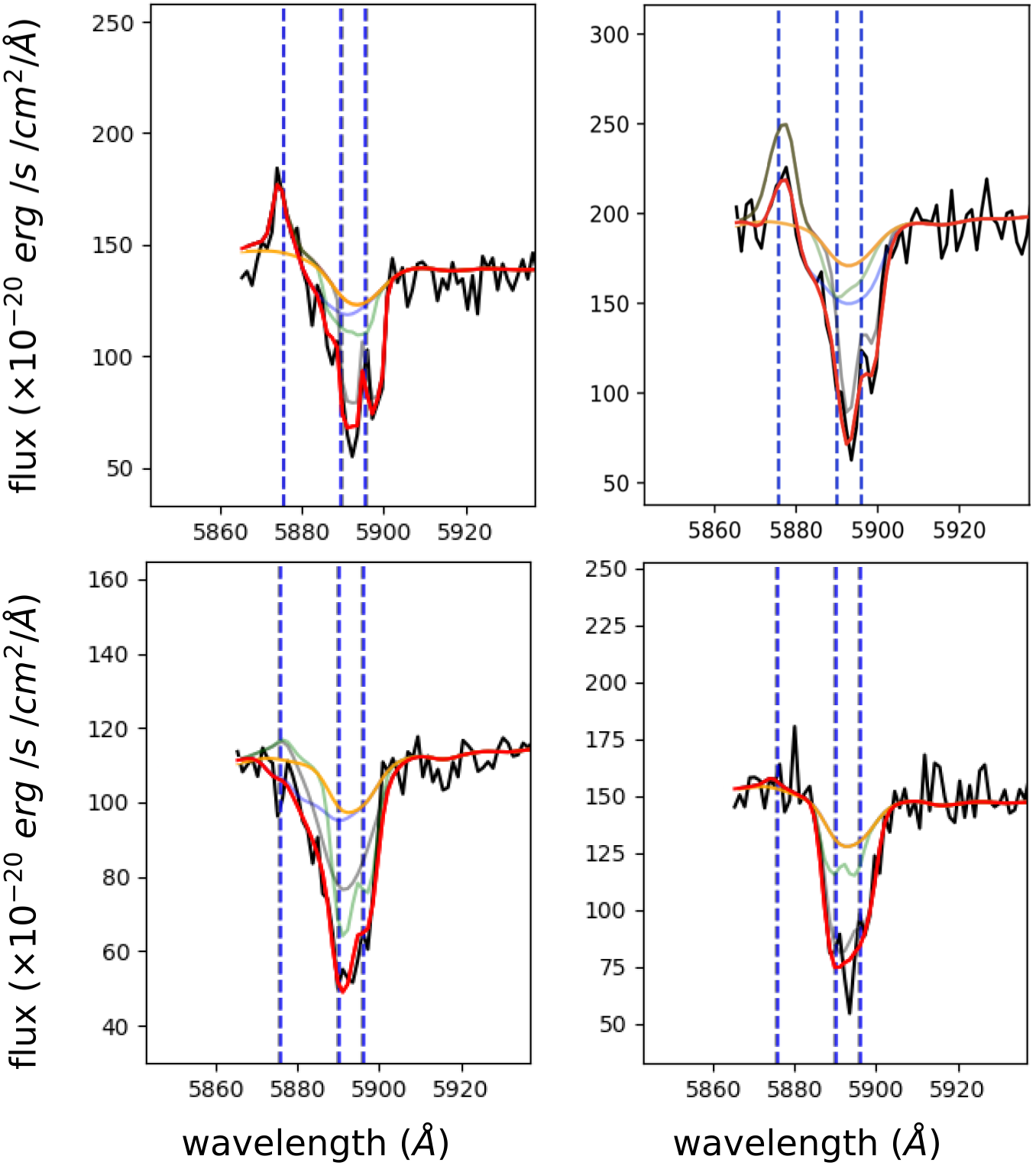}
\includegraphics[width=8.cm,trim=0 0 0 0,clip]{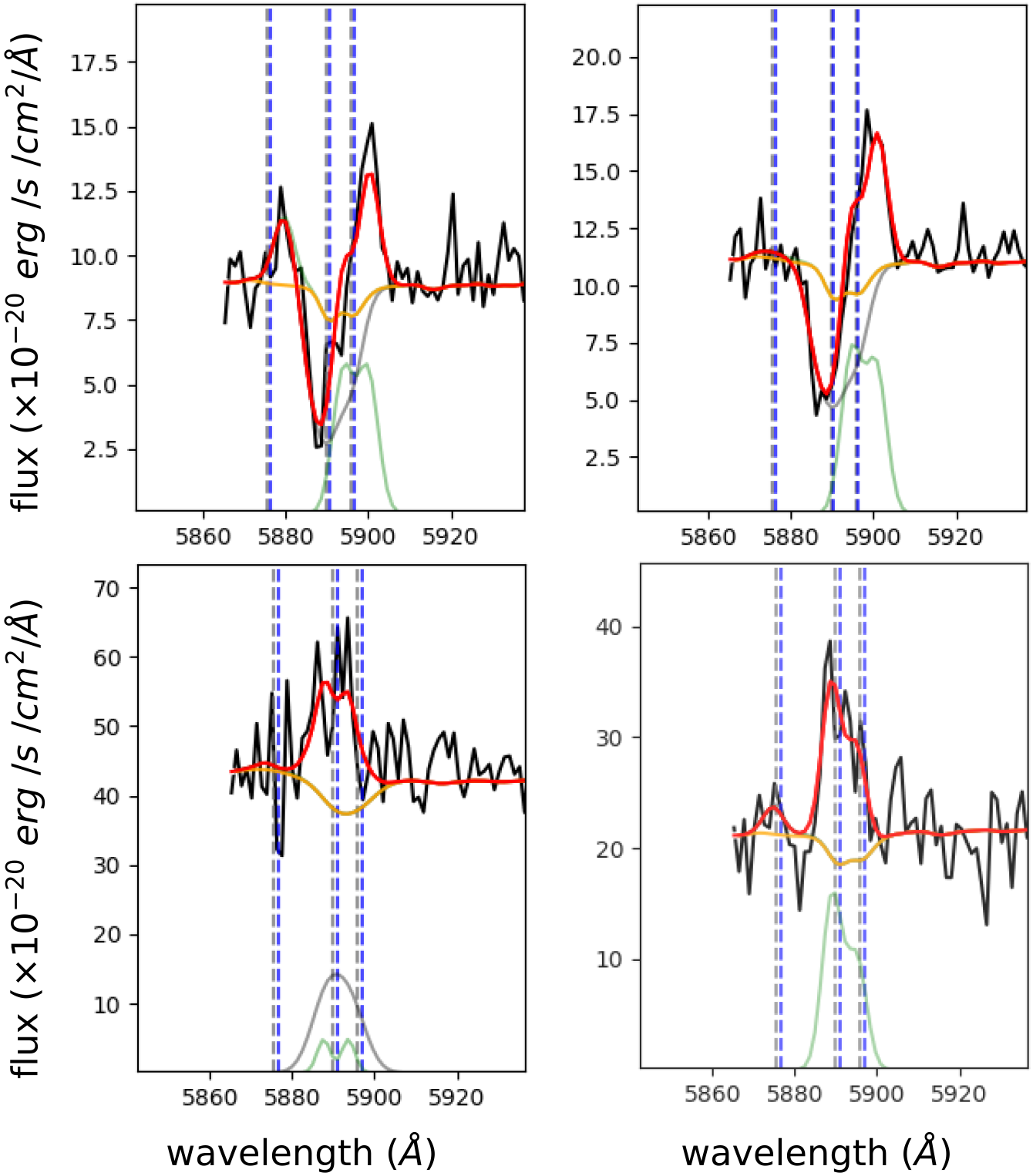}
\caption{\small  Best-fitting models for eight representative spectra in the vicinity of the Na {\small ID} complex. The original spectra are shown in black, the pPXF best-fit models in orange, while the red profiles represent the best-fit models obtained with the simultaneous multi-component approach.  Green, grey and blue profiles are obtained from Eq. \ref{eqrupke} and represent the different kinematic components used to model the Na {\small ID} absorption; in the last four panels, the Na {\small ID} emission line contribution (modelled with Gaussian profiles) is shown with an arbitrary offset in the y-axis.}
\label{naidregions}
\end{figure}

Taking advantage from the emission line decomposition described in the previous section, we performed a new multicomponent Gaussian fit for the entire data cube,  using the parameter tuning for the Voronoi bins within the 11 regions. 
In this new fit we also simultaneously modelled the contribution of the HeI emission and Na {\small ID} complex system, the latter tracing the neutral gas component of the ISM.

Na {\small ID}  is a resonant system and both absorption and emission contributions can be observed across the extension of a galaxy (e.g. \citealt{Prochaska2011}). 
The sodium emission has been observed at very low level in stacked SDSS spectra (\citealt{Chen2010, Concas2019}) and in a few nearby galaxies (\citealt{Rupke2015} and references therein; \citealt{Perna2019, Baron2020}). 
A visual inspection of the Arp220 data cube revealed the presence of significant Na {\small ID} absorption and emission across the MUSE FOV: the former is stronger towards the bright continuum source, the latter is stronger in the northwest regions and in the eastern side of the FOV.

We fitted the observed profiles of the sodium lines in absorption with a model parameterised in the optical depth space (e.g. \citealt{Rupke2002,Rupke2005a}), starting from the equation of radiative transfer (Spitzer 1978) in the case of a homogenous absorber of matter and a partial coverage of the background emission source.  
Following \citet{Sato2009}, the Na {\small ID} doublet profile  is modeled by 
\begin{multline}\label{eqrupke}
I(\lambda) = I_{em}(\lambda)\times f_{ABS}(\lambda)  \\
 \ \ \ \ \ \ \ = I_{em}(\lambda)\times (1- C_f \times [1-exp(-\tau_0 e^{-(\lambda - \lambda_K)^2/(\lambda_Kb/c)^2} - \\ 
2\tau_0 e^{-(\lambda - \lambda_H)^2/(\lambda_H b/c)^2})]) ,
\end{multline}

where $H$ and $K$ indicate the sodium transitions at 5891 and 5896 $\AA$, $C_f$ is the covering factor, $\tau_0$ is the optical depth at the line centre $\lambda_K$, $b$ is the Doppler parameter ($b=FWHM/[2\sqrt{ln(2)}]$) and $c$ is the light velocity. This model assumes that the velocity distribution of absorbing atoms is Maxwellian and that $C_f$ is independent of velocity. Following \citet{Rupke2005a}, we assumed the case of partially overlapping atoms on the line of sight (LOS), so that the total sodium profile can be reproduced by multiple components and $I(\lambda)= I_{em}(\lambda)\times  \Pi_{i=1}^n f_{ABS}^i(\lambda)$, where $ f_{ABS}^i(\lambda)$ is the $i$-th component (as given in Eq. \ref{eqrupke}) used to model the sodium features (see also Sect. 3.1 in \citealt{Rupke2002}). 

The term $ I_{em}(\lambda)$  in Eq. \ref{eqrupke} represents the intrinsic (unabsorbed) intensity, defined as  $I_{*} + I_{HeI}$, where  $I_{*}$ is the best-fit model obtained from pPXF analysis and $I_{HeI}$ is the helium line intensity. 
This emission line is modelled simultaneously to the features in the [O {\small III}]+H$\beta$ and H$\alpha$+[N {\small II}] regions, using Gaussian profiles.

Atomic neutral Na {\small ID} features and ionised lines are modelled simultaneously, using up to four kinematic components. Namely, the kinematic parameters of a given Gaussian set (used to model ionised lines) also define the Na {\small ID} absorption (i.e. $f_{ABS}(\lambda)$) or emission kinematics. In the case of Na {\small ID} emission, Na {\small ID} doublet line ratio  is free to vary between the optically thick ($f (H)/ f (K) = 1$) and thin ($f (H)/ f (K)= 2$) limits (e.g. \citealt{Rupke2015}). 
To avoid degeneracies between positive and negative contributions at the same velocities, during the fitting procedure we considered the following conditions: 
\begin{itemize}
\item  a given kinematic component (i.e. Gaussian set) cannot be associated with both absorption and emission Na {\small ID} contributions; 
\item  the kinematic component associated with Na {\small ID} emission line contribution is, if present, always redshifted with respect to the absorption component(s) (e.g. \citealt{Prochaska2011}). 
\end{itemize}

Figure \ref{naidregions} shows the best-fitting models for some representative spectra in the vicinity of the Na {\small ID} complex. The different panels display the range in emission and/or absorption line profile shapes detected in the MUSE FOV. Although our model condiders several assumptions, it is well able to describe the observed spectra.

\section{Gas kinematics}\label{Sgaskinematics}

In this section we briefly discuss the general results we obtained from the multicomponent fits described above. The properties of the individual components will be discussed in the next sections. Gas kinematics are traced taking into account two non-parametric velocities (e.g. \citealt{Zakamska2014}): $v50$, the velocity associated with the 50\% percentile,  and $W80$, defined as the line width comprising 80\% of the flux (and corresponding to $1.09\times FWHM$ for a Gaussian profile). For the Na {\small ID} system, all velocities are defined using the H component wavelength as a zero-point; correspondingly, $v50$ and $W80$ are computed using the best-fit H components, and not the whole doublet profile\footnote{Using the whole Na {\small ID} profile, $v50$ velocity would be redshifted, on average,  by $\sim +100 \pm 35$ km/s (with respect to the H transition zero-velocity), while $W80$ values would increase on average by $190\pm 50$ km/s.}.  Line features are typically non-Gaussian and broad; therefore, we do not correct line velocities for the instrumental broadening.

\begin{figure*}[hp]
\centering
\includegraphics[width=18.2cm,trim=0 15 10 13,clip]{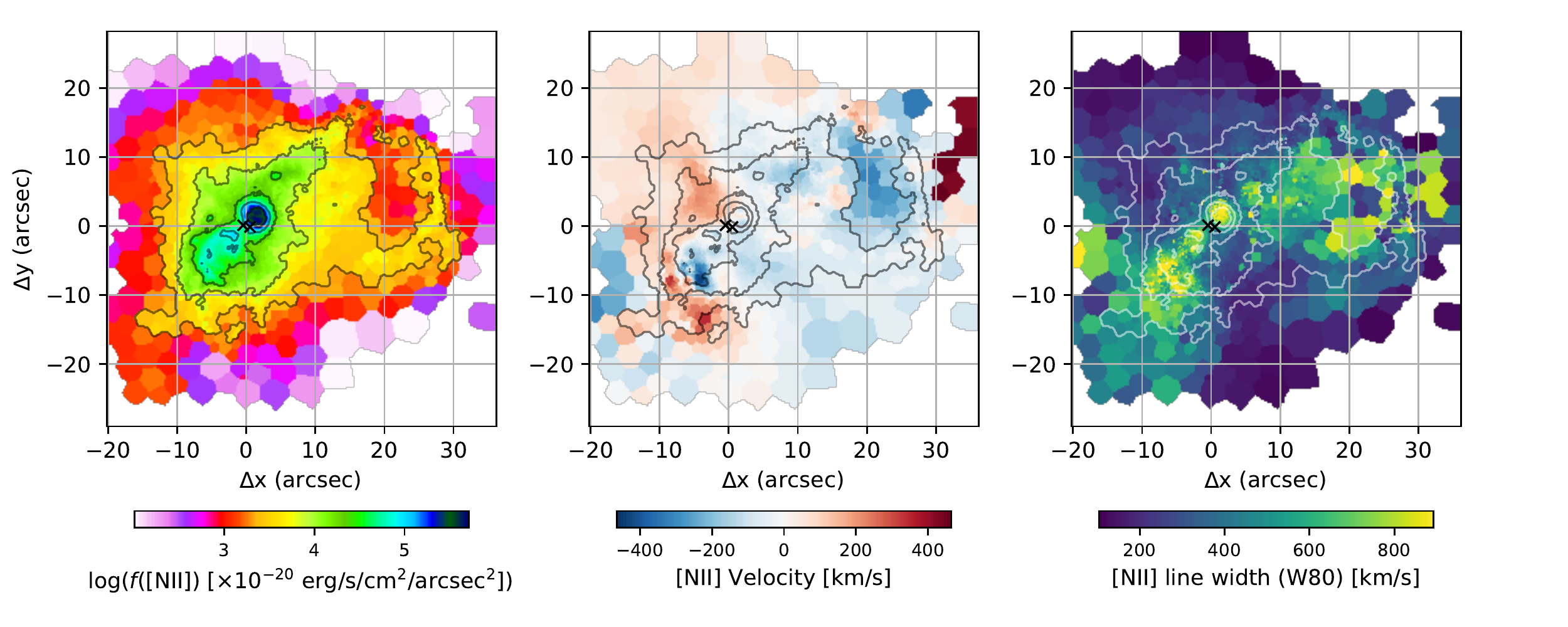}

\caption{\small [N {\small II}] multi-component fit results. {\it Left:} integrated flux; the first solid contour is $3\times 10^{-17}$ erg/s/cm$^2$/arcsec$^2$, and the jump is 0.5 dex. {\it Centre:} [N {\small II}] velocity ($v50$) map.  {\it Right:} [N {\small II}] line width ($W80$) map. Solid contours from the left panel. The crosses mark the two nuclei.
}
\label{NIIimages}
\end{figure*}

In Fig. \ref{NIIimages} we report the [N {\small II}] flux, velocity and line width maps obtained from the total line profiles, i.e. integrating over the different kinematic components required to reproduce the line profiles. 
These maps are  similar to those in Fig. \ref{Ngau1}, obtained with single Gaussian fits, but the present ones are more precise (e.g.  maps in Fig. \ref{Ngau1} overestimates fluxes and widths in the more external regions). The H$\alpha$ flux, velocity and line width maps obtained from multi-component fits are very similar to those of [N {\small II}]. On the contrary, [O {\small III}] and H$\beta$ are noisier and their distributions fuzzy, because of the lower SNR; nonetheless, these faint features reveal the same structures observed in [N {\small II}] maps (Appendix \ref{Alinemaps}). 

\begin{figure*}[hp]
\centering
\includegraphics[width=18.2cm,trim=0 12 10 13,clip]{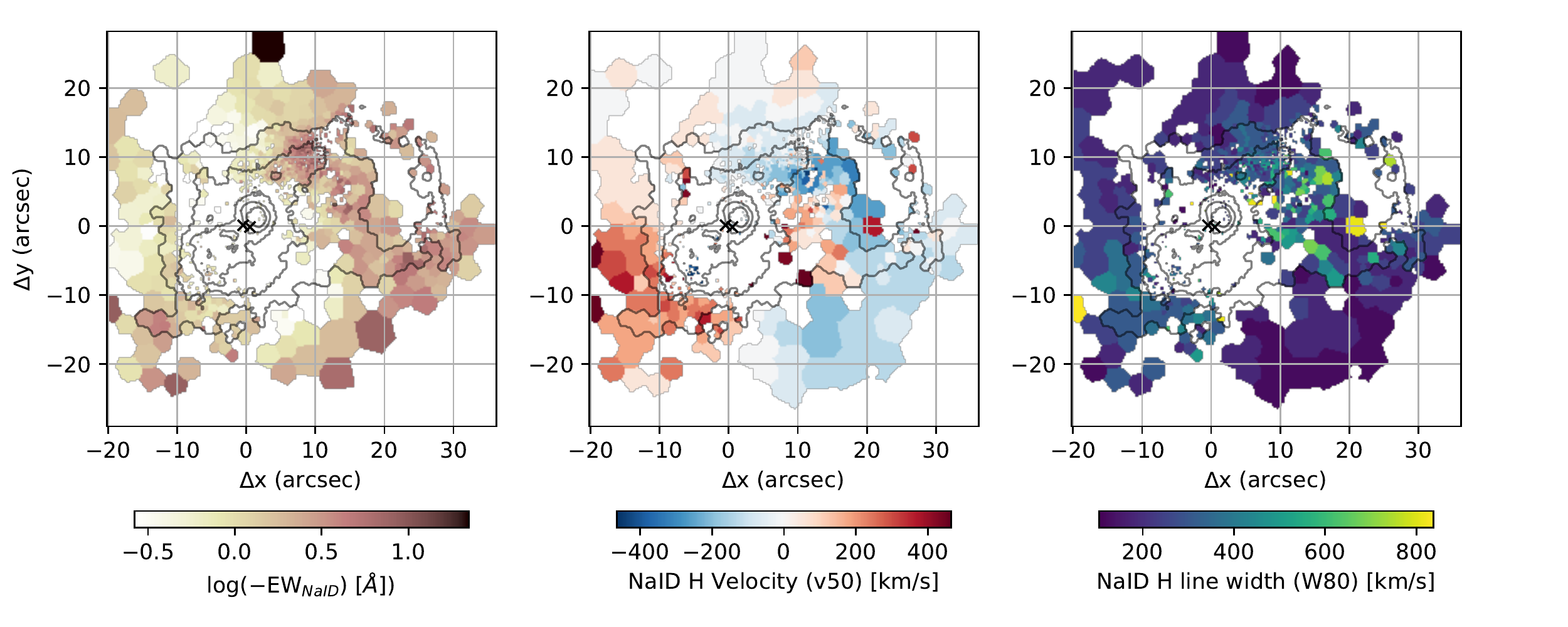}
\caption{\small 
Na {\small ID} emission line maps from the multi-component fit.  {\it Left:} equivalent width. {\it Centre:} Na {\small ID} velocity ($v50$) map. {\it Right:} Na {\small ID} line width ($W80$) map. Solid contours and crosses from Fig. \ref{NIIimages}. 
}
\label{NaIDemimages}
\end{figure*}

\begin{figure*}[hp]
\centering
\includegraphics[width=18.2cm,trim=0 12 10 13,clip]{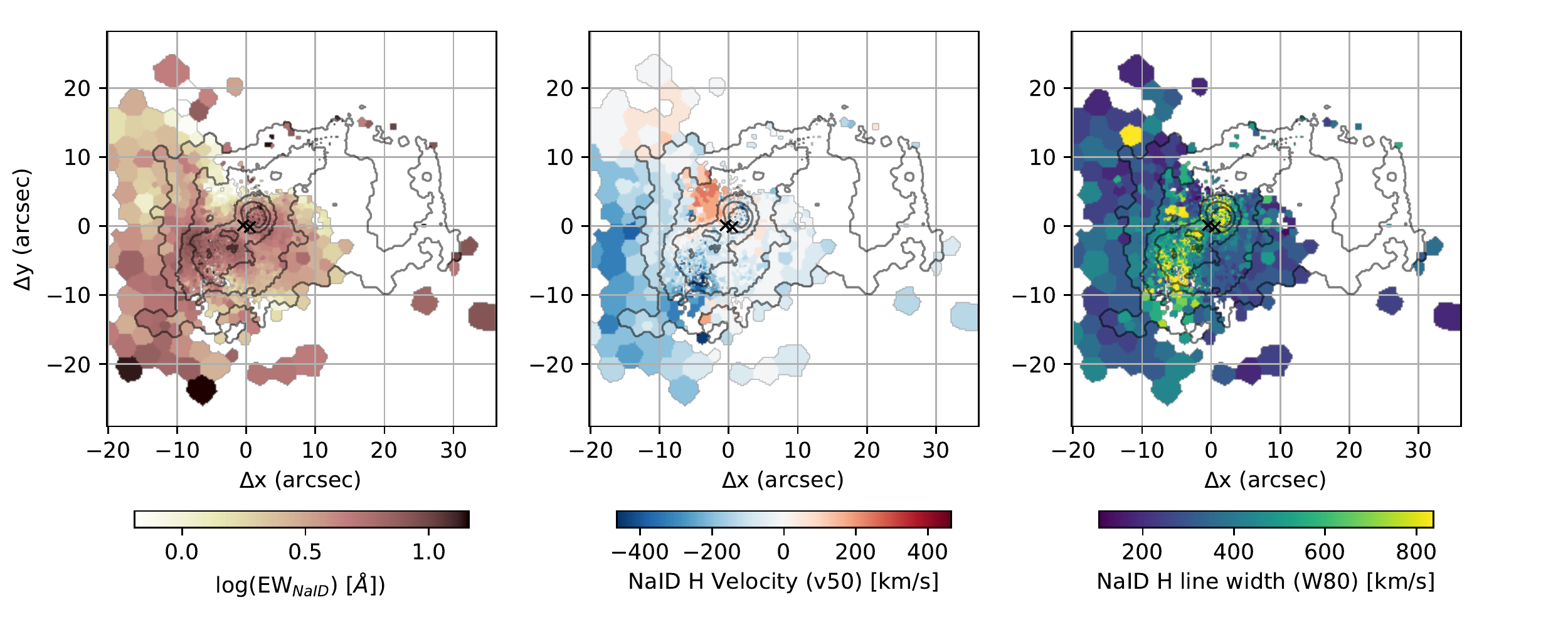}
\caption{\small 
Na {\small ID} absorption line maps from the multi-component fit. {\it Left:} equivalent width. {\it Centre:} Na {\small ID} velocity ($v50$) map. {\it Right:} Na {\small ID} line width ($W80$) map. Solid contours and crosses from Fig. \ref{NIIimages}. 
}
\label{NaIDabsimages}
\end{figure*}

In Figs.  \ref{NaIDemimages} and \ref{NaIDabsimages} we report the equivalent width\footnote{The equivalent width is defined as $\int (1-f_{line}/f_{con}) d\lambda$, where the $f_{line}$ and $f_{con}$ indicate the line and the continuum flux respectively. The EW is defined for absorption line systems and is therefore negative for emission features. Through the paper, we will implicitly refer to its absolute value.} (EW) and kinematic maps derived for the Na {\small ID} emission and absorption contributions. 
The ionised and neutral gas velocity dispersion maps are roughly consistent: they confirm the presence of highly perturbed kinematics along PA $\sim 138^\circ$ and in the west and east lobes, with $W80$ up to $800$ km/s . In particular, the highest $W80$ associated with Na {\small ID} emission are found in the west lobe, while those associated with Na {\small ID} absorption are along  PA $\sim 138^\circ$, within 10$''$ from the two nuclei. 

\begin{figure}[h]
\centering
\includegraphics[width=9.cm,trim=0 0 0 0,clip]{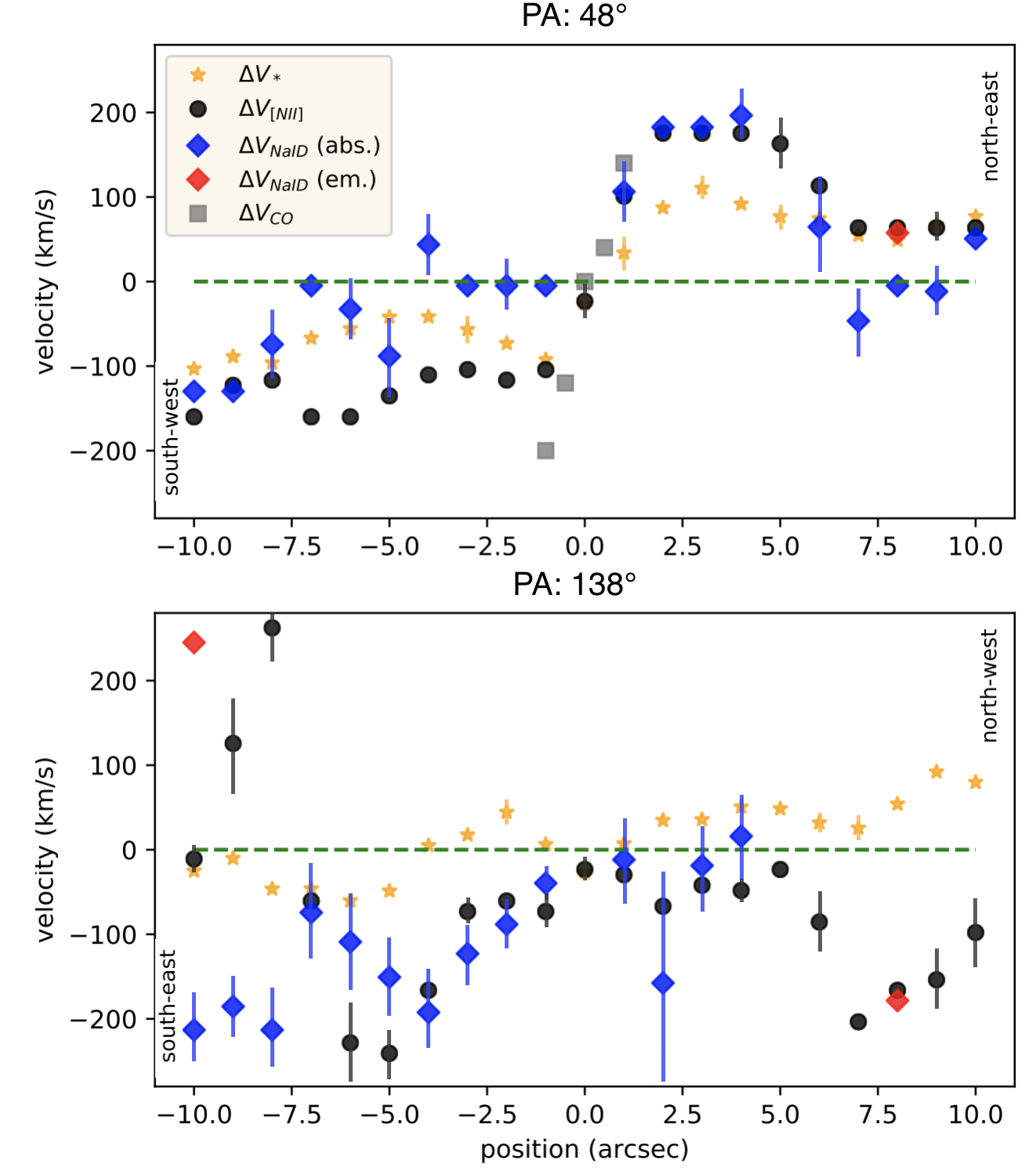}
\caption{\small Position velocity diagram, with positions varying along PA = 48$^\circ$ (top) and PA = 138$^\circ$ (bottom) for the stellar component, [N {\small II}] and Na {\small ID} gas as labeled in the figure. Distances on the x-axis are measured from the intermediate position between the two nuclei, from the bottom to the top of the IFU field.  Stellar and gas velocity measurements are obtained by averaging 3$\times$3 pixels along a given PA, from the maps shown in Figs. \ref{FIG2}, \ref{NIIimages}, \ref{NaIDemimages} and \ref{NaIDabsimages};  CO(2-1) velocities are obtained from Fig. 5 in \citet{Scoville1997}.
}
\label{PAvel}
\end{figure}

The [N {\small II}] and Na {\small ID} velocity maps are instead less consistent: the velocity gradient along PA $\sim 48^\circ$, already revealed by stars, ionised  and molecular gas is also detected in the absorbing Na {\small ID} gas. 
The cold gas absorption dominates over the Na {\small ID} emission within the innermost regions (r $\lesssim 10''$), with EW up to $\sim 10\AA$; it presents strong negative velocities along PA $\sim 138^\circ$, similar to those of the ionised counterpart. In order to ease the comparison between the different emitting and absorbing components, we report in Fig. \ref{PAvel} the velocities we derived from the analysis of MUSE data together with CO(2-1) measurements (from \citealt{Scoville1997}) along PA = 48$^\circ$ and PA = 138$^\circ$. 
Overall, the  two position-velocity trends confirm the presence of a velocity gradient along PA = 48$^\circ$ and multi-phase (i.e. neutral and ionised) approaching gas along  PA = 138$^\circ$.   
We note however that Na {\small ID} velocities along PA = 48$^\circ$ are always redshifted with respect to the stars, especially in the innermost region ($\lesssim 4''$). 
Because of the geometrical configuration that is required to observe absorption features, this absorbing gas could, in principle, trace an inflow of neutral gas towards the innermost regions where the nuclei and associated SBs are located. 
However, the good agreement between the neutral and ionised gas velocities questions this interpretation, as the latter seems to follow a rotation pattern. 
Therefore, an alternative explanation is that the neutral and ionised gas is confined in a rotating disk, with lower velocity dispersions and higher velocity amplitudes with respect to the stars. If this is the case, the motion of stars would take place in a thicker disk, while the gas is confined in a thinner region.


In the more external regions, we observe a significant Na {\small ID} emission (EW $\approx 3\AA$), mostly associated with approaching gas in the west side, and receding gas in the east side. In the latter region, Na {\small ID} is associated with P-Cygni profiles, with approaching absorbing gas and receding emitting material (see also Fig. \ref{naidregions}). The presence of off-nuclear Na {\small ID} emission is consistent with the results presented by \citet{Rupke2015} for IRASF05189-2524, who reported an anti-correlation between Na {\small ID} brightness and optical continuum attenuation. A more quantitative comparison between these quantities in Arp220 will be investigated in an upcoming paper (see Sect. \ref{Sextinction}). 

A visual inspection of the best-fit results reveal that in the innermost nuclear part (i.e. within $\sim 10''$ from the two nuclei), and especially along PA $\sim 138^\circ$, both approaching and receding gas emissions are detected as blueshifted and redshifted components 
(see e.g. Figs. \ref{decomposition} and \ref{decomposition2}; see also the velocity channel maps in Fig. \ref{niisfshocks}), consistent with the presence of powerful outflows.    
On the other hand, the more external regions are associated with simpler profiles;  in particular, the arcs in the north-west lobe are dominated by approaching gas, while the south-west arcs are mostly dominated by redshifted emission. The high $W80$ observed in the latter regions could be therefore explained by the superposition of different streams and tails along the LOS (see e.g. \citealt{Konig2012}), although other mechanisms cannot be excluded.

\begin{figure*}[hp]
\centering
\includegraphics[width=15.cm,trim=0 2 0 1,clip]{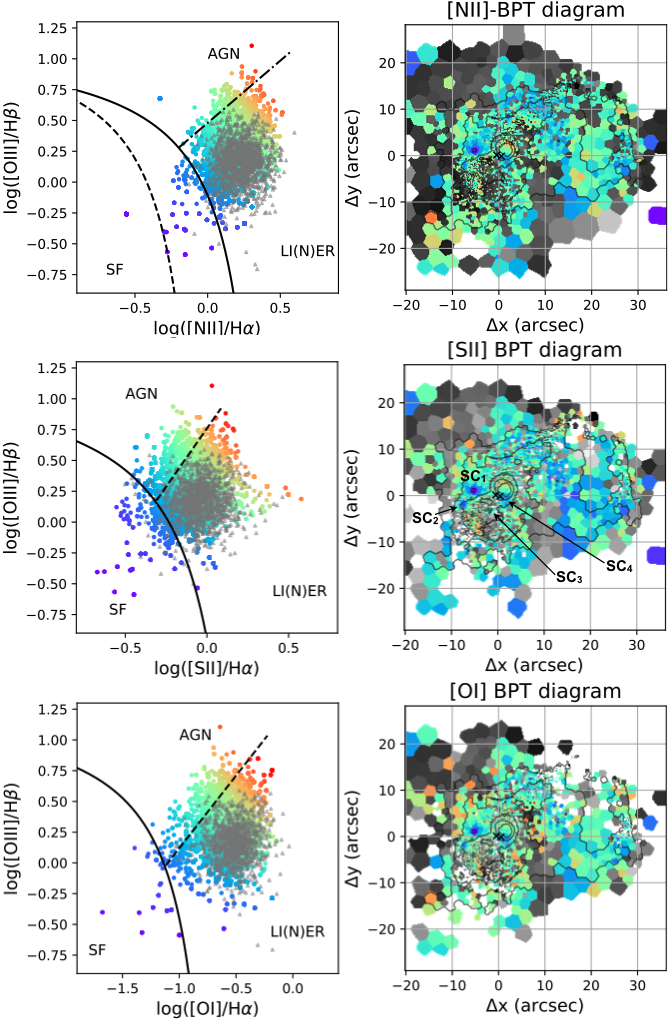}

\caption{\small 
Arp220 resolved BPT diagrams. In left panels we report the [N {\small II}]-BPT (upper), [S {\small II}]-BPT (centre) and [O {\small I}]-BPT (bottom) diagrams for each Voronoi bin with SNR $>$ 3 in each line.  Black curves separate AGN-, SF- and LI(N)ER-like line ratios (see text for details). For each diagnostic, on the right is shown a map marking each Voronoi bin with the colour corresponding to increasing flux ratios (as in left panels).  The colour-codes are defined using the emission-line-ratio (ELR) functions described in the text. Grey points correspond to  [O {\small III}]/H$\beta$  lower limits; light-to-dark grey colours in the maps correspond to increasing [N {\small II}]/H$\alpha$ (top right), [S {\small II}]/H$\alpha$ (middle) and [O {\small I}]/H$\alpha$ (bottom)  ratios. The contours of [N {\small II}] line emission are overplotted in black. Most of the gas emission is dominated by LI(N)ER-like ionisation. In the [S {\small II}]-BPT map, we labeled the position of the four SCs identified in this work.}
\label{BPT}
\end{figure*}

\section{Ionisation conditions}\label{ionisationconditions}

In this section we investigate the dominant ionisation mechanism(s) for the emitting gas across the MUSE FOV using standard emission line ratios  (e.g. \citealt{Baldwin1981,Veilleux1987,Diaz2000}) and line widths (e.g. \citealt{Dopita1995}). The close wavelengths of the line pairs involved in such diagnostics ensure minimum dust reddening effects. The only exception is the \citet{Diaz2000} diagnostic, involving [S {\small II}] and [S {\small III}] transitions, for which we applied the extinction corrections introduced in Sect. \ref{Sextinction}.

\subsection{BPT diagnostics}

Figure \ref{BPT} (top-left panel) shows the [O {\small III}]$\lambda$5007/H$\beta$ versus [N {\small II}]$\lambda$6584/H$\alpha$ flux ratios ([N {\small II}]-BPT diagram hereinafter), derived by integrating the line flux over the entire fitted profiles, for only those Voronoi bins in which all the emission lines  are detected with an SNR $>3$. The points in the diagnostic diagram are colour-coded using a function of the two line ratios, defined as the emission-line-ratio (ELR) function (Eq. 1 in \citealt{Dagostino2019}):

\begin{multline}\label{ELR}
ELR_{X-BPT} =  \frac {log(X) - min(log(X))} {max(log(X)) - min(log(X))} \times \\
 \frac {log(Y) - min(log(Y))} {max(log(Y)) - min(log(Y))}, 
\end{multline}
with $X =$ [N {\small II}]/H$\alpha$ and $Y=$ [O {\small III}]/H$\beta$. The same colours are also reported in the map of Arp220 (top-right panel) to distinguish the different ionisation sources in the FOV: purple-to-blue colours identify regions with the lowest line ratios, while green-to-orange colours identify the Voronoi bins with highest line ratios. 

The flux ratios errors are not reported in the figures; as shown, for instance, in Fig. \ref{decomposition2}, these uncertainties can be significant (up to a few 0.1 dex), both because of the SNR level and the degeneracies in the fit analysis. 

The H$\beta$ line is undetected in several Voronoi bins, because of the strong dust attenuation. Therefore, we decided to include in these diagrams the (grey) points associated with [O {\small III}]/H$\beta$ lower limits, derived using a $3\sigma$ upper limit for the flux of the (undetected) H$\beta$ line. In the Arp220 map, the colour intensity of grey Voronoi bins (from light to dark-grey) is coded as a function of increasing [N {\small II}]/H$\alpha$. 

The curves drawn in the diagrams correspond to the maximum SB curve (\citealt{Kewley2001}) and the empirical relation (\citealt{Kauffman2003}) used to separate purely SF galaxies from composite AGN-SF galaxies and AGN-/LI(N)ER-dominated systems (e.g. \citealt{Kewley2006,Belfiore2016}); the dot-dashed line is from \citet{CidFernandes2010}, and is used to separate between LI(N)ERs and AGN. 
The [N {\small II}]-BPT diagram shows that the SF ionisation is found in the source at $\sim 30''$ west of the nuclei (RA 15:34:54.7, DEC 23:29:58 in Fig. \ref{Hachannels}); composite line ratios are found in a handful of Voronoi bins, mostly associated with a SC at $\sim 5''$ east from the nuclei. 
Most of the Voronoi bins have very high [N {\small II}]/H$\alpha$ and relatively low [O {\small III}]/H$\beta$ ratios, resulting in LI(N)ER-like line ratios. The highest line ratios are found in proximity of the bubble and the high-$v$ structures. 

All [O {\small III}]/H$\beta$ lower limits are in the LI(N)ER region as well. These estimates are  consistent with the measurements derived  by \citet{Lockhart2015} for the gas along PA $\sim 138^\circ$, using HST narrowband filters (with average log([O {\small III}]/H$\beta$) $\approx 0.1$ along PA $\approx 138^\circ$). 

In Fig. \ref{BPT} we also report the  [O {\small III}]/H$\beta$ versus [S {\small II}]$\lambda\lambda$6716,31/H$\alpha$ flux ratios ([S {\small II}]-BPT diagram hereinafter; central panels) and the [O {\small III}]/H$\beta$ versus [O {\small I}]$\lambda$6300/H$\alpha$ flux ratios ([O {\small I}]-BPT hereinafter; bottom panels).  Colour-codes are derived using the ELR function (Eq. \ref{ELR}), with $X =$ [S {\small II}]/H$\alpha$ for the [S {\small II}]-BPT, and $X=$ [O {\small I}]/H$\alpha$ for the [O {\small I}]-BPT.

The lines drawn in the diagrams correspond to the optical classification scheme of \citet{Kewley2006, Kewley2013}, and confirm the ubiquity of LI(N)ER-like line ratios across the MUSE FOV. These two diagrams also show the presence of a second SC at $\sim 8''$ east ($SC_2$ in the [S {\small II}]-BPT map), and low line ratios along a stream in the west regions. 

 It should be kept in mind that the lines in the BPT diagrams do not provide a sharp separation between the different ionisation mechanisms. In particular, LI(N)ER-like line ratios could indicate the presence of  $(i)$ a low-ionisation emission from AGN (e.g. \citealt{Heckman1980, Baron2019}), $(ii)$ SF and/or AGN activity in a high metallicity environment (see e.g. Figs. 1 and 4 in \citealt{Kewley2013}), or  $(iii)$ fast shocks induced by SB-, AGN-driven winds, or galaxy interactions (e.g. \citealt{Allen2008})\footnote{In many local galaxies, the LI(N)ER emission is associated with gas ionised by the hard radiation field of evolved (post-AGB) stars, as dicussed e.g. by \citet{Belfiore2016}. However, this does not likely apply to SB galaxies like Arp220.}. 
 In fact, the location of the separation lines between SF and AGN ionisation in the diagrams strongly depend on the metallicity regime of the ISM (e.g. \citealt{Kewley2013}). 
 On the other hand, general shock models predict line ratios that can cover a very large range in the BPT diagrams (e.g. \citealt{Alarie2019}). 
 The high [N {\small II}]/H$\alpha$,  [S {\small II}]/H$\alpha$ and [O {\small I}]/H$\alpha$ ratios measured in Arp220 allows us to exclude stellar photoionisation in almost every Voronoi bin, but BPTs alone cannot allow us to discriminate between AGN and shock ionisation.  
We therefore investigate in more detail the gas conditions considering additional diagnostics to isolate the cause(s) of the lines emission.

\subsection{Ionisation parameter diagnostics}\label{Sionisationparameter}

Arp220 resolved BPT diagrams could be consistent with the predictions by AGN photoionisation models  with a  ionisation parameter log $U$ $\lesssim -3$ and metallicities 12+log(O/H) $\gtrsim 8.7$ (e.g. \citealt{Groves2004,Davies2016,Baron2019}).
In order to measure the $U$ parameter, 
defined as the number of ionising photons $S_*$ per hydrogen atom density $n_H$ divided by the speed of light $c$, we use of the [S {\small III}]$\lambda\lambda9069,9532$/[S {\small II}]$\lambda\lambda6716,31$ ratio (e.g. \citealt{Diaz2000}). 
Since [S {\small III}]$\lambda$9532 is not covered by the wavelength range observed by MUSE, we adopted a theoretical ratio of [S {\small III}]$\lambda$9532/[S {\small III}]$\lambda$9069 = 2.5 (\citealt{Osterbrock2006}), fixed by atomic physics. 
The [S {\small III}]$\lambda$9069 line has been modelled using the best-fitting kinematic parameters obtained for the optical lines (i.e. the kinematics and the number of components per individual Voronoi bin).

\begin{figure}[h]
\centering
\includegraphics[width=9.cm,trim=0 0 0 0,clip]{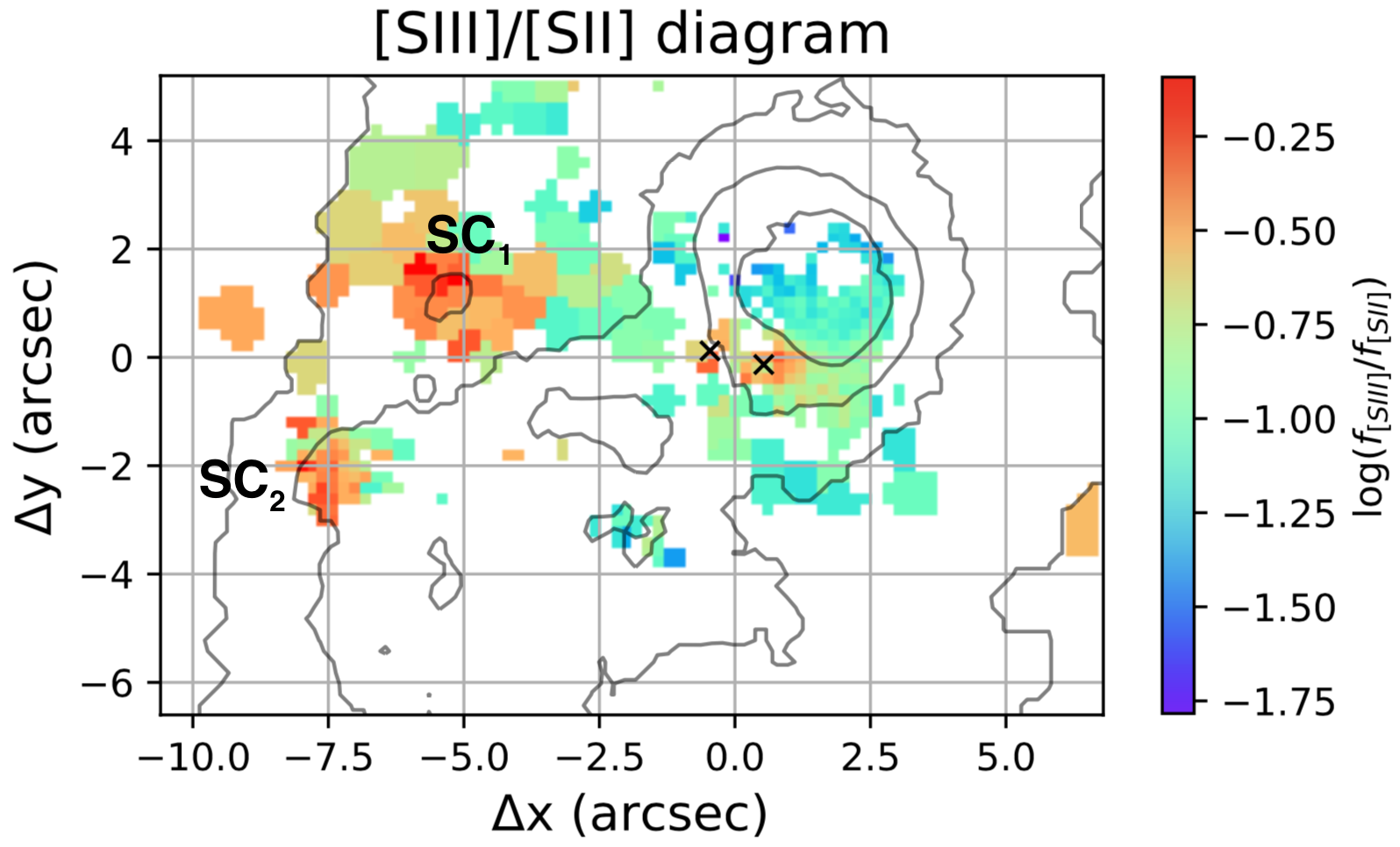}
\caption{\small  Arp220 map of the [S {\small III}]$\lambda\lambda9069,9532$/[S {\small II}]$\lambda\lambda6716,31$ ratios, obtained from the fitted total line profiles. Emission line fluxes have been corrected for extinction (see Sect. \ref{Sextinction}). The contours of [N {\small II}] line emission and the position of the E and W nuclei are overplotted in black.  }
\label{SIIISII}
\end{figure}

In Fig. \ref{SIIISII} we show the [S {\small III}]/[S {\small II}] ratio map of Arp220\footnote{These flux ratios have been obtained using the [O {\small III}]-based Voronoi tesselation; a more uniform coverage of the circum-nuclear regions could be obtained using a [S {\small III}]-based binning, but without affecting outcomes shown in the figure.}. 
The [S {\small III}] emission is detected only in the vicinity of the two nuclei and the clumps $SC_1$ and $SC_2$; on the contrary, [S {\small II}] lines are detected across most the MUSE field.  We found log([S {\small III}]/[S {\small II}]) in the range from  $\approx -0.2$ (log $U \approx -3$, in the nuclei and SCs) to $\approx -1$ (log $U \approx -4.5$, in the circum-nuclear regions). Even lower ionisation parameters might be associated with the regions where [S {\small III}] is undetected. 

Unfortunately, standard metallicity diagnostics are not calibrated for LI(N)ER-like line ratios (\citealt{Dopita2016,Curti2017}), and cannot be used. Therefore, with the information so far collected, we cannot confirm that ISM conditions in Arp220 are consistent with the AGN ionisation expectations mentioned above.

\subsection{Shock diagnostics}\label{shocks}

In order to investigate the possible contribution of shock excitation, we studied the correlation between the gas kinematics and ionisation state. The line ratios produced in photo-ionised regions should be independent of the gas kinematics, while are expected to correlate with the kinematics of the shock-ionised material (e.g. \citealt{Monreal2006,Arribas2014,McElroy2015,Perna2017b,Mingozzi2019}). 

\begin{figure*}[h]
\centering
\includegraphics[width=17.cm,trim=0 30 0 50,clip]{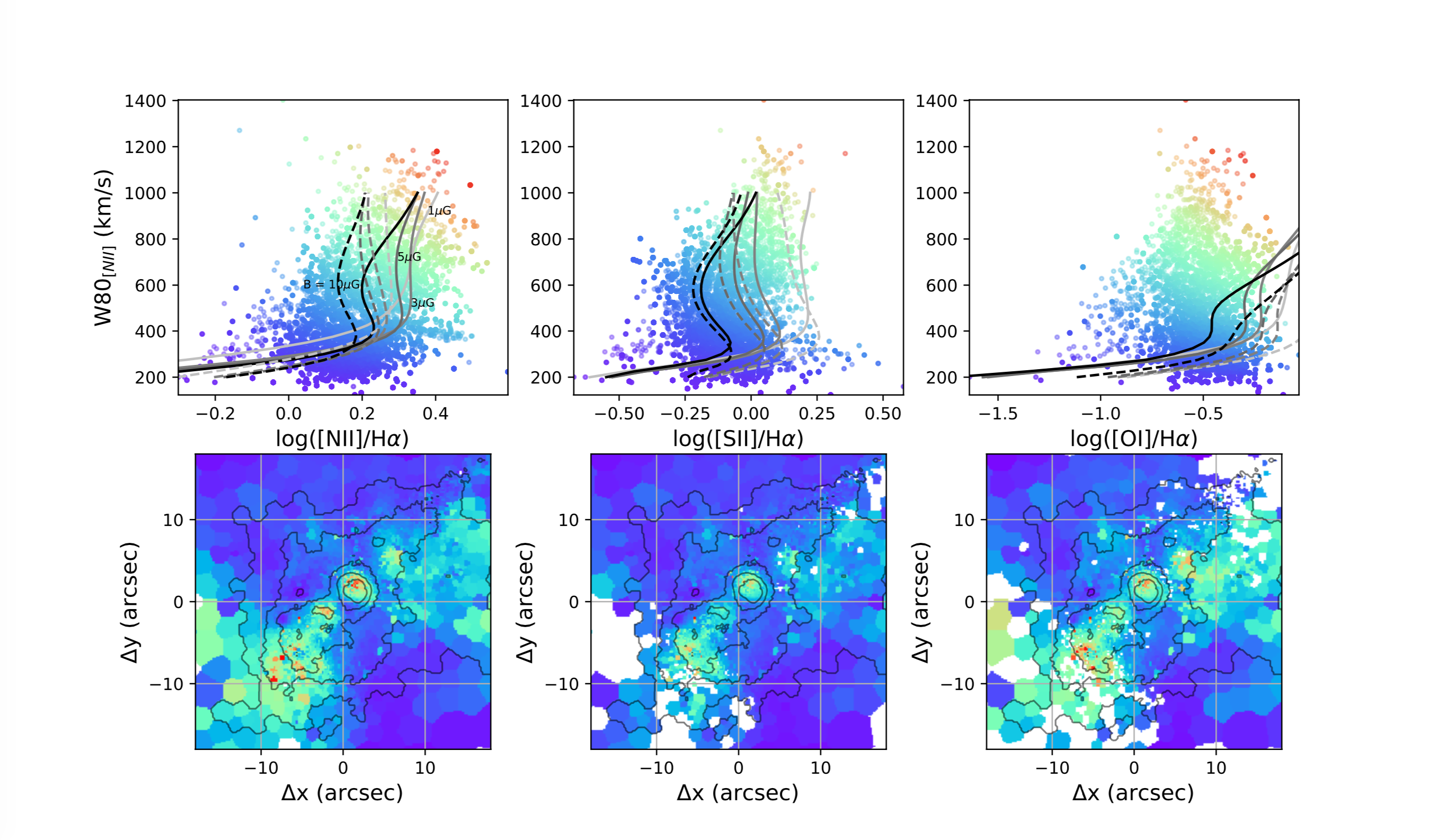}

\caption{\small {\it Top panels:} $W80_{[NII]}$ against log([N {\small II}]/)H$\alpha$,  log([S {\small II}]/)H$\alpha$, and  log([O {\small I}]/)H$\alpha$ from left to right, obtained from the fitted total line profiles. The plotted measurements are colour-coded from purple-to-red going from low to high flux ratios and line widths. Dashed and solid lines represent shock model grids from  {\rm MAPPING V} (\citealt{Sutherland2017, Sutherland2018}). Each line connects predicted line ratios of certain metallicity (1 $Z_\odot$, solid lines; 2 $Z_\odot$, dashed lines), pre-shock electron density (1 cm$^{-3}$) and magnetic field (from 1 to 10 $\mu G$, as labeled in the first panel for 2 $Z_\odot$ models), with changing shock velocities $V_s$ in the range $200- 1000$ km/s. We assumed a one-on-one correlation between $V_s$ and $W80$.  {\it Bottom panels:} Arp220 maps associated with the top panel diagrams, using the same colour-codes. }
\label{w80lineratios}
\end{figure*}

In Fig. \ref{w80lineratios} (top panels) we show $W80_{[NII]}$ against [N {\small II}]/H$\alpha$, [S {\small II}]/H$\alpha$ and [O {\small I}]/H$\alpha$ with colours from purple-to-red going from low to high flux ratios and line widths; in the bottom panels we report the corresponding positions on the Arp220 map. The figure shows that the highest line ratios come from the regions with highest $W80$. 
To  explore the possible correlations between flux ratios and line widths, we report in Fig. \ref{w80lineratios} (top panels) the predicted [N {\small II}]/H$\alpha$, [S {\small II}]/H$\alpha$ and [O {\small I}]/H$\alpha$ as a function of shock velocity ($V_s$) from grids of shocks models calculated with the code {\rm MAPPING V} (\citealt{Sutherland2017, Sutherland2018}). The models have been extracted from the 3MdBs database (\citealt{Alarie2019}), considering the exact replica of \citet{Allen2008} grids with magnetic field values in the range $1 - 10$ $\mu G$, shock velocities in the range $200- 1000$ km/s and  a fixed pre-shock density of 1 cm$^{-3}$, assuming a metallicity of 1 $Z_\odot$ (dashed lines) and 2 $Z_\odot$ (solid lines). 
The measured $W80$ may depend on shock geometry and is not predicted in the shock models, although a positive correlation between $V_s$ and $W80$ is expected. The predicted trends in Fig. \ref{w80lineratios} are therefore shown assuming a one-on-one correlation between the two velocities. 

The  [N {\small II}]/H$\alpha$ and [S {\small II}]/H$\alpha$ diagrams show a clear match between our measurements and the predicted trends, confirming the  close connection between $V_s$ and $W80$ (see also \citealt{Dopita2012} for a similar result). 
Interestingly, the two diagrams suggest that widespread, extended shock excitation may account for most of the gas emission in Arp220.

On the other hand, the [O {\small I}]/H$\alpha$ diagram displays a noticeable discrepancy between predicted trends and observations, possibly because of 
the dependence of the different line ratios on the metallicity regime: 
for a fixed $V_s$ and magnetic field, [S {\small II}]/H$\alpha$ ratios show a negligible dependence on the metallicity, going from 1 to 2 $Z_\odot$. This diagnostic also presents the best match between observations and predictions. On the other hand, the [N {\small II}]/H$\alpha$ - and even more so the [O {\small I}]/H$\alpha$ ratios -  show a clear dependence on the metallicity and less obvious match with shock models predictions. 

An alternative explanation for the different match between line ratios and shock predictions in Fig. \ref{w80lineratios} is that the one-on-one correlation between $W80$ and $V_s$ is not correct. \citet{Ho2014} used the velocity dispersion of the emitting gas to relate the line measurements to $V_s$.  Following \citet{Ho2014}, and using $\sigma$ velocities instead of $W80$,  our measurements  in Fig. \ref{w80lineratios} would shift vertically to $\approx \times 2.35$ smaller values (see Fig. \ref{sigmalineratios}). This would translate in a good match between predictions and measurements in the [S {\small II}] and [O {\small I}] diagrams, at least for the regions with $\sigma > 200$ km/s, but would totally forfeit the match in the [N {\small II}] diagram. In particular, the shock models would significantly under-predict  [N {\small II}]/H$\alpha$  ratios by a factor of $\sim 0.2$ dex in the regions with broad emission lines. Moreover, shock models would not explain the presence of gas with  emission line ratios not compatible with SF ionisation and $\sigma < 200$ km/s. These arguments disfavour a  one-on-one correlation between $\sigma$ and $V_s$; we therefore consider the $W80-V_s$ relation as more reliable.

3MdBs database also provides the predicted emission line luminosities per unit area for different shock models. We used H$\alpha$ predictions from the same models mentioned above, obtaining log(H$\alpha$) within the range [$-4.8, -2.4$] erg/s/cm$^2$, corresponding to expected surface brightness between $5\times 10^{-17}$ erg/s/cm$^2$/arcsec$^2$ (for $V_s = 100$ km/s) and $1\times 10^{-15}$ erg/s/cm$^2$/arcsec$^2$ (for $V_s = 1000$ km/s). These values are fully consistent with the observed H$\alpha$ surface brightness shown in Fig. \ref{Hachannels},  hence confirming that the ionised gas in Arp220 can be associated with shocks, according to MAPPINGS V predictions.

The relation between the excitation degree and the line width of ionised gas can be interpreted as evidence for tidally- or outflow-induced shocks. \citet{Monreal2006, Monreal2010} and \citet{Rich2015} shown that, while isolated U/LIRGs are mainly associated with ionisation caused by young stars, an increasingly important emission component from shock ionisation is found in interacting pairs and more advanced mergers (see also e.g. \citealt{Joshi2019,Mortazavi2019}). This indicates that tidal forces play a key role in the origin of the ionising shocks in U/LIRGs. 

In particular, \citet{Monreal2006, Monreal2010} and \citet{Rich2015} found that interacting and merging U/LIRGs have line widths $W80 \sim$ 250 km/s (up to 450 km/s in rare cases). These velocities are very similar to the lower $W80$ values observed in Arp220 (e.g. Fig. \ref{w80lineratios}). On the other hand, the Arp220 line widths measured along PA $\sim 138^\circ$ are significantly higher,  
and more similar to those observed in SB- and AGN-driven winds (e.g. \citealt{Rupke2013,Cazzoli2016, Perna2017b, Hinkle2019}).
The similarities between Arp220 and other systems kinematics could suggest that tidal forces are responsible of shock excitation in the emitting material with low velocity dispersion, while the extreme line widths along PA $\approx 138^\circ$ favour the presence of outflow-induced shocks. 

Summarising, the positive correlation between the excitation degree and the line widths, and the good match with theoretical predictions from shock models suggest that the ionised gas in Arp220 is strongly exposed to shock excitation. Indeed, this mechanisms could be the dominant process responsible for the optical lines emission.

\section{Electron density}\label{Sne}

Plasma properties represent further key ingredients to characterise the ISM conditions. 
Electron density ($N_e$) and temperature ($T_e$) can be derived using diagnostic line ratios involving forbidden line transitions (\citealt{Osterbrock2006}; see e.g. \citealt{Perna2017b,Perna2019, Rose2018, Santoro2018, Mingozzi2019}). The only available diagnostic lines detected within the MUSE spectral range are the [S {\small II}]$\lambda\lambda$6716,31,  sensitive to densities in the range $10^2\lesssim N_e/$cm$^{-3}$ $\lesssim 10^{3.5}$. 

The $N_e$ estimates are usually quite uncertain: [S {\small II}] doublet  ratio estimates are strongly affected by the degeneracies in the fit because the involved emission lines are faint and severely blended. Probably because of this, the [S {\small II}] ratios derived from our best-fit modelling do not show any significant variation across the MUSE FOV. We therefore report the median value derived from the $24''\times 24''$ innermost regions (as shown in Fig. \ref{w80lineratios}), [S {\small II}]$\lambda$6716/[S {\small II}]$\lambda$6731 $=1.30_{-0.27}^{+0.12}$, corresponding to $N_e = 170_{-110}^{+440}$ cm$^{-3}$. 

Compared to the integrated $N_e$ values in the local star forming galaxies from the SDSS Survey (e.g. \citealt{Sanders2016}), our results indicate that the $N_e$ in Arp220 is a factor of $\sim 7$ higher.  
Our results are instead consistent with recent studies by \citet{Rupke2017} and \citet{Mingozzi2019}, who find spatially averaged values of 150 cm$^{-3}$ in samples of nearby AGN with outflows observed with IFS (see also \citealt{Kakkad2018}), and with mean densities measured in other local ULIRGs (e.g. \citealt{Arribas2014}).
Moreover, our $N_e$ estimate  is similar to the typical values measured in star forming galaxies at $z > 1$ (e.g. \citealt{Harshan2020} and references therein). 

Finally, we note that the measured [S {\small II}] ratios are consistent with those expected for the  {\rm MAPPING V} shock models presented in the previous section, for a pre-shock  $N_e =1$ cm$^{-3}$. 
Therefore, the plasma properties of Arp220 might be explained by the presence of tidally-induced shocks and outflows, which can pressurise the ISM. 

\section{Dust attenuation}\label{Sextinction}

Another key ingredient to characterise the ISM properties is the dust. We used the Balmer decrement ratios  $H\alpha /H\beta$ to estimate the dust attenuation across the FOV. Assuming a ratio of 2.85 (Case B recombination), a gas temperature  $T_e =10^4$ K, a dusty screen, and the Milky Way extinction law (\citealt{Cardelli1989}; CCM law hereinafter), the colour excess is given by: 

\begin{equation}\label{eqEBVgas}
E(B-V)_{gas} = 2.33 \times log((H\alpha /H\beta)/2.85).
\end{equation}

The relative strength of the Balmer lines depends only weakly on local conditions.
Variation in $T_e$ by a factor of two would result in $\sim 0.1$ mag difference in the colour excess (\citealt{Dopita2003}); even lower variations are expected over four order of magnitude in electron density ($N_e = 10^2 - 10^6$ cm$^{-3}$; \citealt{Osterbrock2006}). The Balmer decrement varies little in the case of collisional heating, as the H$\alpha$ emission is enhanced with respect to the Case B (e.g. \citealt{Osterbrock2006}): for instance, the  average value of the  Balmer decrement measured in broad-line AGN and SN remnants, where we expect extreme plasma conditions, is $\approx 3$ (\citealt{Dong2008,Raymond1979}). Recently, \citet{Sutherland2017} shown that shocks can cause a significant deviation from the Case B value (up to $H\alpha /H\beta \approx 5$) in the presence of low-$v$ shocks ($V_s < 30$ km/s), as in the case of Herbig-Haro outflows (\citealt{Dopita2017}). However, these processes are not expected to dominate in Arp220 (with inferred $V_s > 200$ km/s); hence, Eq. \ref{eqEBVgas} provides good dust attenuation estimates for the ISM in Arp220. 

\begin{figure}[t]
\centering
\includegraphics[width=9.3cm,trim=0 476 0 0,clip]{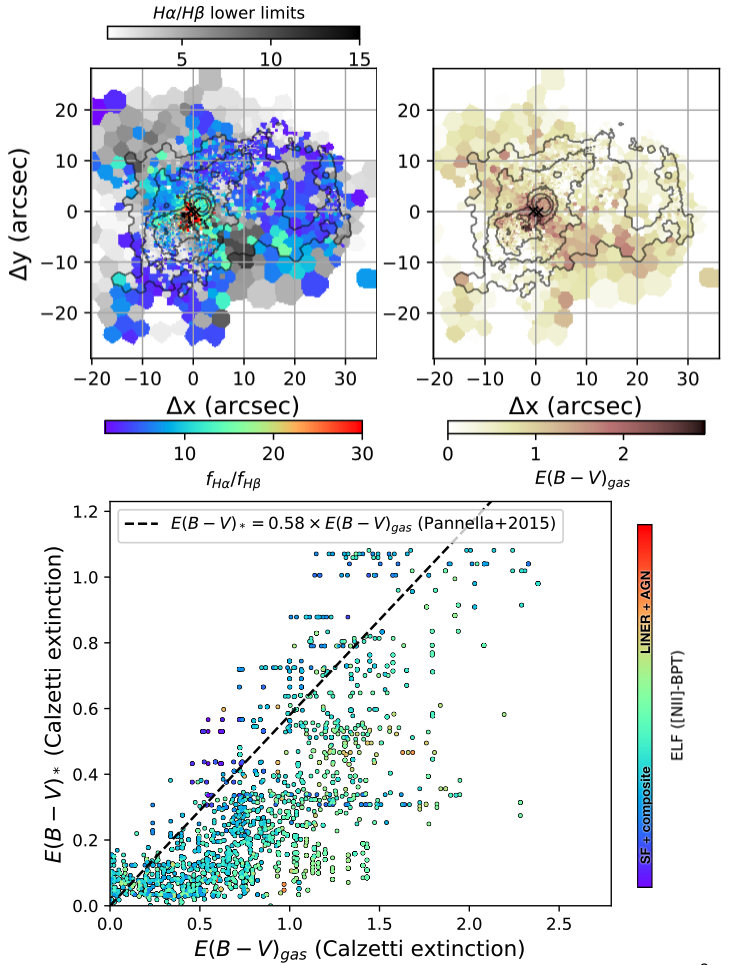}

\caption{\small {\it Left}: Balmer decrement measurements (purple-to-red) and lower limits (grey-to-black) across the MUSE FOV. H$\alpha$ and H$\beta$ fluxes are derived from the fitted total line profiles; 3$\sigma$ upper limits for the H$\beta$ flux are used to derive Balmer decrement lower limits.   {\it Right}: $E(B-V)_{gas}$ map of Arp220, derived from the $H\alpha/H\beta$ ratios in top-left panel, using the CCM extinction law. }
\label{EBV}
\end{figure}

In Fig. \ref{EBV} (right) we illustrate the $E(B-V)_{gas}$  map of Arp220 obtained for the fitted total line profiles (without separating the kinematic components).  
Given the good match between [O {\small III}]/H$\beta$ lower limits and the flux ratio measurements reported in \citet{Lockhart2015}, we also consider the $H\alpha /H\beta$ lower limits in deriving $E(B-V)_{gas}$ values when the H$\beta$ detection is below the 3$\sigma$ threshold. 
Both $H\alpha /H\beta$ measurements and lower limits are therefore shown in Fig. \ref{EBV}, left. The highest line ratios and $E(B-V)_{gas}$ are found in the innermost nuclear regions and along PA $\sim 80^\circ$, resembling the position and extension of the dust lanes observed in the HST image (Fig. \ref{HST}).
The two nuclei are associated with  $E(B-V)_{gas} = 2.5\pm 0.4$ (E) and $2.2\pm 0.3$ (W nucleus), roughly consistent with the nuclear regions column densities derived from X-ray data analysis in \citet{Paggi2017}, log($N_H/$cm$^2$) $\approx 22.3$ (assuming a thermal X-ray emission), corresponding to $E(B-V)_{gas} \approx 2.9$, using the \citet{Guver2009} relation.


\section{Neutral gas covering factor and Hydrogen column densities}\label{Snh}
 
In this section we investigate the correlation between $E(B-V)_{gas}$ and cool gas absorption traced by Na {\small ID} transitions. A more detailed comparison between dust attenuation and Na {\small ID} absorption will be presented in a forthcoming paper, combining the results from the analysis presented in this paper with those from Catal\'an-Torrecilla et al. (in prep.).

A positive correlation between the EW of absorbing gas and $E(B-V)_{gas}$ has been used as an indirect evidence of dust in outflows (e.g. \citealt{Shapley2003,Reichard2003,Perna2019}). In particular, \citet{Rupke2015} reported a positive correlation between the EW of outflowing Na {\small ID} and the stellar continuum attenuation in the ULIRG IRASF05189-2524. Following the arguments presented in their work, under particular conditions, the colour excess and $EW(NaID_{abs})$  can be used as a proxy of $N_H$. In fact,  the attenuation calculated from $E(B-V)_{gas}$ can be related to $N_H$, assuming  a constant dust-to-gas ratio (e.g. in the MW, $N_H/E(B-V) = 6.9\times 10^{21}$ cm$^{-2}$; \citealt{Guver2009}). Similarly, when the absorbing neutral material has a low optical depth ($\tau < 1$) and a uniform and total coverage of the continuum source ($C_f=1$), the $EW(NaID_{abs})$ is proportional to $N_H$ (e.g. \citealt{Rupke2005b}). 
Following \citet{Cazzoli2016}, we considered the average $N_H - EW(NaID_{abs})$ relation within the two extreme relationships found by \citet{Turatto2003}, derived from SNe reddening measurements, to obtain the empirical equation 

\begin{equation}\label{rel1}
E(B-V)_{gas} = -0.02 + 0.29\times EW(NaID_{abs}).
\end{equation}

\begin{figure}[h]
\centering
\includegraphics[width=9.cm,trim=0 0 0 0,clip]{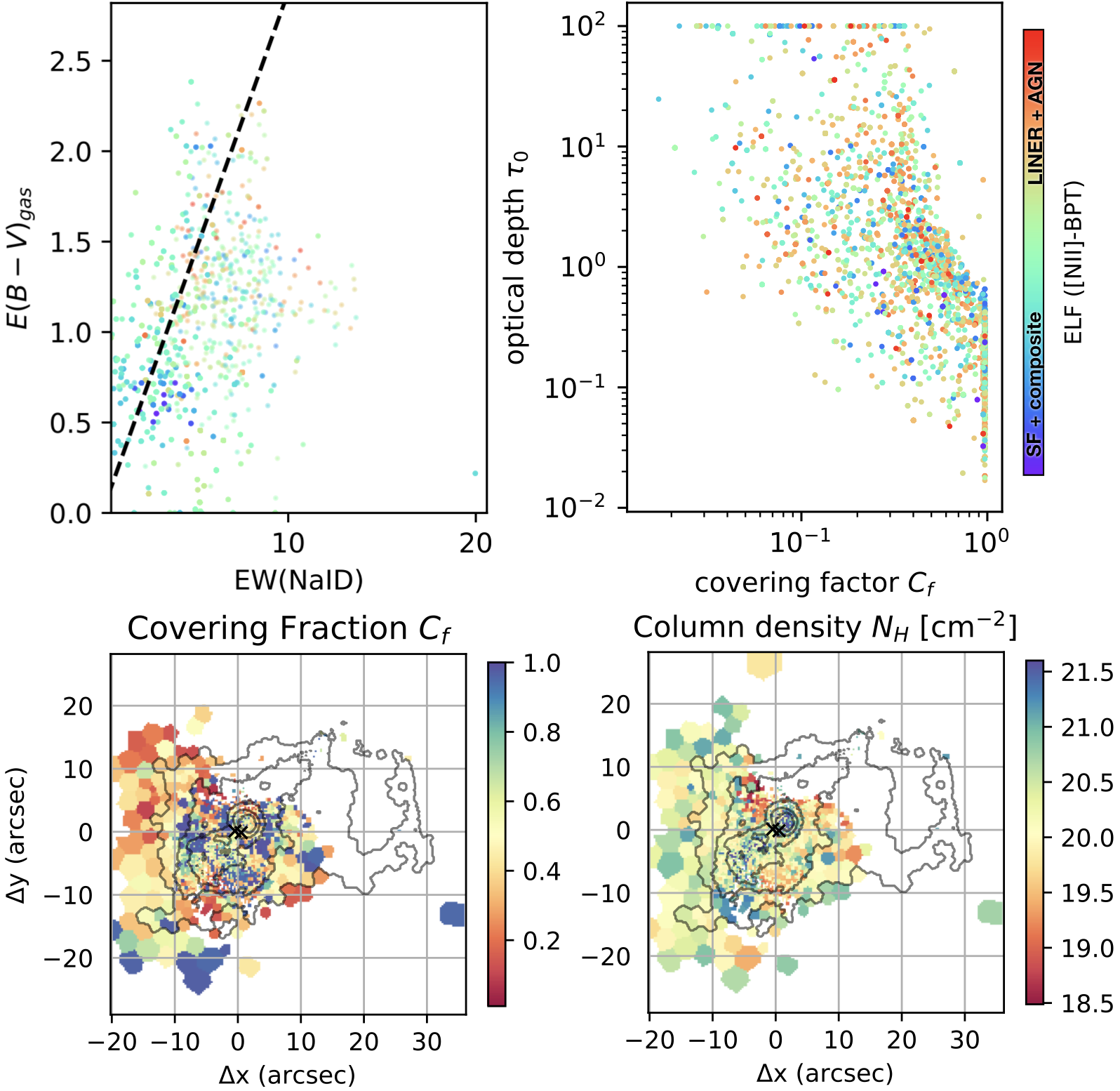}

\caption{\small  
{\it Top-left:}  $E(B-V)_{gas}$ as a function of the Na {\small ID} equivalent width. The dashed line represents Eq. \ref{rel1}, derived under the assumption that both quantities are proportional to the hydrogen column density $N_H$. {\it Top-right:} covering factor vs. optical depth for each kinematic component used to model the Na {\small ID} profiles. The points in the top panels are colour-coded according to their position in the [N {\small II}]-BPT diagram (Fig. \ref{BPT}, top right). {\it Bottom-left:} Covering factor map, obtained from the total Na {\small ID} profiles following the \citet{Rupke2005b} prescription. {\it Bottom-right:} Column density map, obtained summing $N_H$ derived from each kinematic component required to model Na {\small ID} profiles.}
\label{EBVvsEW}
\end{figure}

In Fig. \ref{EBVvsEW} (top left) we report the measured $EW(NaID_{abs})$ as a function of  $E(B-V)_{gas}$. The measurements are colour-coded using the ELR function of the [N {\small II}]-BPT diagram; the dashed line displays Eq. \ref{rel1}. This figure shows that our measurements follow the predicted relation, regardless the ionisation mechanisms responsible of the line emission. 
In support of the observed correlation, we note that our fit results tend to favour low $\tau$ and high $C_f$: the median covering factor and the optical depths observed across the MUSE FOV are $C_f = 0.65_{-0.43}^{+0.32}$ and $\tau = 0.93_{-0.78}^{+7.1}$ (see Fig. \ref{EBVvsEW}, top right).

Finally, following the prescriptions reported in \citet{Rupke2005b}, we derived the sodium abundance taking into account the ionisation fraction ($y=0.9$) and the depletion of Na atoms onto dust and the abundance as

\begin{equation}
log(N_{Na}/N_H)=log(1-y)+A+B,
\end{equation}

 where $A$ = log($N_{Na}/N_H$)$_{gal}$ is the Na abundance in the galaxy and $B =$ log($N_{Na}/N_H$) $-$ log($N_{Na}/N_H$)$_{gal} =-0.95$ is the depletion (the canonical Galactic value, \citealt{Savage1996}).  Using the galaxy abundance derived from the Eq. 12 in \citet{Rupke2005b}, $A = -5.4$, 
we derived the column densities reported in Fig. \ref{EBVvsEW} (bottom-right). These measurements are obtained from the sum of all the kinematic components contributions, and are consistent with column densities generally derived from outflowing Na {\small ID} gas (e.g. \citealt{Rupke2015,Cazzoli2016,Perna2019,Catalan2020}; see also \citealt{Veilleux2020}). Column densities of the order of $0.5 - 5\times 10^{21} cm^{-2}$ have been also derived from X-ray Chandra data within the innermost kpc by \citet{Huo2004} and \citet{Grimes2005}.

\section{Outflow characterisation}\label{Soutflow}

In this section we derive the general properties of the outflowing gas, taking advantage of the information so far collected. We propose two possible outflow configurations, namely a collimated and a wide-angle biconical outflow, and derive their energetics assuming simple mass-conserving wind models.

\subsection{Outflow structure}\label{Sstructure}

\begin{figure*}[h]
\centering
\includegraphics[width=19.cm,trim=0 0 0 0,clip]{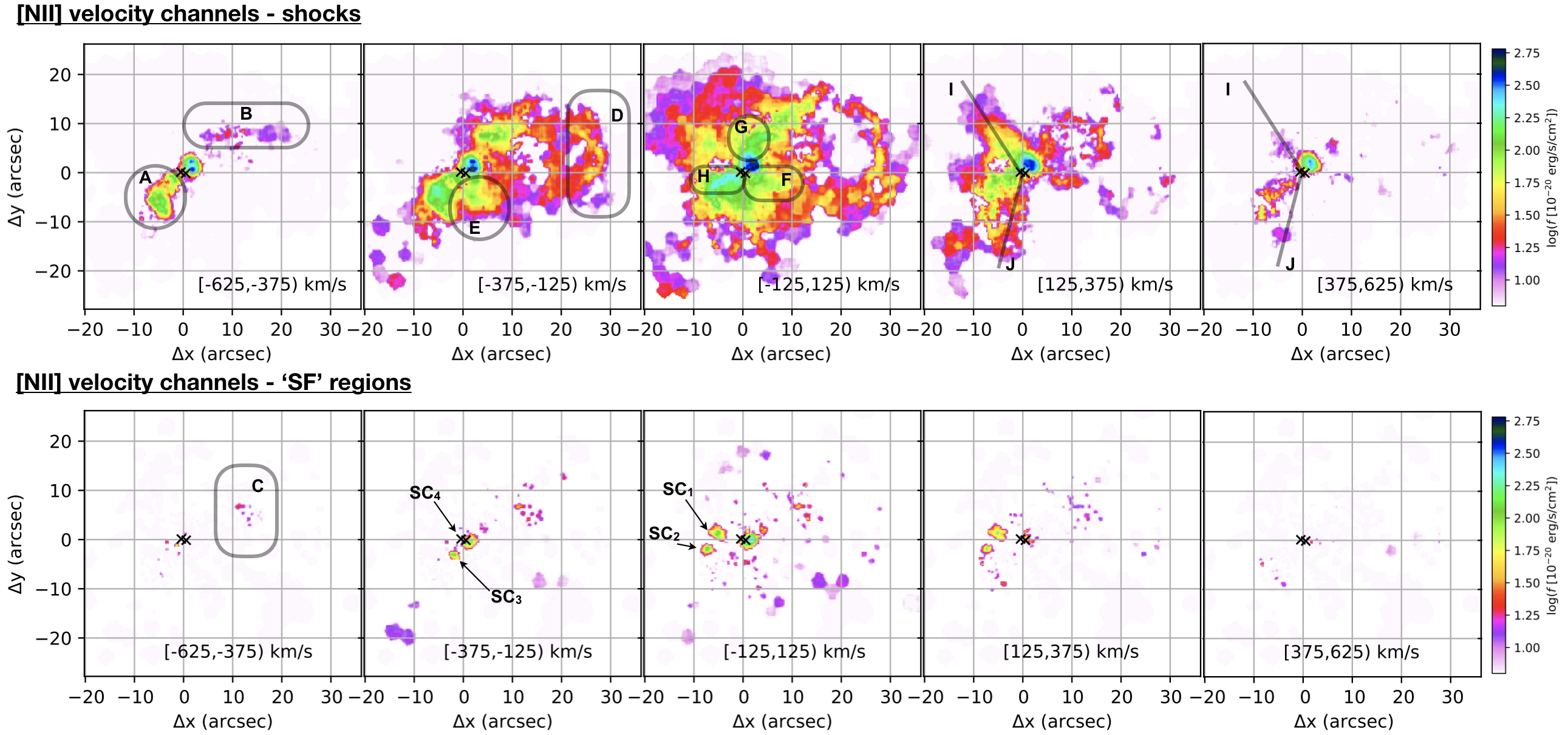}

\caption{\small  
[N {\small II}] velocity channels extracted from the data-cubes containing the  best-fit profiles of shock  (top) and SF ionised (bottom) gas. Letters from $A$ to $J$ locate the main kinematic structures discussed in Sect. \ref{Soutflow}; in the bottom panels, we also mark the position of the four SCs.}
\label{niisfshocks}
\end{figure*}

[N {\small II}]-, [S {\small II}]- and [O {\small I}]-BPT diagrams are used to distinguish between SF and shock ionisation in each individual Gaussian set used to model the line profiles. In particular, a Gaussian set is associated with SF ionisation if at least two BPT diagrams indicate HII-like ratios; on the other hand, a Gaussian set is associated with shock ionisation when line ratios in all BPT diagrams are in the LI(N)ER regions. SF and shock Gaussian sets are then used to construct two data-cubes containing the  best-fit SF- and shock-[N {\small II}] profiles respectively. These data-cubes are used to generate the [N {\small II}] velocity channel maps presented in Fig. \ref{niisfshocks}, for both shock excitation (top panels) and SF ionisation (bottom). In the following, we describe the main features observed in individual velocity channels:

\begin{itemize}
\item  [-]
[$-625,-375$] km/s: fast approaching gas is mostly associated with shocks. This shock-excited gas is preferentially found in the innermost nuclear regions, where we observe two bright clumps within the bubble ($\sim 2''$ north-west from the nuclei), and along two main filaments ($A$ and $B$ regions in the panel). In the SF emission map, we observe some diffuse gas in the north-west quadrant ($C$ region), as well as a few Voronoi bins of faint emission towards south-east.

\item [-]
$[-375,-125]$ km/s: the distribution of shock-induced emission is very similar to the one in the previous velocity channel; the western part of the lobe is identified ($D$ region); we also find high-$v$ gas in the south-west quadrant  ($E$ region).  In the SF map, we identify the two clumps of SF labeled as $SC_3$ and $SC_4$, as well as other fainter emission preferentially aligned along PA $\approx 138^\circ$ (i.e. the outflow direction). 
 
\item  [-] [$-125,+125$] km/s: low-$v$ shock-induced emission fills the bottom part of the lobe structure  ($F$ region), and apparently joins the bubble to the upper part of the west lobe ($G$); similarly, in the south-east quadrant, low-$v$ emission connects the nuclear regions to the high-$v$ shocked gas ($H$). The SF gas is preferentially associated with innermost nuclear regions ($SC_1$, $SC_2$ and $SC_4$), but presents several clumps isotropically distributed across the MUSE FOV. 

\item  [-] [$+125,+375$] km/s: shock-induced emission fills a cone with a large opening angle toward the east direction (within the angle defined by $I$ and $J$ lines). SF gas is mostly associated with $SC_1$ and $SC_2$, but again fainter emission can be located along PA $\approx 138^\circ$.   
 
 \item  [-] [$+375,+625$] km/s: shock-induced emission reproduces the main features observed in the previous channels; SF emission is mostly undetected, except a few Voronoi bins of faint emission towards south-east. 
 
 \end{itemize}

\begin{figure}[h]
\centering
\includegraphics[width=8.6cm,trim=0 0 0 0,clip]{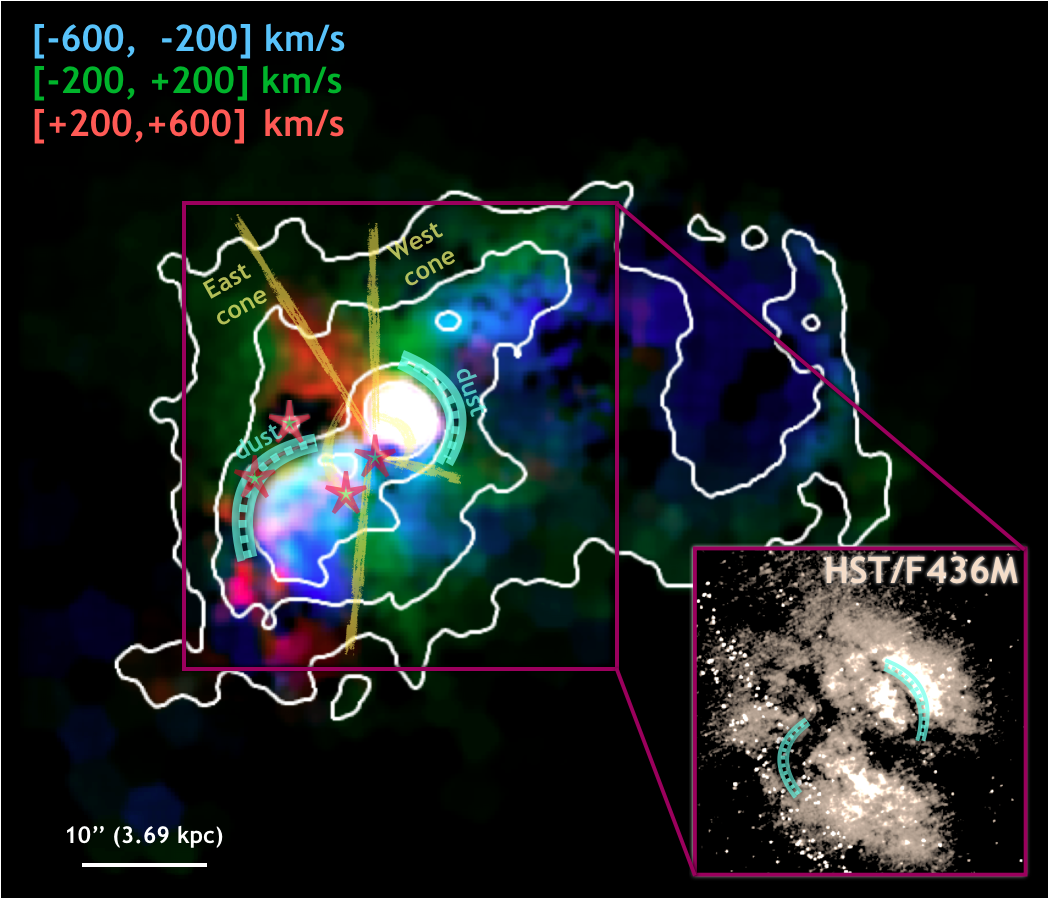}

\caption{\small  
Schematic view of the main structures in Arp220, revealed by MUSE.  The three-colour image shows the [N {\small II}] shock-emission, reconstructed from multi-component best-fit results. Superimposed yellow lines highlight the biconical outflow. Star symbols identify the four main SCs identified in Sect. \ref{Soutflow}. Cyan curves locate the main dust filaments possibly affecting the outflow geometry. The same curves are reported in the inset, showing the blue stellar continuum emission as well as the dusty structures in the innermost nuclear regions of Arp220.}
\label{RGBoutflow}
\end{figure}

Summarising, shock emission is mostly following a few kinematic structures. The most extreme velocities ($|v| \gtrsim 600$ km/s) in the ionised gas are measured along PA $\sim 138^\circ$; their spatial correlation with high-$v$ Na {\small ID} and with X-ray emission, and their alignment with the `minor axis' of the stellar disk suggest the presence of a collimated outflow driven by nuclear SB and/or AGN activity. It is still unclear, however, if this ejected gas is part of a more structured outflow morphology, e.g. also involving some of the structures presented in Fig. \ref{niisfshocks}. 

In the following, we propose that the high-$v$ gas in Arp220 might also be associated with a wide-angle biconical outflow. We anticipate however that the collimated outflow scenario will be preferred, because of the arguments presented at the end of this section.  
Figure \ref{RGBoutflow} shows a three-colour image of [N {\small II}] shock emission, where high-$v$ approaching and receding gas are shown in blue and red, respectively, while low-$v$ gas is in green. In the figure, we also marked with cyan curves the regions associated with the main dust lanes, which in our proposed scenario affect the outflow path; the same regions can be easily identified in the inset, showing the blue stellar continuum emission - and the dust filaments with flux deficiency -  from HST/F436M observations. 

\underline{In the east part of the FOV}, there could be a  one-sided and wide-angled kpc-scale conical outflow, expanding in a clumpy medium.
The fast receding gas well defines the edges of a cone ($I$ and $J$ lines in Fig. \ref{niisfshocks}, top-right); along the cone axis there are both low-$v$ blueshifted and redshifted components ($H$ region in Fig. \ref{niisfshocks}). This gas is brighter than the one at the edges of the cone (seen at faster receding velocities), indicating that the latter likely resides behind the kpc-scale disk. 
Instead, the gas in the $H$ region is likely propagating close to the plane of sky, as it is associated with velocities from $\approx -600$ to $\approx +600$ km/s. This gas could be confined by the dust lane extending along the south-east direction from the nuclei and then toward south-west at $r \sim 10''$ (see Fig. \ref{RGBoutflow}). The dust distribution is possibly deviating the outflow direction toward a path closer to the LOS, generating the comma-shaped region ($A$) with high-$v$ approaching gas (up to $\approx -1000$ km/s).   

This scenario is also consistent with the neutral outflow kinematics traced by Na {\small ID} gas (Figs. \ref{NaIDabsimages} and \ref{NaIDemimages}). Blueshifted absorbing material with high $W80$ is revealed along PA $\sim 138^\circ$ in the south-east quadrant; importantly, P-Cygni profiles are found in the more external regions, along the lower edge of the cone, where we observe both blueshifted ($v < -300$ km/s) and redshifted ($v> 200$ km/s) ionised gas. Na {\small ID} emission is therefore tracing the more distant receding outflowing gas, which is less attenuated by the dust with respect to the innermost regions; Na {\small ID} absorption is instead related to the foreground approaching outflowing material.

In support of this scenario, we note that the distribution of receding and approaching gas in Arp220 is very similar to the one in Circinus galaxy and NGC4945, once we consider a different (opposite) LOS (see e.g.  \citealt{Mingozzi2019}, Fig. 2; \citealt{Venturi2017}, Fig. 2). The presence of higher velocities at the edges of the cone can be explained by projection effects, assuming they are mostly perpendicular to the plane of the sky. Instead, the higher velocity dispersions along the cone axis, due to the presence of both approaching and receding ionised gas, can be explained considering a cone axis close to the plane of the sky. 
 
\underline{In the west part of the FOV}, the shock emission is dominated by the bubble. This structure could originate from the counterpart of the eastern outflow cone, confined by the dust lanes observed in the HST images (see Fig. \ref{RGBoutflow}). 
The lobe structure could be originated by outflowing gas which  at larger distances starts to flow back to the interacting system. Alternatively, 
it might be originated from the interaction of the merging galaxies, as suggested by the presence of tidal tails in the merger simulations presented in \citet{Konig2012}.

The information so far collected, however, does not allow us to robustly discriminate between tidally- and outflow-induced shocks at low to intermediate ($\sim 500$ km/s) velocities; moreover, the location of dust and ejected gas along the LOS are unconstrained and we do not know if the dust is actually deviating the outflow path. Therefore, 
in the following, we will refer to the collimated outflow geometry as the preferred scenario.

\subsection{Location of the outflow origin}\label{Sorigin}

Even with the present AO-assisted IFS data, we cannot accurately locate the origin of the kpc-scale outflow we detect in atomic ionised and neutral gas. 
However, a few considerations can be drawn taking into account the high-resolution ($\sim 0.1''$) ALMA observations tracing the HCN (1-0) and HCO+ (1-0) emission lines in the surroundings of the two nuclei, presented in \citet{BarcosMunoz2016} and \citet{BarcosMunoz2018}. The authors report the detection of a spatially resolved molecular outflow associated with the W nucleus, with an extension of $\lesssim 120$ pc along the north-south direction. 
The kpc-scale atomic outflow we present in this paper is unlikely related to the molecular wind in the W nucleus, as it is preferentially oriented along PA $\sim 138^\circ$. 

Recently, \citet{Wheeler2020} reported evidence for a collimated outflow in the E nucleus, traced by CO(3-2) at high resolution ($\sim 0.2''$) with ALMA observations. This outflow is oriented along the kinematic minor axis of the nuclear molecular disk, and extends out to $\lesssim 100$ pc from the E nucleus. These findings might suggest 
that such nuclear wind is related to the kpc-scale atomic outflow presented in this paper. This scenario, however, has to be confirmed with follow-up observations, e.g. tracing atomic gas kinematics through IR transitions, which are less affected by dust extinction emission line, with high angular resolution.

 \subsection{Outflow energetics}

The construction of a detailed 3D kinematic model for the Arp220 outflow is beyond the scope of this paper. In this section, we derive order-of-magnitude estimates of the outflow energetics assuming simple mass-conserving wind models. Inclination-corrected velocities and distances are obtained assuming that the outflow is oriented in the plane of the E nucleus disk (see Sect. \ref{Sorigin}), with an inclination of $70^\circ$ to the LOS (e.g. \citealt{Scoville2017}). 

We calculated the outflow mass rate of the ionised gas in each Voronoi bin from the de-reddened flux of the H$\alpha$ components associated with non-SF ionisation, assuming the Case B recombination in fully ionised gas with $T_e = 10^4$ K (see e.g. \citealt{Cresci2017}), and a uniform electron density across the MUSE FOV, $N_e = 170$ cm$^{-3}$ (Sect. \ref{Sne}). 
We performed the outflow mass rate calculation for each Voronoi bin using the relation $\dot M = M_{out}v_{out}/ R_{out}$, with $v_{out} = v50$ (e.g. \citealt{Harrison2014}), and considering the respective local properties (e.g. velocities, H$\alpha$ flux, distance from the nuclear regions). 
We thus obtained the total ionised outflow mass rate and the kinetic ($\dot K = 1/2 \dot M (v_{out}^2 + 3\sigma^2)$) and moment power ($\dot P =\dot M v_{out}$) by summing the values from the single Voronoi bins: $\dot M = 20$ M$_\odot$/yr, $\dot K = 2 \times 10^{42}$ erg/s and $\dot P = 4\times 10^{34}$  dyne\footnote{These values are consistent (within a few \%) with those derived from integrated quantities, considering an outflow extension of 6 kpc (see also \citealt{Venturi2018}).}. For each outflow energetic, we also derived a confidence interval (CI), considering most and least conservative assumptions, taking into account an $N_e$  in the range [60, 440] cm$^{-3}$ (Sect. \ref{Sne}), and that H$\alpha$ emission at $|v| < 375$ km/s could or could not participate in the outflow (Sect. \ref{Sstructure}): we obtained $\dot M \in [5, 60]$ M$_\odot$/yr, $\dot K \in [0.5,7]\times 10^{42}$ erg/s and $\dot P \in [0.9,9]\times 10^{34}$ dyne. 

We calculated the outflow energetics of the neutral gas traced by Na {\small ID} absorption assuming the same wind model, following the arguments presented in \citet{Shih2010}. We excluded, also in this case, the kinematic components which ionised gas is associated with SF, and performed a Gaussian smooth (with $\sigma = 2$ pixels) to the $N_H$ map (Fig. \ref{EBVvsEW}) to remove a few outlier measurements due to (probably) unphysical values associated with $\tau \gg 1$\footnote{This is the same as removing all Voronoi bins with $\tau > 7$ from the energetics computation.}. To further limit the possible inclusion of gas not participating in the outflow, we considered only the kinematic components with $v < -50$ km/s (e.g. \citealt{Rupke2005a}). We assumed a single radius for the wind, $R_{out} = 6$ kpc (roughly corresponding to the distance of the comma-shaped region), 
and used the $N_H$, $C_f$ and velocities from individual Voronoi bins.  
The total outflow energetics are: $\dot M= 27$ M$_\odot$/yr, $\dot K = 4\times 10^{42}$ erg/s and $\dot P = 10^{35}$ dyne.     
As for the ionised component, we also derived a CI considering most and least conservative assumptions: $\dot M  \in [2,40]$ M$_\odot$/yr, $\dot K = [0.7,6]\times 10^{42}$ erg/s and $\dot P = [0.1,2]\times 10^{35}$ dyne. The minimum values are computed  considering only the Voronoi bins with $W80 > 700$ km/s (roughly corresponding to the selection of the Na {\small ID} profiles extended at $v < - 375$ km/s, as for the ionised component), while the maximum values are obtained considering an outflow radius of  9 kpc (see e.g. Fig. \ref{NaIDabsimages}).  
All outflow energetics so far computed are reported in Table \ref{toutflow}.

Our results suggest that the neutral 
and ionised outflows in Arp220 are likely to have similar mass rates and energetics, 
consistent with other AGN- and SB-driven outflows at low-$z$ (e.g. \citealt{Rupke2017, Fluetsch2019, Fluetsch2020}). In Table \ref{toutflow} we also report the relative contributions associated with the main outflow features in Arp220, the bubble and the comma-shaped region along PA $\sim 138^\circ$. A significant portion of the ionised outflow mass, energy and momentum rates are associated with the bubble, which is  closer to the central engine of the outflow. As suggested by e.g. \citet{Venturi2018}, a decrease of outflow energetics with distance might either imply that the outflow slows down at larger distances, or that the SB (AGN) pushing the wind has become more powerful recently. Another possible explanation could be related to projection effects. In order to better characterise the outflow properties, a more detailed wind model is therefore required. 

The values reported in Table \ref{toutflow} also suggest that the possible contribution of gravitational induced shocks should be minor, as a significant fraction of outflow kinetic energy derived from the ionised and neutral gas components are associated with the two brightest structures in Arp220, the bubble and the comma-shaped region, which also present the most extreme kinematics (hardly attributable to tidally-induced shocks). 

\subsection{Outflow launch mechanism(s)}\label{outflowlaunching}

In order to investigate the possible origin of the outflow, we compared our inferred values of the total outflow energetics with the expected kinetic and momentum power ascribed to stellar processes. 
The expected rate of energy injection from SNe explotion is $\sim 10^{44}$ erg/s, and has been derived assuming the standard SN energy,  $K_{SN} = 10^{51}$ erg, and considering the Arp220 SN rate of $4$ yr$^{-1}$ (\citealt{Varenius2019}). This expected kinetic power is very similar to the one derived assuming a proportionality between $\dot K$ and SFR (\citealt{Veilleux2005}), $\dot K_{SF} \sim 1.8\times 10^{44}$ erg/s,  taking into account the Arp220 SFR $= 250$ M$_\odot$/yr (\citealt{Nardini2010}). The measured $\dot K$ suggests a  $\approx 4\%$ coupling between the stellar processes and the wind energy, consistent with the one expected for SB-driven outflows (e.g. \citealt{Chevalier1974,Schneider2020}). 
 Similarly, the mass-loading factor $\mu = \dot M_{out}/SFR = 0.2$ is similar to those measured in other local ULIRGs (\citealt{Arribas2014, Chisholm2017,Cresci2017}) and high-$z$ star-forming galaxies (\citealt{Genzel2014,Newman2012,Perna2018}) with SB-driven outflows.
 However, the momentum rate generally observed in SB-driven outflows is $\gtrsim 10$ times smaller than the one derived for Arp220 (see e.g. \citealt{Fluetsch2019}, Fig. 20; \citealt{Cicone2014}, Fig. 16).

We also compared our inferred outflow properties with the expected kinetic and momentum power expected for AGN-driven outflows. 
We tentatively estimated the expected rate of energy from CT AGN considering the AGN luminosities inferred by \citet{Paggi2017} and \citet{Nardini2010}, from X-ray and IR data respectively (see Sect. \ref{Sanchillary}). We derived the bolometric $L_{AGN} > 10^{44}$ erg/s (W) and $> 3\times 10^{44}$ erg/s (E) from the two lower limit $L_{2-10\ keV}$ of the two nuclei, considering a bolometric correction $K_X \sim 11.5$ (\citealt{Duras2020}). These estimates translate in the upper limit $\dot K/L_{AGN} < 0.007$ for the W nucleus, and  $\dot K/L_{AGN} < 0.02$ for the E nucleus. On the other hand, the average IR-based bolometric AGN luminosity, of the order of $\sim 1.7\times 10^{45}$ erg/s, allows us to obtain a $\dot K/L_{AGN} \sim 0.004$, and a momentum rate ratio $\dot P/(L_{AGN}/c) \sim 0.74$, consistent with other AGN-driven outflows at low-z (e.g. \citealt{Cicone2014, Fluetsch2019, Fluetsch2020}), as well as with theoretical predictions (e.g. \citealt{Harrison2018}).

Therefore, the inferred outflow energetics are overall consistent with both AGN- and SB-driven expectations, and cannot be used to infer the outflow origin, i.e. whether they are AGN-driven or starburst-driven. 
Finally, we also stress that the energetics should be considered as rough estimates because of the assumed simple wind model, which could not well describe the complex kinematics observed in Arp220. Moreover, further deeper multi-wavelength observations are required to better constrain the AGN bolometric luminosity, as well as the IR luminosity and SFR for the two individual nuclei (e.g. \citealt{Paggi2017,Dwek2020}).

\begin{table}
\footnotesize
\begin{minipage}[!h]{1\linewidth}
\setlength{\tabcolsep}{6pt}
\setlength{\extrarowheight}{4pt}
\centering
\caption{Outflow properties}
\begin{tabular}{lcccc}
&total  &CI & Bubble & Comma\\
\toprule
\multicolumn{5}{c}{ionised component}\\
\hline
$\dot M_{out}$ ($M_\odot/yr$) & $21$ &$[4.7-60]$ & $19$& $0.004$\\
$\dot K$ ($\times 10^{42}$ erg/s) & $2.3$ &$ [0.5-7]$ & $2.2$&$0.001$\\
$\dot P$ ($\times 10^{34}$ dyne) & $3.9$ &$ [0.9-9]$ & $3.5$&$0.002$\\ 
\hline
\multicolumn{5}{c}{neutral component}\\
\hline
$\dot M_{out}$ ($M_\odot$/yr)      & $27$ &$ [2-40]$ & $2$     &   $1$\\
$\dot K$ ($\times 10^{42}$ erg/s) & $4.2$ &$ [0.7-20] $       & $1.5$        &   $2.6$\\
$\dot P$ ($\times 10^{34}$ dyne) & $9.7$ &$ [0.9-15]$        & $0.2$        &   $0.4$\\ 
\hline
\multicolumn{5}{c}{neutral+ionised gas}\\
\hline
$\dot M_{out}$ ($M_\odot$/yr)      & $48$ &$ [7-100]$ & $21$     &   $1$\\
$\dot K$ ($\times 10^{42}$ erg/s) & $6.5$ &$ [1-27]$     & $3.7$       &       $2.6$\\
$\dot P$ ($\times 10^{34}$ dyne) &   $12.6$ &$ [2-24]$     & $3.7$       &         $0.4$\\ 
$\mu$ & $0.2$ &$ [0.03-0.4]$& --&--\\
$\dot K/\dot K_{SF}$ & $0.04$ &$ [0.006-0.15]$& --&--\\
$\dot P/(L_{SF}/c)$ & $21$ &$ [3-40]$ & -- &--\\
$\dot K/ L_{AGN}^{IR}$ & $0.004$ &$ [0.001-0.02]$ & --&--\\
$\dot P/(L_{AGN}^{IR}/c)$ & $0.74$ &$ [0.1-1.4]$ & -- &--\\
\hline
\toprule
\end{tabular}
\label{toutflow}
\end{minipage}
Notes:  
For the total energetics, we report in brackets the confidence intervals (CI) obtained considering the most and least conservative assumptions, as explained in the text. 
Measured AGN and SF bolometric luminosities from \citet{Nardini2010}. 

\end{table}
 
 \subsection{Negative feedback}\label{Snegative}

In general, if outflow velocities are high enough to escape the potential of the galaxy, then galactic outflows can effectively clear the galaxy of its gas content (``ejective feedback''; e.g. \citealt{Nelson2019}). However, this mechanism could be inefficient in Arp220, as a significant fraction of the outflowing material could collide with tidal streams and infalling gas. 

In such conditions, the ``preventive feedback'' (e.g. \citealt{Pillepich2018,Cresci2018}) could represent a more efficient mechanism: 
in this scenario, the formation of new stars is limited by the injection of outflow energy, which heats the gas preventing it from cooling.  The ubiquitous detection of powerful, kpc-scale outflows in merging systems (eg. \citealt{Feruglio2015,Rupke2013,Rupke2015,Saito2018,Perna2019}) could suggest a significant impact of AGN and SB outflows already from the early phases of the formation of massive galaxies (see also e.g. \citealt{Arribas2014}).

\section{SF activity and evidence for positive feedback}\label{Ssf}

\begin{figure}[ht]
\centering
\includegraphics[width=8.cm,trim=0 0 0 0,clip]{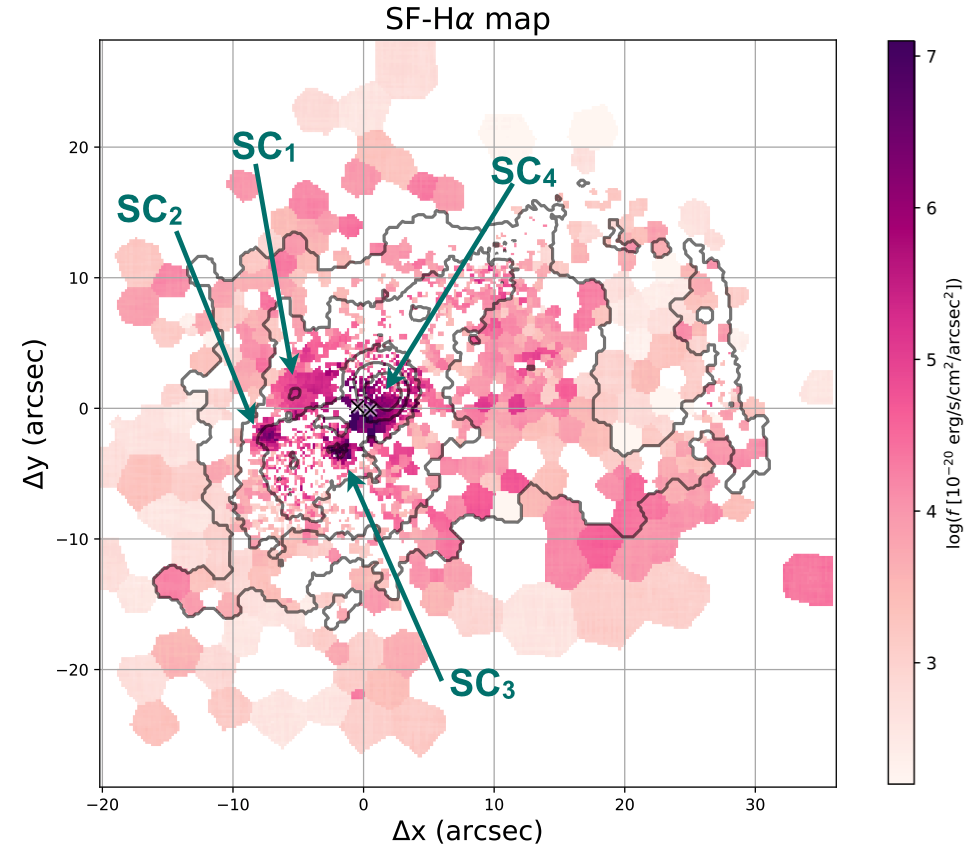}

\caption{\small  
Distribution of gas ionised by young and massive stars, traced by H$\alpha$ emission. All fluxes have been corrected for dust attenuation.  See Sect. \ref{Ssf} for details.}
\label{Hasf}
\end{figure}

In Fig. \ref{Hasf} we report the  distribution of gas ionised by young and massive stars, traced by H$\alpha$ emission. This map has been constructed on the basis of our multicomponent best-fit results, selecting the  Gaussian components associated with SF ionisation, according to at least two BPT diagnostics (Fig. \ref{BPT}). All fluxes in the figure have been corrected for dust attenuation, using the $E(B-V)_{gas}$ measurements and a CCM extinction law (Sect. \ref{Sextinction}). The map highlights the presence of at least four bright clumps, already identified in Sects. \ref{ionisationconditions} and \ref{Soutflow}, with diameters from $\sim 400$ pc (SC$_2$) to $\sim 700$ pc (SC$_4$), and associated with SFR from $\approx 0.1$ M$_\odot$/yr to $2.8 $ M$_\odot$/yr (using the \citealt{Kennicutt1998} relation). The main properties of these SCs,  reported in Table \ref{tclumps}, follow the expected empirical relationships between velocity dispersion, size and luminosity of stellar clumps in local and high-$z$ galaxies (see e.g. Fig. 8 in \citealt{Arribas2014}). 

The integrated spectra of the four clumps are reported in Appendix \ref{Ascs}. We fitted these spectra to derive the main properties of the atomic gas ionised by young stars, following the prescriptions presented in previous sections. In particular, we mention here that the N2-based (e.g. \citealt{Curti2017}) metallicities in the main stellar clumps are $12 + log(O/H) \gtrsim 8.6$, consistent with those of ULIRGs SF regions ($8.5 < 12 + log(O/H) < 8.9$, from \citealt{PereiraSantaella2017}). 
In summary, not only the BPT diagnostic diagrams suggest SF-like ionisation, but also the properties of these SCs (i.e size, velocity dispersion, luminosity, metallicity) correspond to those observed in other stellar clumps in local ULIRGs. 

The derived SFR associated with these clumps are  two order of magnitude lower than the one derived from $L_{IR}$. The integrated H$\alpha$ emission in Fig. \ref{Hasf} corresponds to a total SFR $\lesssim 10$ M$_\odot$/yr. This suggests the presence of additional highly obscured SF regions in the innermost nuclear part of Arp220. We stress that the detection of SF components in MUSE data is made even more difficult because of shocks, whose emission dominates over the SF ionisation in optical lines.

\subsection{Positive feedback: SF by outflow-induced pressure}

The stellar clumps $SC_2$ and $SC_4$ are located on the edge of the outflow regions (see star symbols in Fig. \ref{RGBoutflow}); moreover, the gas velocities within these SCs  significantly deviate from those of the main stellar component. Similar spatial configuration between outflowing gas and SF regions has been already reported in other systems, both at low-$z$ (e.g. \citealt{Cresci2015b,Maiolino2017,Shin2019,Cicone2020}) and at $z > 1$ (e.g. \citealt{Cresci2015,Carniani2016}; but see also \citealt{Scholtz2020}). These indications have been interpreted as possible evidence for SF triggered by outflows (e.g. \citealt{Cresci2018}): in this scenario, the outflows compress the ISM at its edges, enhancing the formation of new stars with high velocities. 

\subsection{Positive feedback: SF within the outflow}

A different scenario can be proposed for $SC_3$. This clump presents the most extreme gas velocities ($V\approx -230$ km/s), clearly decoupled from the global stellar component (with velocities in the range $V_* \in  [-130,+130]$ km/s over the entire MUSE FOV). Moreover, unlike $SC_2$ and $SC_4$, this clump is well within the outflow region. 
We also note that, when SF emission is split in velocity channels (Figure  \ref{niisfshocks}, lower panels), in addition to the SCs, the presence of faint and clumpy SF is revealed. This emission is preferentially found along the outflow direction (i.e. PA $\sim 138^\circ$) for the most extreme velocities, showing additional evidence that SF may be associated to the outflow. Its detection, however, have lower significance than in $SC_3$, as it is strongly dependent on our best-fit analysis: stronger degeneracies between individual kinematic components might be present where the line profiles are more complex.

Our results are consistent with those reported for other nearby sources (\citealt{Maiolino2017,Gallagher2019,Rodriguez2019}), and suggest that AGN and SB winds can ignite SF within the outflow itself, consistent with predictions from models and recent numerical simulations (e.g. \citealt{Ishibashi2013,Zubovas2013,Zubovas2014,Decataldo2019,Yu2020}). 
Quantitatively, this represents only a small amount ($\sim  2$\%) of the SFR  inferred from the total $L_{IR}$, slightly at odds with the results by \citet{Gallagher2019}, who detected widespread SF inside MaNGA galaxies outflows with rates between 5\% and 30\%  to the total SFR. We argue that this mild inconsistency could be  explained considering the difficulties in detecting SF ionisation in Arp220, due to the ubiquitous presence of shocks, whose emission dominates in the emission line profiles.

\begin{table*}
\footnotesize
\begin{minipage}[!h]{1\linewidth}
\setlength{\tabcolsep}{4pt}
\setlength{\extrarowheight}{4pt}
\centering
\caption{SF clumps properties}
\begin{tabular}{l|cccccccccc}
 &$r_x \times r_y$ &SFR & $V_{H\alpha}$ & $V_*$ & $\sigma_{H\alpha}$ & $\sigma_*$ & $E(B-V)_{gas}$ & 12+log(O/H)$_{C17}$ &log $U_{D16}$ &  $N_e$\\
              & & (M$_\odot$/yr) & (km/s) & (km/s) & (km/s) & (km/s) & & & &  (cm$^{-3}$) \\
\toprule
$SC_1$ &  $1.6''\times 1.5''$ &0.1& $+127\pm 4$ & $+77\pm 7$  & $62\pm 2$ & $120\pm 13$& $0.77\pm 0.03$ & $8.8\pm 0.1$   &$-3.1\pm 0.2$ &  $60 \pm 20$\\
$SC_2$ &  $1.1''\times 1.1''$ &0.2& $+130\pm 8$ & $-19\pm 8$ & $65\pm 5$ & $90\pm 31$  & $1.60\pm 0.10$ &  $8.7\pm 0.2$  & $-3.0\pm 0.2$ & $55\pm 10$\\
$SC_3$ &  $1.7''\times 1.3''$ &0.5& $-232\pm 4$  & $+32\pm 7$& $70\pm 6$ & $130\pm 10$ & $1.29\pm 0.03$ & $8.6\pm 0.2$   & $-3.7\pm 0.2$ & $120_{-70}^{+20}$\\
$SC_4$ & $2.3''\times 1.7''$  &2.8& $-65\pm   4$  & $-9\pm 7$   & $90\pm 3$ & $175\pm 10$ & $1.17\pm 0.02$ & $8.8\pm 0.1$   & $-3.1\pm 0.2$ &  $240\pm 20$\\ 
\hline
\toprule
\end{tabular}
\label{tclumps}
\end{minipage}
Notes: $r_x$ and $r_y$ are derived from 2D Gaussian fits.  Stellar velocities and velocity dispersions have been obtained from pPXF analysis, considering the integrated spectra over individual SCs (Appendix \ref{Ascs}). H$\alpha$ velocity and velocity dispersion, colour excess $E(B-V)_{gas}$, metallicity and log $U_{D16}$ have been derived from the spectral fit analysis of the integrated spectra, for the only kinematic component associated with SF ionisation. Stellar and gas velocity dispersions are not corrected for the instrumental broadening. Log $U_{D16}$ refers to the ionisation parameters derived following the approach described in \citet{Dopita2016}; the ionisation parameters derived from [S {\small III}]/[S {\small II}] ratios are consistent with those reported in the table, within a few $\sim 0.1$ dex. 
The metallicities have been derived following \citet{Curti2017}, from N2 line ratios; slightly higher values (still consistent within uncertainties) can be derived following \citet{Dopita2016}. 
\end{table*}

\subsection{Evidence for in-situ SF within the SCs}

One potential concern of the positive feedback scenario could be that 
the  main clumps of SF-ionised gas have been isolated on the basis of BPT line ratios. Such diagnostics do not allow a direct evidence of the presence of young stars within the SCs: stars in the galaxy disk can potentially ionise their gas (i.e. from the outside), and produce the line ratios observed in the SCs  integrated spectra (Fig. \ref{aSCspectra}). Therefore, we need to discriminate between in-situ and external photoionisation for individual clumps.

In the scenario of in-situ SF, one would expect that the ionising flux in SCs would be similar to those in standard SF regions, while the electron density $N_e$ would be similar of even higher than those in the unperturbed ISM (e.g. Sect. 3.2 in \citealt{Gallagher2019}). 
We derived the ionisation parameter for each clump, analysing their integrated spectra and isolating the Gaussian components associated with SF. 
The derived log $U$, reported in Table \ref{tclumps}, are consistent with those reported by \citet{Maiolino2017}, \citet{Gallagher2019} and \citet{Rodriguez2019} for stellar clumps in other AGN driven outflows, and standard SDSS star-forming regions.  
The electron densities in individual stellar clumps are compatible (or even higher) than the median $N_e$ measured across the MUSE FOV. Therefore, both log $U$ and $N_e$ are  consistent with the scenario of in-situ photoionisation of the gas.

All results so far reported are compatible with the finding of SF triggered by the outflow reported in the literature: although we cannot exclude that these SCs are simple star-forming regions in the interacting system, their peculiar kinematics and their location with respect to the outflowing gas support the scenario of positive feedback. 
Even if the SFR in the outflow is low compared with the global SFR in the whole system, one should take into account that stars formed inside the outflow have kinematics completely different from those formed in the galaxy discs, in the sense that they have highly radial orbits and therefore they can potentially contribute significantly to the formation and growth of the spheroidal component of the galaxy. More detailed investigation is however required to confirm the origin of the SCs in Arp220.

\section{Conclusions}\label{Sconclusions}

We have presented recent MUSE-AO observations of Arp220, a prototypical ULIRG and late-stage merger with dominant SF in the centre (\citealt{Nardini2010}) and kpc-scale warm gas emission in plumes and lobes (e.g. \citealt{McDowell2003,Arribas2001,Colina2004}). We produced high-resolution ($\sim 0.56''$, 210 pc) maps of stellar and gas kinematics, and studied the state of the ionised and neutral gas. 

The main results inferred from the modelling of the stellar kinematics, and the characterisation of the systemic  ISM (i.e. not perturbed by the outflow) are summarised as follows.
\begin{itemize}
\item [-] We observed a velocity gradient along north-east -- south-west direction (PA $\sim 48^\circ$) in the  stellar (gas) velocity maps, with amplitudes of $\approx \pm 100$ km/s ($\approx \pm 200$ km/s). However, gas and stars are still strongly disturbed and have not yet settled in a galactic plane. High-$v$ tidal structures at projected distances $>10''$ (3.7 kpc) are observed both in stellar kinematics (with velocity amplitudes up to $\pm 130$ km/s) and in ionised and neutral gas (up to $\pm 300$ km/s). 

\item [-] Spatially resolved BPT diagnostics have been used to locate SF regions. A significant fraction of stellar H$\alpha$ emission comes from four clumpy regions within the innermost nuclear regions ($r < 10''$); additional diffuse H$\alpha$ emission is found across the MUSE FOV. The total SFR inferred from stellar H$\alpha$ ($\lesssim 10$ M$_\odot$/yr) is one order of magnitude lower than the IR-based SFR $\sim 250$ M$_\odot$/yr. This suggests that most of the SF activity in Arp220 is highly obscured.

\item [-] We measured [S {\small II}]-based average electron densities of the order of $\approx 170$ cm$^{-3}$.  
This result suggests similar conditions in the ISM gas for local ULIRGs and high-$z$ star-forming galaxies.

\item [-] The $E(B-V)_{gas}$ and the equivalent width of Na {\small ID} absorbing gas can be used as a proxy of the hydrogen column density in Arp220, as suggested by the correlation between  $E(B-V)_{gas}$  and $EW(NaID_{abs})$. This correlation is observed in SF regions as well as in those with clear evidence of outflows; moreover, it is also consistent with the empirical relations by \citet{Turatto2003}, derived from ISM lines in SNe spectra.

\end{itemize}

The high resolution, wide field MUSE-AO observations have also allowed us to characterise in detail the  kpc- scale outflow in Arp220. Our main findings can be summarised as follows.

\begin{itemize}

\item [-] We revealed a close correspondence between the X-ray emission and the presence of extremely broad ionised and neutral gas line features along the south-east -- north-west direction (PA $\sim 138^\circ$), i.e along the `minor axis' of the disturbed  kpc-scale disk revealed in the stellar kinematics analysis.

\item [-] Tidally-induced shocks and outflows are the main responsible for ISM ionisation, as inferred from spatially resolved BPT diagnostic diagrams. This result was supplemented with comparisons of the measured line ratios and line widths with the predictions of shock models from {\rm MAPPING V}. These models suggest that diffuse gas is ionised by tidally-induced shocks with $V_s$ of few $100$ km/s, while the gas along PA $\sim 138^\circ$ is associated with velocities up to $\approx 1000$ km/s, reasonably due to SB- or AGN-driven outflows.

\item [-] We derived the outflow energetics assuming a simple mass conserving wind model: combining the atomic neutral and ionised gas components,  we obtained a mass rate $\dot M \sim 50$ M$_\odot$/yr, a kinetic power $\dot K \sim 10^{43}$ erg/s, a momentum power $\dot P \sim 10^{35}$ dyne and a mass-loading factor $\mu \sim 0.2$. These properties do not allow us to distinguish the origin of the outflows, i.e. whether they are SB- or AGN-driven.  Nevertheless, the inferred energetics suggest the outflow can strongly affect the evolution of the system, either through negative feedback, i.e. expelling the gas (but see Sect. \ref{Snegative}), and through preventive feedback, i.e. preventing the gas cooling and the formation of new stars.

\item [-] 
We reported evidence for positive feedback in Arp220: we locate two clumps of SF at the outflow edges ($SC_2$ and $SC_4$), with velocities clearly decoupled from the global stellar component, and SFR of 0.2 M$_\odot$/yr ($SC_2$) and 2.8 M$_\odot$/yr ($SC_4$). Our findings are consistent with the positive feedback scenario, according to which such clumps are forming from the compression of ISM at the outflow edges. Furthermore, we located an additional clump within the outflow, $SC_3$, with the most extreme gas velocities ($-230$ km/s from the Arp220 systemic), and SFR $\sim 0.5$ M$_\odot$/yr. Such peculiar properties suggest a different and even more fascinating scenario of positive feedback: the formation of $SC_3$ may have happened as the outflow material cooled down and fragmented leading to the formation of new stars within the outflow, as recently reported in other systems (e.g. \citealt{Maiolino2017}). Interestingly, as also reported by these authors, even if the SF in these clumps is small relative to the global star formation, their peculiar kinematics can potentially contribute significantly to the formation of the spheroidal component of a galaxy (e.g. \citealt{Yu2020}).

\end{itemize}

Our analysis showed that detailed multi-phase studies are required to  characterise the outflows. 
Our findings emphasise the capabilities of MUSE, which allow the simultaneous characterisation of both neutral and ionised gas in nearby galaxies. However, for a comprehensive understanding of the outflow phenomena all different gas phases  must be  carefully investigated. 
So far, no indication of large-scale molecular outflows has been reported. This could be due to the fact that high-$v$ molecular emission is usually more than ten times fainter than that from non-outflowing gas, and large integrations are needed even with the most sensitive millimetre/sub-millimeter interferometers (e.g. \citealt{Cicone2018}). 

Nuclear molecular outflows have been detected in both Arp220 nuclei.  
Hence, we tried to combine the information from MUSE-AO data analysis with the high-resolution ($\sim 0.1''$) ALMA observations tracing molecular gas emission in the surrounding of the E and W nuclei (e.g. \citealt{BarcosMunoz2016,BarcosMunoz2019, Wheeler2020}). We suggest that the galactic scale atomic outflow is emerging from the E nucleus of Arp220, taking into account simple geometrical arguments. A direct evidence for the location of the kpc-scale outflow, however, requires dedicated follow-up observations.   
JWST/NIRSpec IFS capabilities  will allow a comprehensive characterisation of the ionised and warm molecular (e.g. H$_2 1-0$) phases of the ISM, covering the spectral range from 0.6 to 5.3 $\mu$m, with a sub-arcsec resolution and a sampling of $0.1''$, hence testing our proposed scenario for the location of the origin of the kpc-scale atomic outflow. Furthermore, JWST/MIRI observations will be key to find direct evidence of dust-obscured AGN in the two nuclei.

\vspace{2cm}

{\small 
{\it Acknowledgments:}

We thank the referee for an expert review of our paper.
MP thanks A. Pensabene, M. Mingozzi, G. Venturi, S. Quai, G. Vietri, A. Puglisi, L. Costantin, D. Baron,  G. Cresci, M. Brusa and R. Marques-Chaves for fruitful discussions regarding different aspects presented in this manuscript. MP is supported by the Programa Atracci\'on de Talento de la Comunidad de Madrid via grant 2018-T2/TIC-11715. MP, SA, CTC and LC acknowledge support from the Spanish Ministerio de Econom\'ia y Competitividad through the grant ESP2017-83197-P, and PID2019-106280GB-I00. MPS acknowledges support from the Comunidad de Madrid through the Atracci\'on de Talento Investigador Grant 2018-T1/TIC-11035. SC acknowledges financial support from the State Agency for Research of the Spanish MCIU through the ``Center of Excellence Severo Ochoa'' award to the Instituto de Astrofísica de Andalucía (SEV-2017-0709). AF and RM acknowledge ERC Advanced Grant 695671 ``QUENCH'' and support by the Science and Technology Facilities Council (STFC). EB acknowledges the support from Comunidad de Madrid through the Atracci\'on de Talento grant 2017-T1/TIC-5213. JPL acknowledges financial support by the Spanish MICINN under grant AYA2017-85170-R. 

}


\begin{appendix}

\onecolumn

\section{Background galaxies and AGN}\label{Abkg}

\begin{figure*}[h]
\centering
\includegraphics[width=17.3cm,trim=0 0 0 0,clip]{{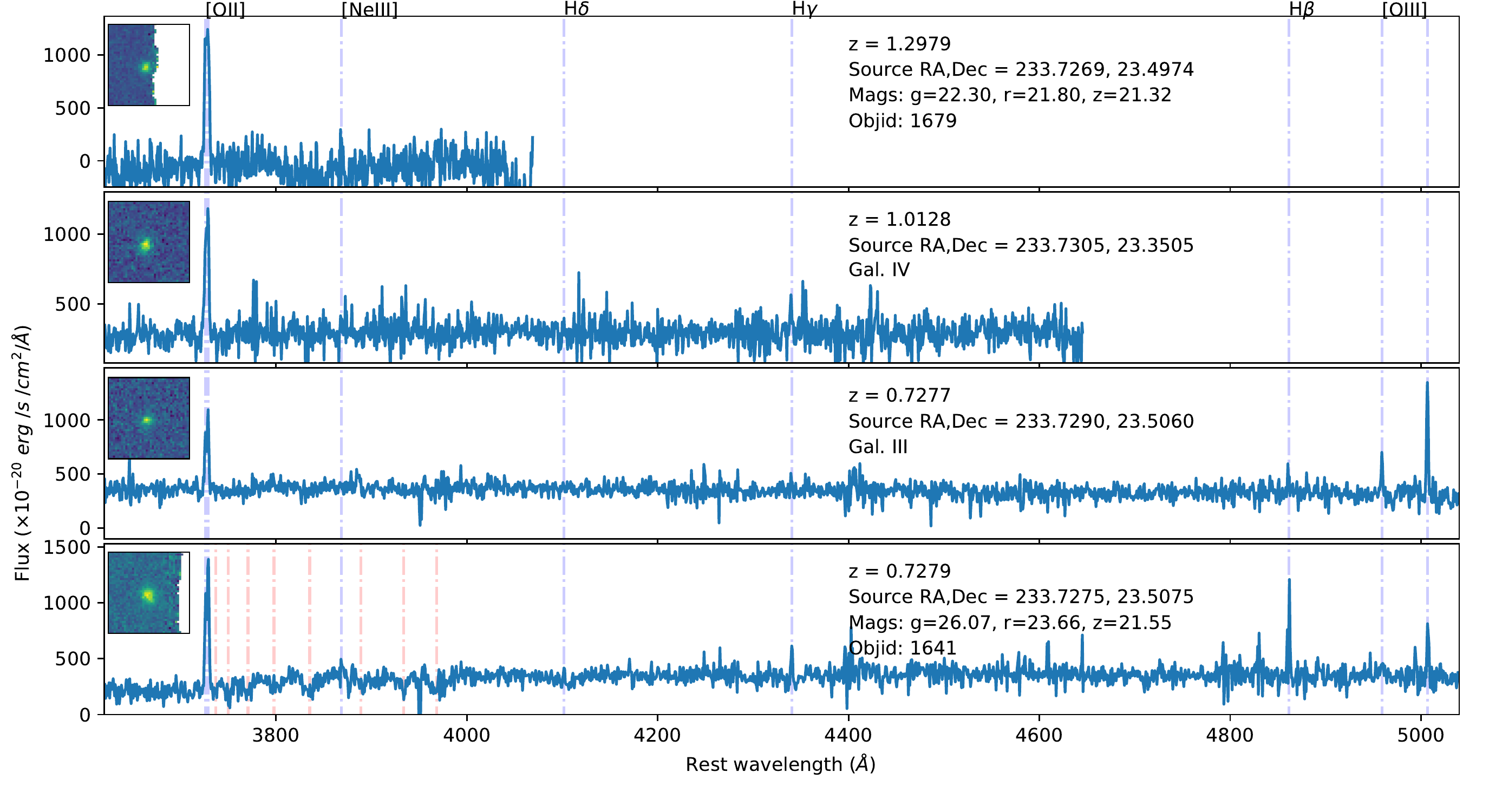}}
\caption{\small Spectra extracted from circular apertures ($r = 0.6''$) centred on the sources shown in the insets. The vertical blue lines mark the emission lines; in the last panel, Balmer and CaII absorption features around 3800$\AA$ are marked with red lines. For each target, we report the spectroscopic redshift, the coordinates and, for the sources in the  DECaLS Survey (\citealt{Dey2018}), the optical magnitudes. The insets show the [OII] emission maps (with a FOV of $8''$$\times8''$). 
}
\label{aFIG1}
\end{figure*}

\begin{figure*}[h]
\centering
\includegraphics[width=17.3cm,trim=0 0 0 0,clip]{{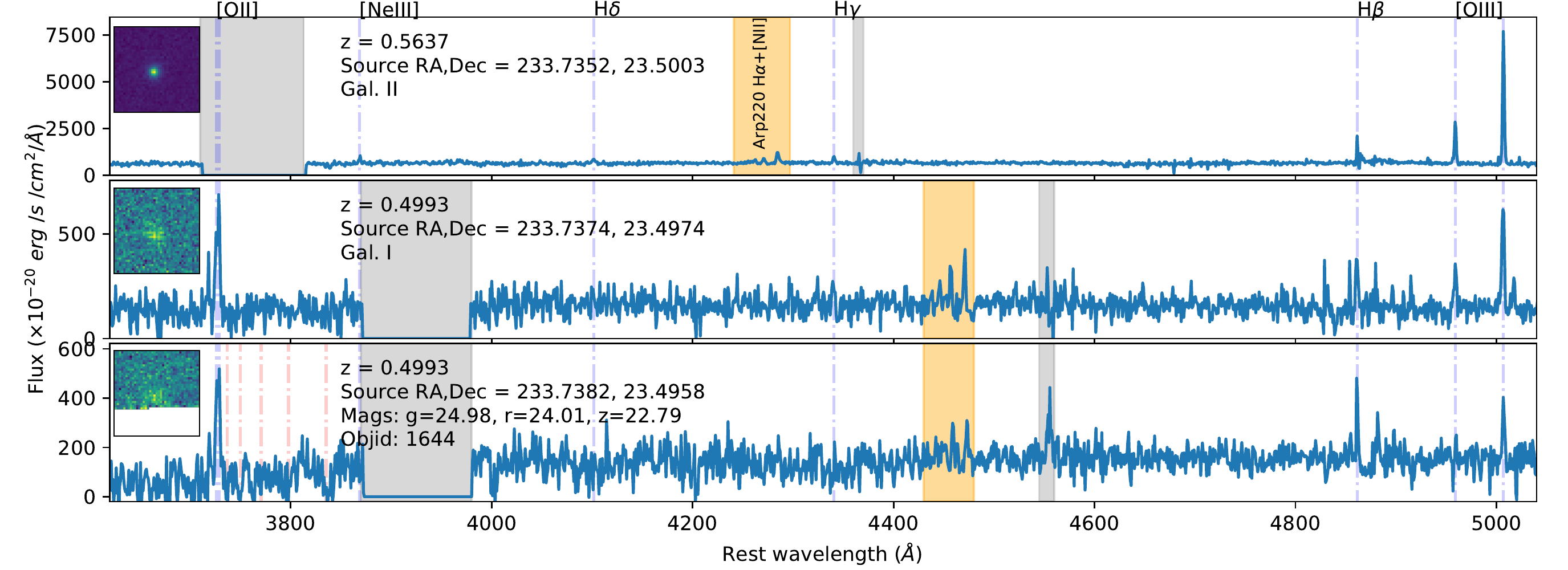}}
\caption{\small Spectra extracted from circular apertures  ($r = 0.6''$)  centred on the sources shown in the top left insets. The orange shaded regions highlight the H$\alpha$+[N {\small II}] system associated with Arp220; the grey regions mark the channels with strong contamination caused by Na Lasers. 
The vertical blue lines mark the brightest emission lines; in the bottom panel, Balmer and CaII absorption features around 3800$\AA$ are marked with red lines. For each source, we report the spectroscopic redshift, the coordinates and, for the sources in the  DECaLS Survey (\citealt{Dey2018}), the optical magnitudes. The insets show the [O {\small III}] (for the first source) and [OII] emission maps for each source. 
}
\label{aFIG2}
\end{figure*}

\begin{figure*}[h]
\centering
\includegraphics[width=17.3cm,trim=0 70 0 60,clip]{{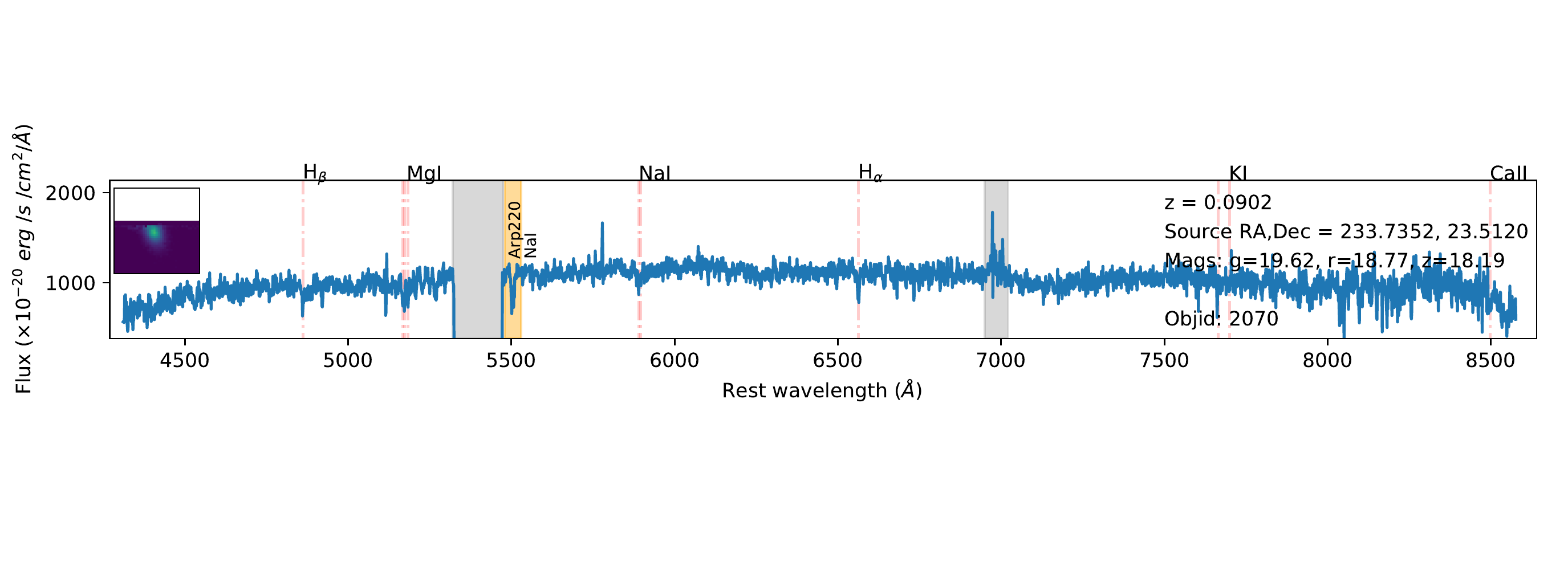}}
\caption{\small Spectrum extracted from a circular aperture with  ($r = 0.6''$)  centred on the source shown in the top left insets. Stellar and ISM absorption features are indicated with red lines. The orange shaded region highlight the Na {\small ID} system associated with Arp220; the grey regions mark the channels with strong contamination caused by Na Lasers.  The inset shows the continuum emission at $\approx 5300\AA$. 
}
\label{aFIG3}
\end{figure*}

In this section, we report the discovery of 8 background galaxies in the MUSE FOV, with spectroscopic redshifts ranging from 0.09 to 1.3 (see Table \ref{tab1}). Four of them are already present in the DECaLS Survey DR7 catalogue (\citealt{Dey2018}); in this work, therefore, these sources are identified  with DECaLS object numbers. The remaining sources are labeled as 'Gal. \#', with \# from I to IV.   
In Figs. \ref{aFIG1}, \ref{aFIG2}, \ref{aFIG3} we present the spectra extracted from circular apertures of radius 0.6$''$. We refer the reader to \citet{Ohyama1999} and \citet{Arp2001} for optical and X-ray detection of further (brighter) sources around Arp220 on larger spatial scales. 

The sources at highest redshifts ($z \approx 1$) are shown in the first figure. They can be clearly identified thanks to the bright [OII] doublet at 3729$\AA$, and a few additional fainter emission lines. For the source id1641 (bottom panel) we also identified several stellar absorption features. For id1679 (top panel), only a single emission line is detected; it shows a broad (FWHM $\sim 400$ km/s) and double peaked profile (with a separation of $\approx 200$ km/s). We also report a tentative detection of a faint feature at $\sim 8891\AA$ (with $\sim 2\sigma$ significance). Assuming that the strong feature is the [OII] doublet, the faint feature can be associated with [NeIII]$\lambda 3869$ emission. We therefore provided a tentative spectroscopic redshift $z = 1.2979$ for this target. 

Gal. III and ID 1641 ($3^{rd}$ and $4^{th}$ panels) are at the same redshift, $z \approx 0.7278$, and are separated by  $7.6''$ ($\sim 60$ kpc). The different [O {\small III}]/[OII] line ratios suggest a different degree of ionisation in the two systems, with Gal. III being associated with a higher ionisation state. This is also suggested by the different [O {\small III}]/H$\beta$ ratios.

In Fig. \ref{aFIG2} we report the spectra of the three sources at $z \approx 0.5$. For all of them, we clearly identified several emission lines.
Gal. II shows strong [O {\small III}] doublet lines, as well as faint H$\beta$, H$\gamma$, H$\delta$ and [NeIII]; the high [O {\small III}]/H$\beta$ flux ratio could suggest the presence of an AGN in this target. ID 1644 and Gal. I  are at the same redshift, $z = 0.4993$, and are separated by $\sim 6''$ ($\sim 35$ kpc). The [O {\small III}]/[OII] line ratios are consistent within the errors, and compatible with those of SDSS galaxies (\citealt{Kauffman2003b}); the [O {\small III}]/H$\beta$ ratio cannot be constrained, due to the presence of bad sky-subtraction residuals around $4861\AA$. 

In Fig. \ref{aFIG3} we report the spectrum of the nearest source, ID 2070, at $z = 0.0901$. It shows several stellar features in absorption, in addition to the Na {\small ID} absorption transitions related to foreground cold gas associated with the Arp220 system. Its spectroscopic redshift has been derived with pPXF, following the same prescriptions introduced in Sect. \ref{Sanalysis}. For all other targets, spectroscopic redshifts are derived fitting individual emission lines with Gaussian profiles, following the prescriptions presented in Sect. \ref{Sanalysis}. 


\clearpage

\section{Monte-Carlo analysis for measurements errors on $\sigma_*$}\label{Asigmastar}

\begin{figure*}[h]
\centering
\includegraphics[width=14.cm,trim=0 0 0 0,clip]{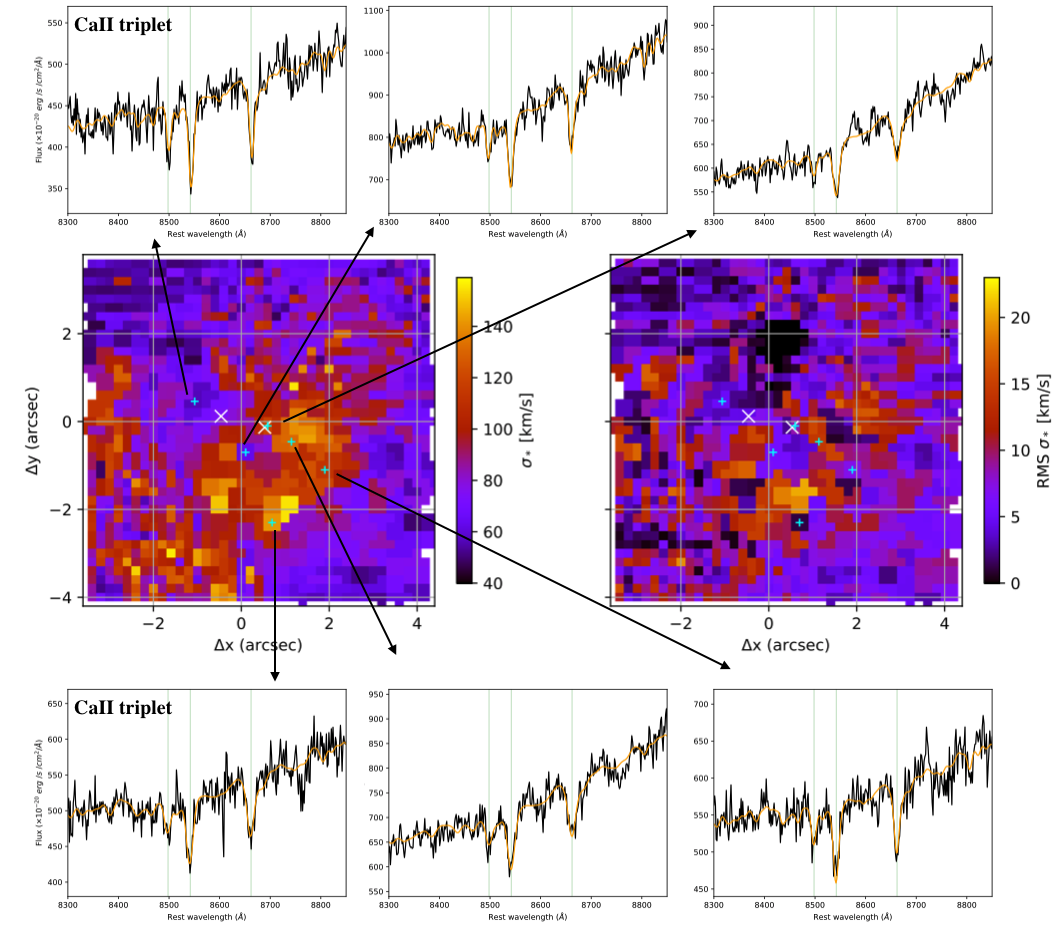}
\caption{\small Mean (left) and the standard error (right) of the $\sigma_*$ measurements obtained from MC trials, in the vicinity of the ring-like feature close to the two Arp220 nuclei. Spectra extracted from six different Voronoi bins (labeled with cyan $+$ symbols) are also reported in the insets: black (orange) curves indicate the spectra (pPXF best-fit models), while green vertical lines mark the position of the CaII triplet lines.
}
\label{MCsigmastar}
\end{figure*}

 Figure \ref{MCsigmastar} shows the mean (left) and the standard error (right) of the $\sigma_*$ measurements obtained from MC trials, in the vicinity of the ring-like feature close to the two nuclei. Spectra extracted from six different Voronoi bins are also reported in the insets, to show the quality of data and pPXF best-fit models. 

\section{Emission line decomposition}\label{Adecomposition}

\begin{figure*}[t]
\centering
\includegraphics[width=17.1 cm,trim= 0 0 0 0,clip]{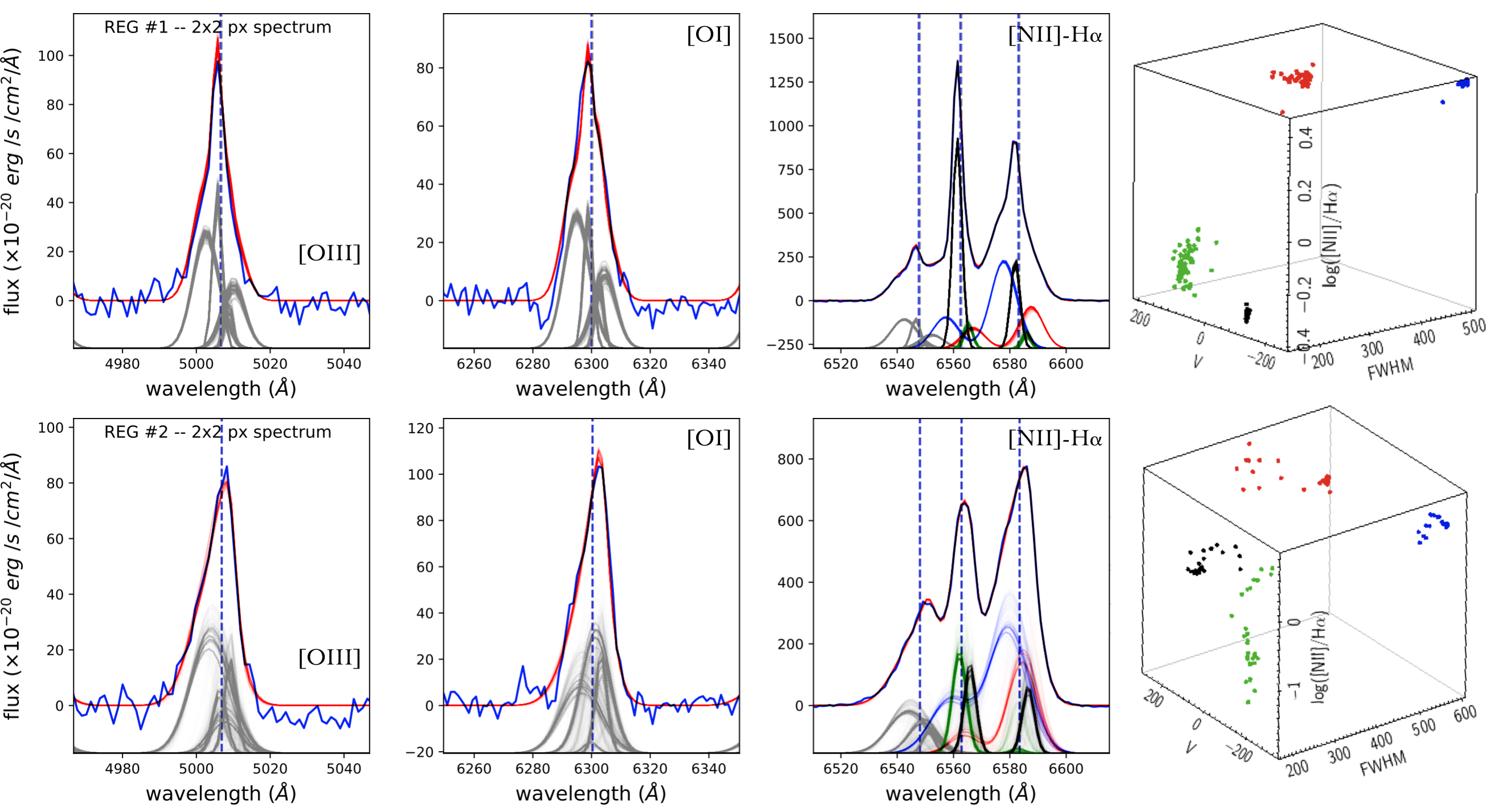}
\includegraphics[width=17.1 cm,trim= 0 0 0 0,clip]{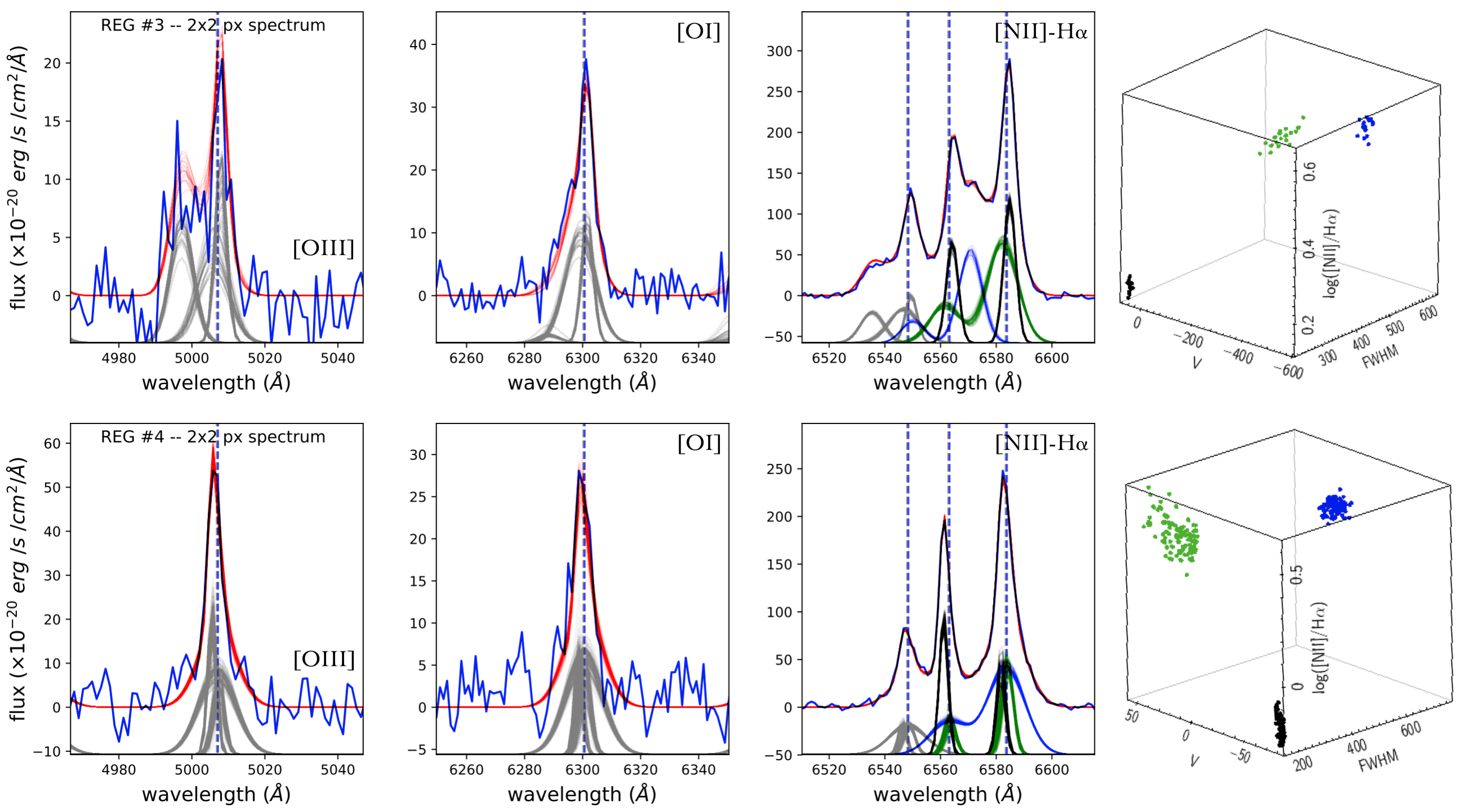}

\caption{\small  Illustration of the profile decomposition method. See Fig. \ref{decomposition} for details.}
\label{adecomposition}
\end{figure*}

\begin{figure*}[t]
\centering
\includegraphics[width=17.1 cm,trim= 0 0 0 0,clip]{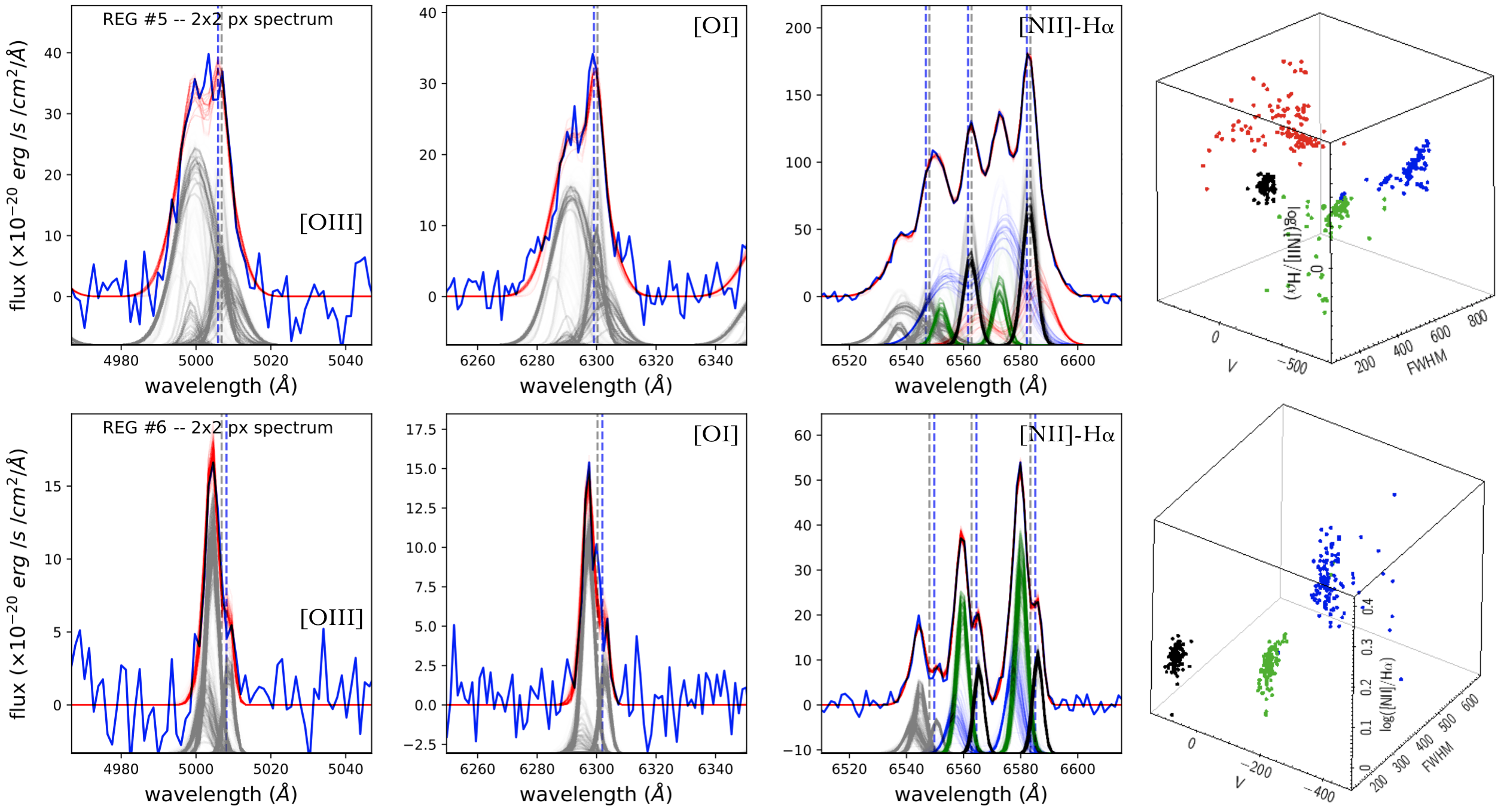}
\includegraphics[width=17.1 cm,trim= 0 0 0 0,clip]{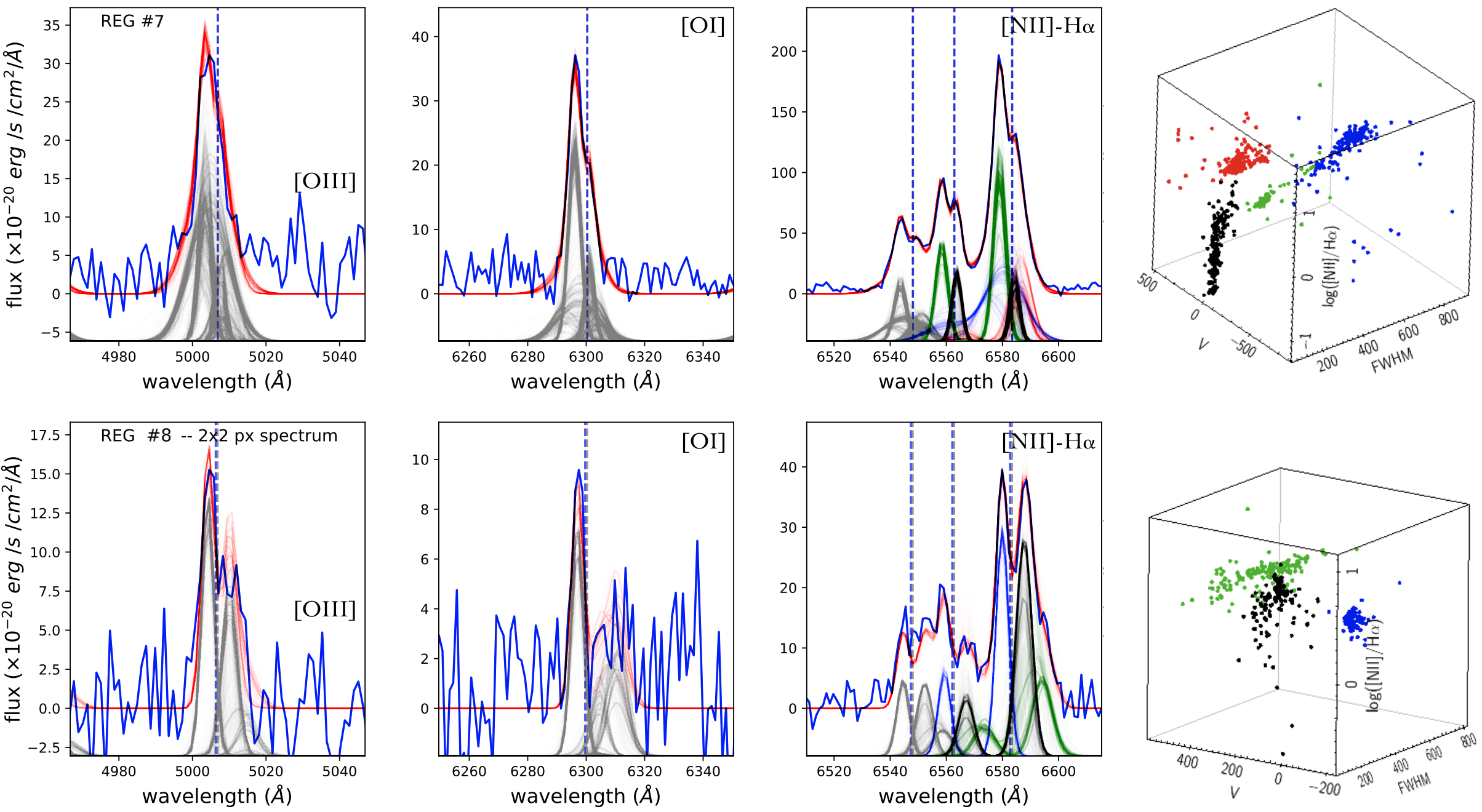}
\includegraphics[width=17.1 cm,trim= 0 0 0 0,clip]{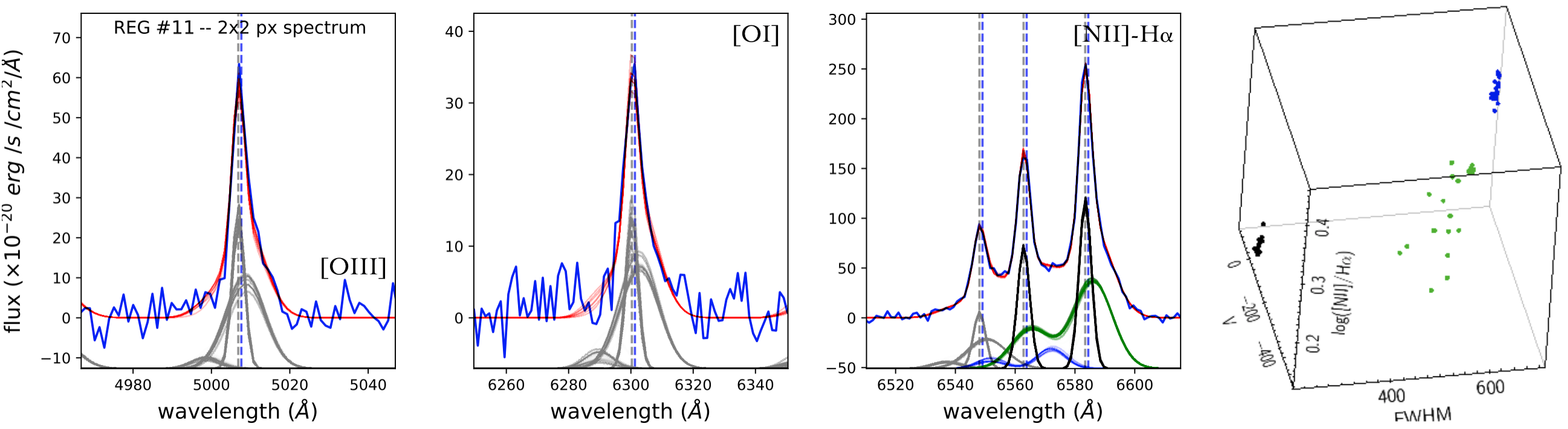}
\caption{\small  Illustration of the profile decomposition method. See Fig. \ref{decomposition} for details.}
\label{adecomposition2}
\end{figure*}

\begin{figure*}[t]
\centering
\includegraphics[width=8.1 cm,trim= 0 0 0 0,clip]{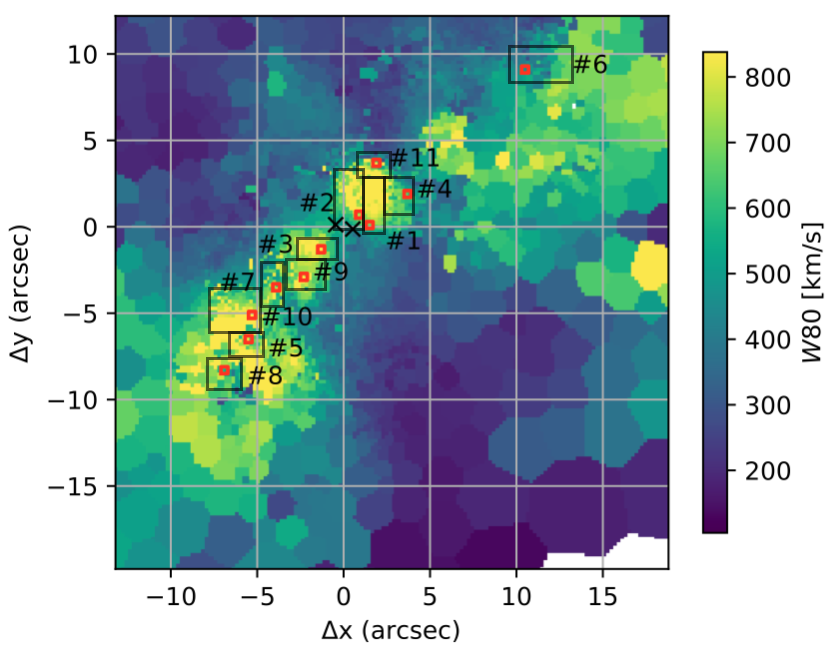}
\caption{\small  Arp220 $W80$ map showing, with red squares, the location of the  2x2 pixel spectra reported in Figs. \ref{adecomposition} and \ref{adecomposition2}. The spatial extent of the associated regions with well determined kinematics are indicatively represented with black squares. }
\label{w80decomposition}
\end{figure*}

Figures \ref{adecomposition} and \ref{adecomposition2} show the 2x2 pixel spectra with a clear distinction between different kinematic components in the emission line profiles. Their spatial location in the MUSE FOV is shown in Fig. \ref{w80decomposition}, together with the extend of the associated regions. 

\section{Balmer lines and [O {\small III}]$\lambda5007$ best-fit maps}\label{Alinemaps}

\begin{figure*}[t]
\centering
\includegraphics[width=17.1 cm,trim= 0 0 0 0,clip]{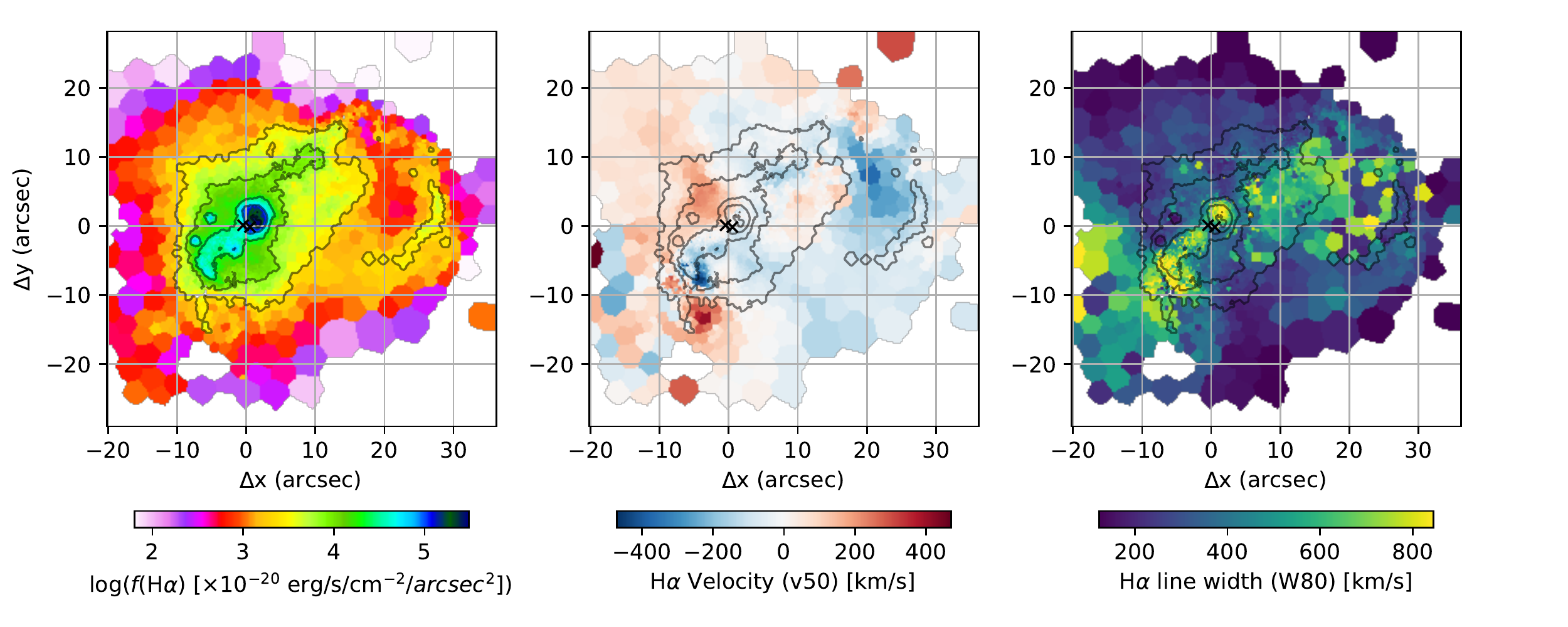}
\includegraphics[width=17.1 cm,trim= 0 0 0 0,clip]{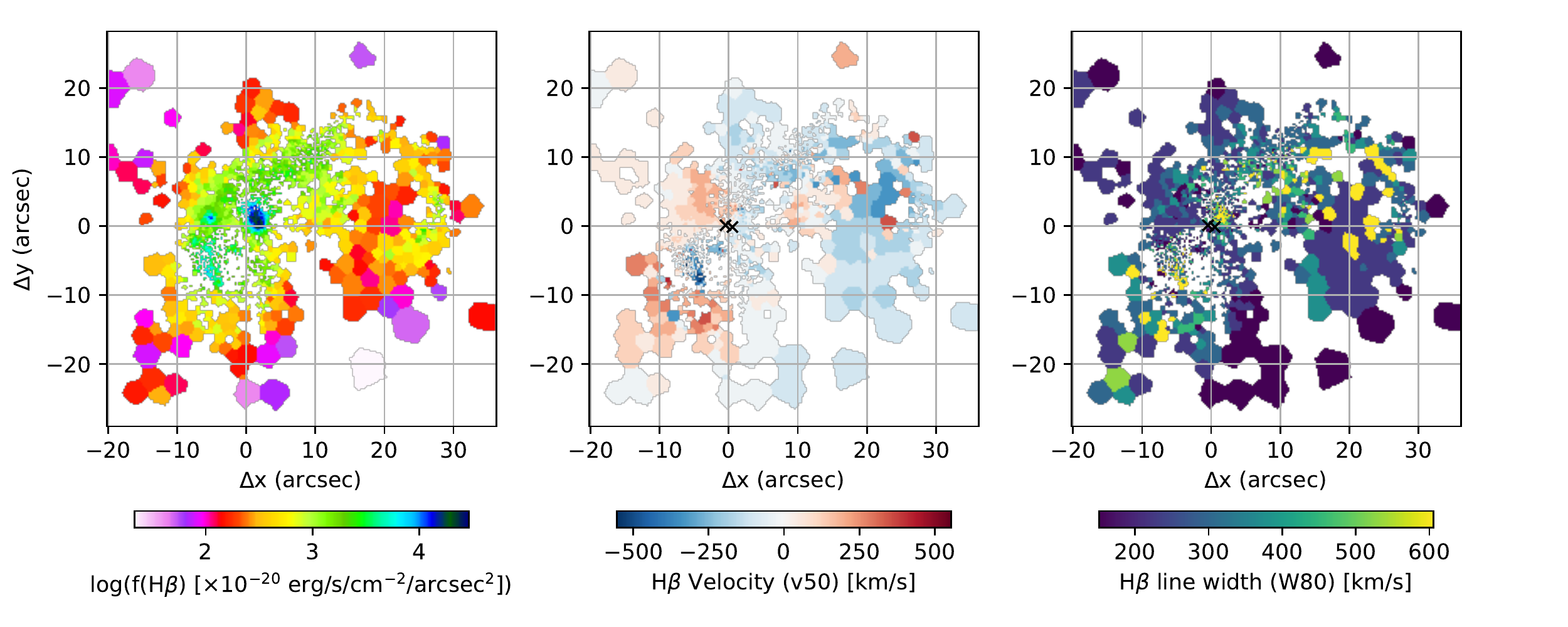}
\includegraphics[width=17.1 cm,trim= 0 0 0 0,clip]{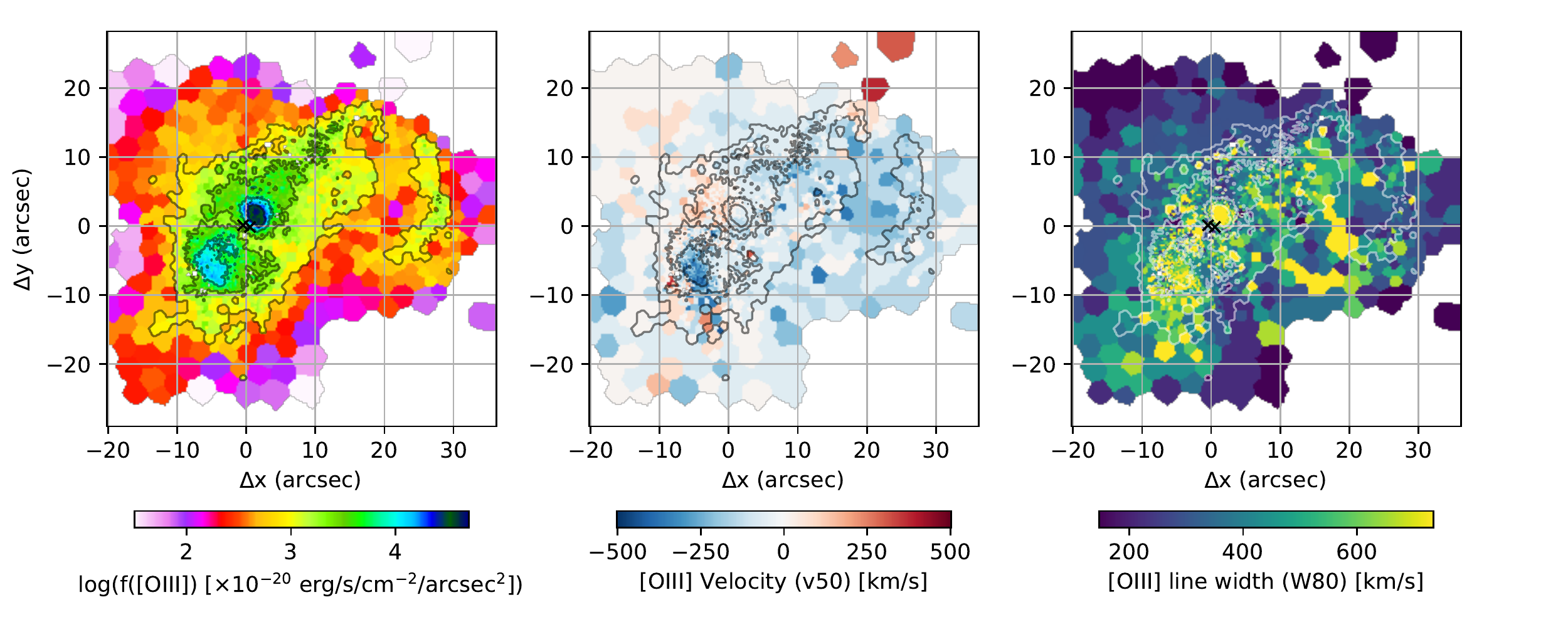}
\caption{\small  H$\alpha$ (top), H$\beta$ (centre) and [O {\small III}]$\lambda 5007$ (bottom) multi-component fit results. See Fig. \ref{NIIimages} for details. }
\label{hahboiiimaps}
\end{figure*}

In Fig. \ref{hahboiiimaps} we show the multicomponent fit results for the H$\beta$, [O {\small III}]$\lambda$5007, H$\alpha$, to be compared with the [N {\small II}]$\lambda6583$ maps in Fig. \ref{NIIimages}. The H$\alpha$ maps are very similar to those of [N {\small II}]. The main differences are found in proximity of the SCs at $\sim 7''$ from the Arp220 nuclei, associated with brighter H$\alpha$ fluxes and narrower Balmer $W80$. 
On the contrary, [O {\small III}] and H$\beta$ are noisier and their distributions fuzzy, because of the lower SNR; nonetheless, these faint features reveal the same kinematic structures observed in [N {\small II}] maps.

\section{Shock diagnostics and $\sigma_{[NII]}- V_s$ relation}\label{Ashock}

\begin{figure*}[h]
\centering
\includegraphics[width=17.cm,trim=0 00 0 0,clip]{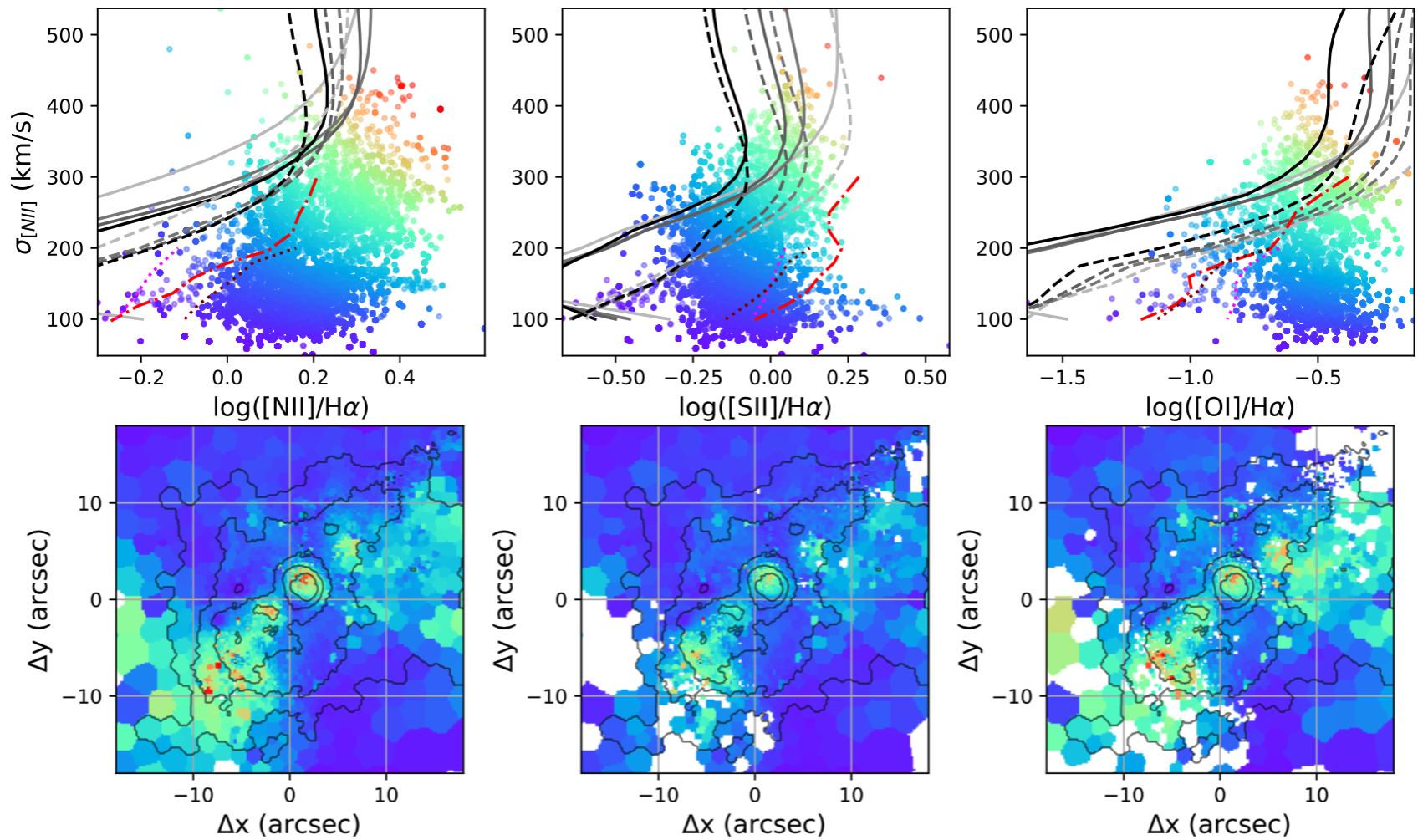}

\caption{\small  {\it Top panels:} $\sigma_{[NII]}$ against log([N {\small II}]/)H$\alpha$,  log([S {\small II}]/)H$\alpha$, and  log([O {\small I}]/)H$\alpha$ from left to right, obtained from the fitted total line profiles. The plotted measurements are colour-coded from purple-to-red going from low to high flux ratios and line widths. Dashed and solid lines represent shock model grids from  {\rm MAPPING V} (\citealt{Sutherland2017, Sutherland2018}; see Fig. \ref{w80lineratios} for details). We assumed a one-on-one correlation between $V_s$ and $\sigma$.  For comparison, we also display the shock models predictions derived by \citet[red dot-dashed lines]{Ho2014} , and by \citet[magenta and purple curves, considering a 80\% shock fraction]{Rich2011}; all these predictions, derived assuming a pre-shock density of 10 cm$^{-3}$ and different metallicities and magnetic field strengths (see Sect 6.2.1 in \citealt{Ho2014}, and Sect. 7.1 in \citealt{Rich2011} for details) do not match the majority of Arp220 measurements.  {\it Bottom panels:} Arp220 maps associated with the top panel diagrams, using the same colour-codes. }
\label{sigmalineratios}
\end{figure*}

Figure \ref{sigmalineratios} shows the correlation between velocity dispersion and emission line ratios, and the comparison with shock models predictions from MAPPING V, assuming a one-on-one relation between $\sigma_{[NII]}$ and $V_s$. With this assumption, shock models predictions well match the measurements in the [S {\small II}] and [O {\small I}] diagrams, at least for the regions with $\sigma_{[NII]} > 200$ km/s. However, they {\it i)} significantly under-predict [N {\small II}]/H$\alpha$ ratios by a factor of $\sim$ 0.2 dex, and {\it ii)} do not explain the presence of gas with velocity dispersion  $< 200$ km/s and [N {\small II}]/H$\alpha$, [S {\small II}]/H$\alpha$ and [O {\small I}]/H$\alpha$ emission line ratios not compatible with SF ionisation. Taking into account these two arguments, we considered the assumption about a one-on-one $W80-V_s$ relation as more reliable (see Fig. \ref{w80lineratios}). 

\section{Star-Formation clumps}\label{Ascs}

\begin{figure*}[t]
\centering
\includegraphics[width=17.1 cm,trim= 0 0 0 0,clip]{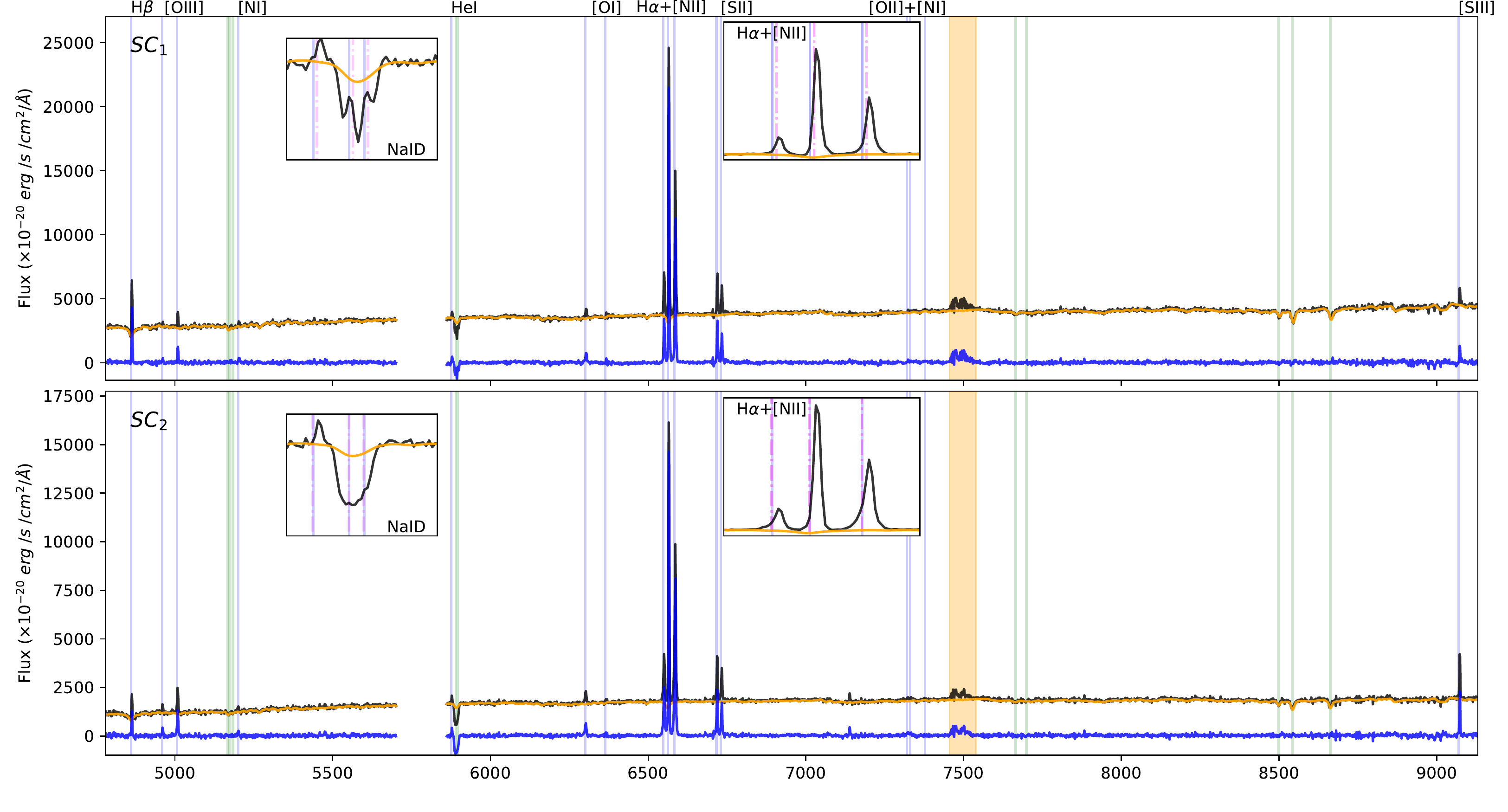}
\includegraphics[width=17.1 cm,trim= 0 0 0 0,clip]{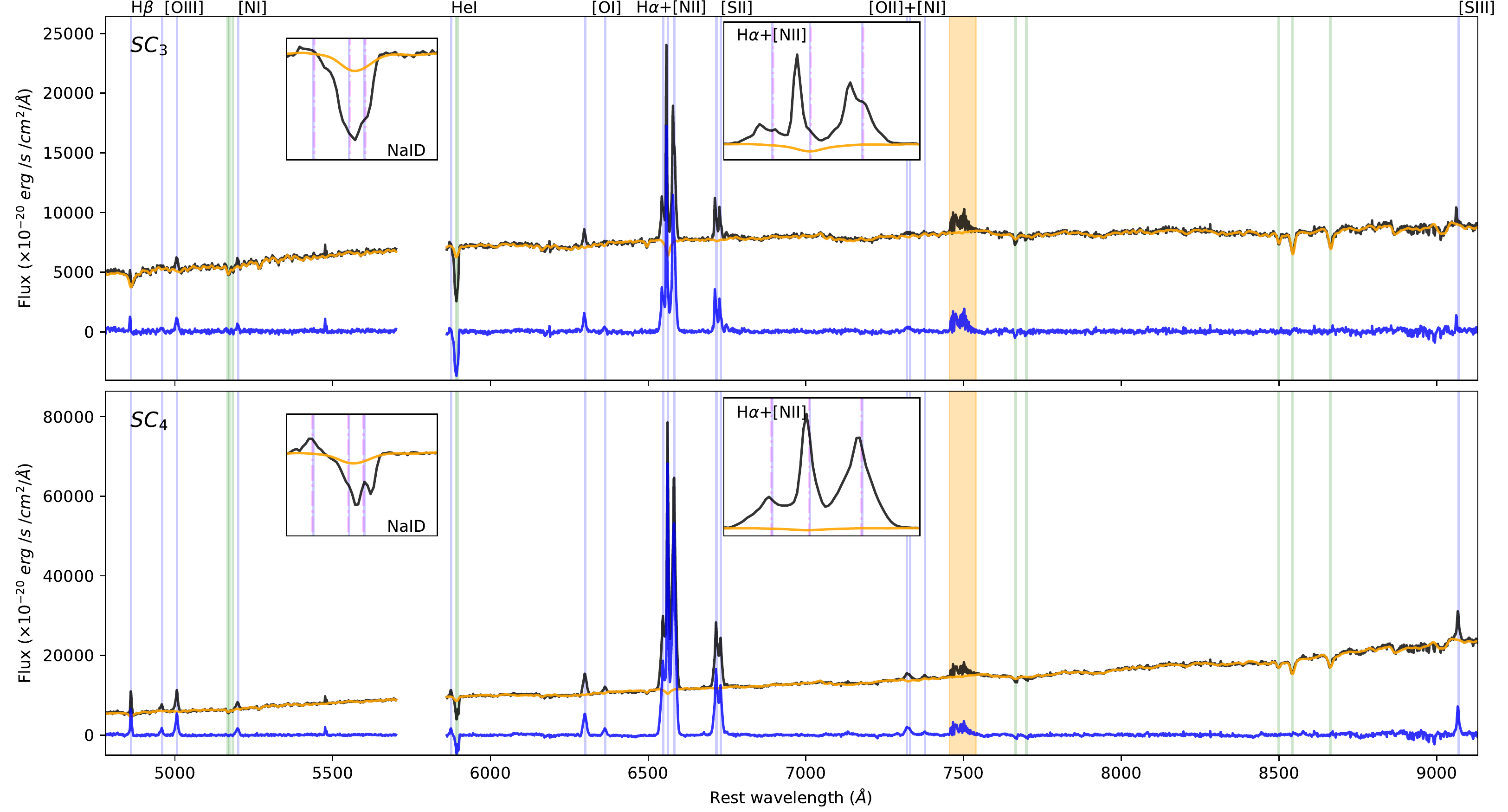}
\caption{\small  Integrated spectra of the four star forming clumps (SCs) identified in the MUSE FOV (black curves). The corresponding pPXF best-fit model profiles are shown with orange curves. The pure emission/absorption ISM spectra (blue curves) are obtained subtracting the best-fit stellar contribution from the original spectra. The insets show the spectra and stellar models around Na {\small ID} and the H$\alpha$+[N {\small II}] complex. The blue vertical lines mark the wavelengths of the emission lines detected in the spectra; the green lines mark the position of stellar absorption systems (i.e. from left to right: MgI triplet, Na {\small ID} and KI doublets, CaII triplet); in the two insets, dash-dotted magenta lines indicate the median stellar velocity in individual SCs. The region excluded from the pPXF fits and corresponding to the most intense sky line residuals are highlighted as orange shaded areas; the portion of the spectra around 5700$\AA$ is missing, because of a filter blocking the laser contamination.}
\label{aSCspectra}
\end{figure*}

In Fig. \ref{aSCspectra} we report the integrated spectra of the four star-forming clumps identified in the MUSE FOV.  All but $SC_1$ show strong [N {\small II}] emission and Na {\small ID} absorption with asymmetric and broad profiles. This finding is consistent with the fact that such SCs are very close to - or along - the outflow path (see e.g. Fig. \ref{RGBoutflow}). 

\end{appendix}

\end{document}